\PassOptionsToPackage{numbers,sort&compress}{natbib}

\documentclass[twocolumn]{bioRxiv}

\usepackage{lipsum}

\usepackage{multirow} % allows merging of cells in tables

\usepackage[symbol]{footmisc}  % For using symbols like † in footnotes
\setcitestyle{numbers,sort&compress}

\begin{document}

%% comment out any files that you don't want to include as sections
 \leadauthor{Chowdhary \& Lebert}

\title{Panoramic Voltage-Sensitive Optical Mapping of Contracting Hearts using Cooperative Multi-View Motion Tracking with 12 to 24 Cameras}
\shorttitle{3D Optical Mapping}

\author[1]{Shrey Chowdhary$^*$ \orcidlink {0009-0000-9401-7257}}
\author[1]{Jan Lebert$^*$ \orcidlink {0000-0001-8754-4964} }
\author[1,2]{Shai Dickman}
\author[1,2]{Charles Gordon}
\author[1,3,\Letter]{Jan Christoph \orcidlink {0000-0003-3594-4717}}
\affil[1]{Cardiovascular Research Institute, University of California, San Francisco, 555 Mission Bay Blvd S, San Francisco, CA 94158, USA}
\affil[2]{Department of Electrical Engineering and Computer Science, University of California, Berkeley, 253 Cory Hall, Berkeley, CA 94720-1770, USA}
\affil[3]{Division of Cardiology, University of California, San Francisco, 400 Parnassus Ave, San Francisco, CA 94143, USA}

\maketitle

\begin{abstract}
Voltage-sensitive fluorescence imaging is widely used to image action potential waves in the heart. 
However, while the electrical waves trigger mechanical contraction, imaging needs to be performed with pharmacologically contraction-inhibited hearts, limiting studies of the coupling between cardiac electrophysiology and tissue mechanics. 
Here, we introduce a high-resolution multi-camera optical mapping system with which we image action potential waves at high resolutions across the entire ventricular surface of the beating and strongly deforming heart. 
We imaged intact isolated rabbit hearts inside a soccer-ball shaped imaging chamber facilitating even illumination and panoramic imaging. 
Using 12 high-speed cameras, ratiometric voltage-sensitive imaging, and three-dimensional (3D) multi-view motion tracking, we reconstructed the entire 3D deforming ventricular surface and performed corresponding voltage-sensitive measurements during sinus rhythm, paced rhythm, and ventricular fibrillation. 
Our imaging setup defines a new state-of-the-art in the field and can be used to study the heart’s electromechanical physiology during health and disease at unprecedented resolutions. 
For instance, we measured electrical activation times and observed mechanical strain waves following electrical activation fronts during pacing, observed electromechanical vortices during ventricular fibrillation, and measured action potential duration and contractile changes in response to pharmacological blockage of potassium ion channels.
\end{abstract}


\begin{corrauthor}
jan.christoph\at ucsf.edu
\end{corrauthor}

\footnote{These authors contributed equally to this work.}

\section*{Introduction}\label{s:introduction}

The heartbeat is a highly dynamic process involving bioelectrical and biomechanical phenomena:
each cardiac muscle cell exhibits a sudden rapid change or depolarization of its transmembrane potential before it contracts.
The depolarization is the initial phase of the cardiac action potential, followed by a characteristic plateau phase during which calcium is released into the cell's contractile machinery producing contractile force, and a repolarization phase after which the cell relaxes.
On the organ scale, the action potential spreads from cell to cell and forms an electrical wave that propagates rapidly across the heart muscle, subsequently activating different portions of the heart to contract.
Understanding these intricate dynamics is crucial for understanding various forms of heart disease, and for developing better diagnostics and therapies to treat them.
While the normal heartbeat is characterized by a rapid depolarization occuring nearly everywhere at once via the Purkinje system, abnormal rhythms can be triggered by various electrical activation patterns, such as focal patterns originating from only one location during premature ventricular complexes, or rotating or {\it reentrant} patterns during tachyarrhythmias, such as atrial or ventricular fibrillation.
More generally, abnormalities in the electrical activation can cause contraction abnormalities and vice versa as both mutually influence each other. 
For instance, myocardial infarcts are not only associated with stiffer myocardial tissue and reduced contractility, but they can also alter electrical pathways and induce arrhythmias. 
More, mechanical stretch can influence the action potential via mechano-sensitive ion channels, which in turn affects mechanical contraction, even in healthy tissue.
The action potential shape and duration plays an important role in assessing the heartbeat not only because it controls the timing and duration of the heart's contraction via calcium, but also reflects the behavior of ion channels and the heart's metabolic state.
For instance, a long action potential in long-QT syndrome patients is associated with a delayed muscle relaxation, inefficient mechanical function, and an increased arrhythmia risk, due to changes in the function of ion channels affecting the repolarization phase.

\begin{figure*}[ht]
  \centering
  \includegraphics[clip, trim=0.0cm 0.0cm 0.0cm 0.0cm, width=0.98\textwidth]{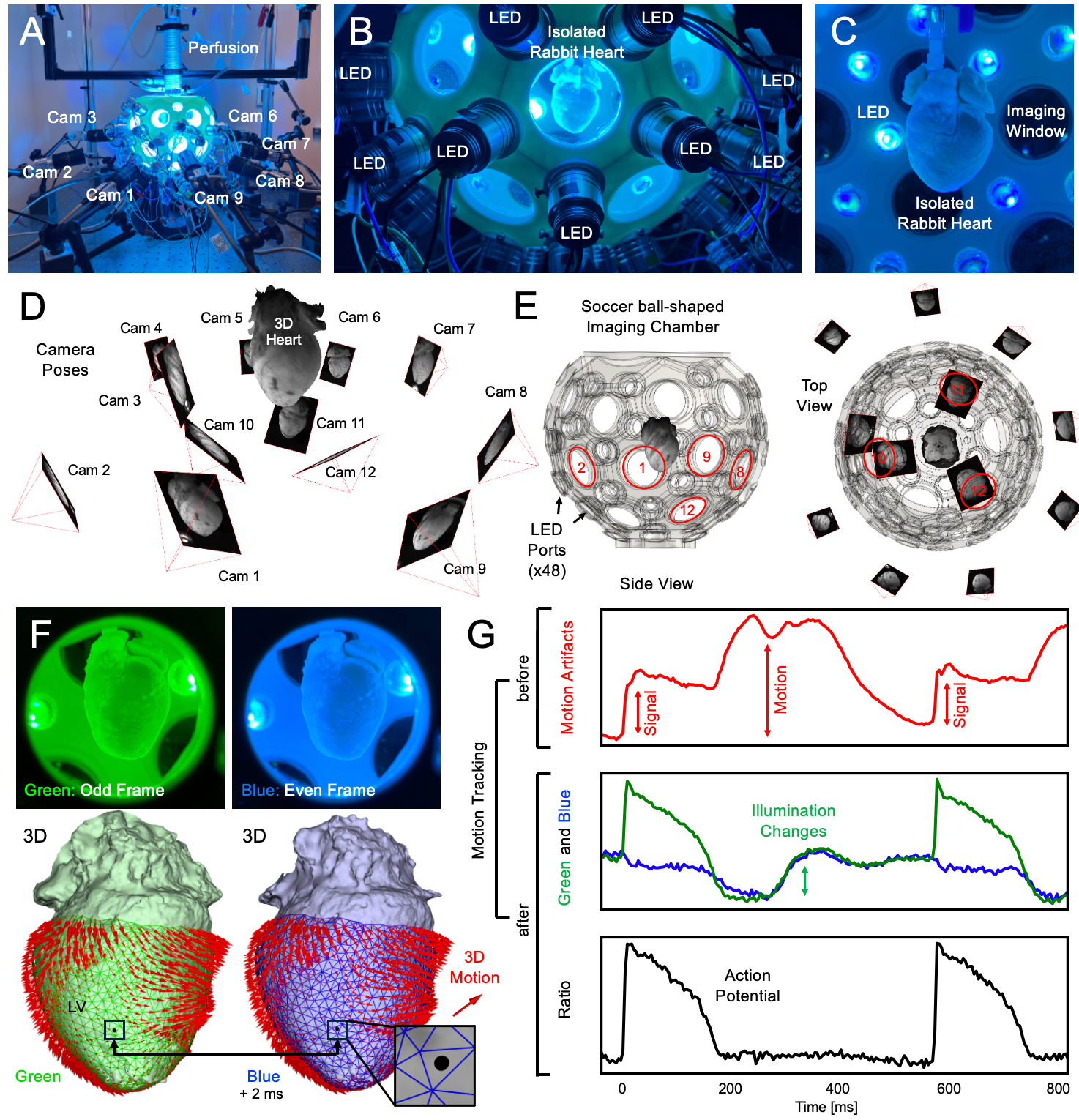}
  \caption{
Panoramic voltage-sensitive optical imaging of action potential waves across the entire surface of contracting and strongly deforming isolated hearts.
\textbf{A} Spherical 3D-printed soccer ball-shaped imaging chamber with 24 windows, 12 synchronized high-speed video cameras (acA720-520um, Basler) and constant-pressure Langendorff-perfusion system.
\textbf{B} View of isolated rabbit heart inside imaging chamber, see also Supplementary Video \ref{video:SV1}.
\textbf{C} Close-up of isolated rabbit heart immersed in Tyrode solution inside imaging chamber.
\textbf{D} Camera poses (position and orientation) and 3D reconstruction of rabbit heart (not to scale) obtained with 12 cameras, see also Supplementary Video \ref{video:SV2}. Imaging was performed at 500 fps with 8 mm and 12 mm lenses with shorter and longer working distances, respectively.
\textbf{E} Technical drawing (side \& top view) of imaging chamber together with camera poses (right). Horizontal imaging was performed through windows 1-9 and apical imaging through windows 10-12.
\textbf{F} Continuous green or pulsed green-blue excitation in odd (1,3,5, ...) and even (2,4,6,...) frames, respectively, with up to 48 light-emitting diodes (LEDs), see also Fig.~\ref{fig:SupplementRatiometry} for more details. In pulsed mode, we performed 2 separate 3D reconstructions using the green and blue video data and subsequently tracked the heart's motion (red vectors, frame-to-frame displacements) over time using the blue video data, see Methods for details and Supplementary Videos \ref{video:green-blue-video-3D} and \ref{video:green-blue-video-3D-texture}.
Hearts were stained with voltage-sensitive fluorescent dye (Di-4-ANEPPS). 
\textbf{G} Optical traces (normalized dimensionless units) obtained from a $3 \times 3$ pixel region at the center of the same triangle (black dot in panel F) before (top, green video) and after 3D motion tracking (center, green and blue videos), and after both 3D tracking and ratiometric compensation (bottom), see also Supplementary Video \ref{video:SV4}. While the green trace (obtained with green excitation) carries the signal, the blue trace is neutral and serves as a measurement for local illumination changes which arise due to relative motion between the tissue and the LEDs.
}
  \label{fig:Setup}
\end{figure*}

For decades, it has been a challenge to image these dynamics throughout the heart, mainly because it requires a high-resolution measurement approach capable of measuring electrophysiology and mechanics simultaneously. 
Optical techniques employing fluorescence-based measurements are an ideal high-resolution modality for this task. Nevertheless, measuring action potential waves at high resolutions in the beating heart has been a long-standing technical challenge in cardiovascular research \cite{Salama1987,Zhang2016,Christoph2017,Garrot2017,Christoph2018Nature,Christoph2018Frontiers,Kappadan2020,Lebert2022,Zhang2023,Kappadan2023}.
Electrical mapping techniques, such as multi-electrode arrays or catheter-based electrode mapping \cite{Nash2006,Narayan2024}, are viable options to map electrical impulse phenomena in the heart and are frequently used in the clinical setting. 
However, they are associated with low spatial resolution, require physical contact with the tissue, and generally do not allow simultaneous mechanical measurements of tissue strain. 
More recently, it became possible to measure electrical impulse phenomena and mechanical stretch simultaneously using flexible nanowire electrode arrays or stretchable patches \cite{Chen2023, Fullenkamp2024}. 
However, while these techniques integrate electrical and mechanical measurements and mitigate adverse effects caused by stiffer electrodes, they yet suffer from low spatial resolution.
Lastly, action potential waves can be measured contactless at high spatial resolutions across the surface of isolated hearts using voltage-sensitive fluorescence imaging or {\it optical mapping} \cite{Salama1987,Gray1995,Swift2021,Efimov2004,Kay2004,Kay2006,Rogers2007}.
With this technique, the main challenge is that the optical signals arising on the heart's surface are miniscule, and in order to be able to measure action potential waves, it is necessary to combine sensitive optical with precise numerical measurement techniques \cite{Christoph2018Frontiers,Lebert2022}.
Due to its sensitivity, optical mapping is routinely performed in pharmacologically contraction-inhibited hearts \cite{Gray1995,Swift2021,Efimov2004}. 
Voltage-sensitive optical mapping with multi-camera systems was performed to image action potential waves across the entire heart surface of motion-inhibited hearts \cite{Bray2000,Kay2004,Kay2006,Rogers2007}.
Optical mapping of action potential waves on the contracting heart surface is much more challenging than in non-contracting hearts, and has been demonstrated in only a small number of studies \cite{Zhang2016,Christoph2017,Garrot2017,Christoph2018Nature,Christoph2018Frontiers,Kappadan2020,Lebert2022,Zhang2023} and in even fewer studies when performed with multi-camera optical mapping reconstructing the three-dimensional (3D) deforming heart surface \cite{Zhang2016,Christoph2017,Zhang2023}.

In this study, we present the first fully panoramic measurements of action potential waves on the entire strongly contracting ventricular surface of isolated hearts.
The 3D measurement data was obtained with a multi-camera optical mapping system comprising 12 high-speed cameras, see Fig.~\ref{fig:Setup}. 
The measurements became possible due to the recent availability of low-cost cameras, a custom-designed spherical imaging chamber, pulsed illumination for ratiometric imaging, and a 3D multi-view motion tracking algorithm.
With the spherical, soccer ball-shaped imaging chamber it is possible to illuminate and image the heart evenly from all sides with up to 48 light-emitting diodes and up to 24 cameras simultaneously, respectively.
Using 12 cameras, we imaged action potential waves propagating across the entire contracting ventricular surface of isolated intact rabbit hearts during sinus rhythm, paced rhythms, and ventricular fibrillation.
We mapped the electrical activation and measured epicardial deformation and tissue strain at high spatial and temporal resolutions (spatial resolution electrical: 0.5-1.0 million pixels, \SI{120}{\micro\metre} per pixel, spatial resolution mechanical: $\sim$ 0.5 - \SI{3}{\milli \meter} per triangle, temporal resolution: 2ms).
Our imaging system is not only the first fully panoramic electromechanical optical mapping system covering the entire \SI{360}{\degree} heart surface, but it also provides unprecedented spatial resolutions and has several key improvements and practical advantages over previously reported 3D electromechanical optical mapping systems \cite{Zhang2016,Christoph2017}.
With this system, we imaged action potential waves and mechanical deformation simultaneously at high resolutions during regular and irregular rhythms in isolated rabbit hearts.
  
\subsection*{Technical Background}

Optical mapping is a highly sensitive measurement, which involves measuring small fractional changes in fluorescence exhibited by dyes or genetic indicators in response to a physiological change, such as a change in the transmembrane potential during the cardiac action potential. 
Motion in cardiac optical mapping studies is typically associated with so-called motion artifacts, which arise quickly even with the slightest motion \cite{Rohde2005, Svrcek2009, Christoph2018Frontiers}.
With the normal motion of the heart, and without numerical motion correction, motion artifacts completely compromise the measurements and prohibit further analysis of the data.
Numerical motion tracking can effectively circumvent motion artifacts \cite{Zhang2016,Christoph2017,Christoph2018Nature,Christoph2018Frontiers,Kappadan2020,Lebert2022,Khwaounjoo2015,Rodriguez2015}, and is particularly effective when further combined with ratiometric imaging \cite{Zhang2016,Kappadan2020}.
Nevertheless, eliminating motion artifacts and measuring artifact-free action potentials on the contracting heart surface is a non-trivial task and challenge that has yet to be solved.

Numerical motion tracking and 3D object reconstruction are routine tasks in the field of computer vision.
Yet, despite their widespread use in engineering and other fields, computer vision techniques are still not fully utilized in cardiac optical mapping studies, likely because they require particular technical knowledge which makes them challenging to adopt by cardiovascular researchers.
On the other hand, 3D optical mapping of contracting hearts is a more challenging problem than most other computer vision tasks, which may have also hindered computer vision experts from entering the field.
The technical challenges include: 
The tissue needs to be tracked precisely, at times with sub-pixel precision, to be able to measure the optical signals without motion artifacts \cite{Christoph2018Frontiers}. 
The optical signals are small and even slight tracking errors can lead to large measurement errors.
The fluorescence changes rapidly over time as it reports physiological function, which can irritate tracking algorithms and introduce tracking artifacts \cite{Christoph2018Frontiers,Lebert2022}.
Noise, inhomogeneous illumination and other factors can cause further tracking and measurement artifacts.
Overall, the video data comprises very particular image features determined by the fluorescent indicators, filters, and low-resolution cameras.
The spatial resolution of the videos is typically ranging in the order of $100 \times 100$ to $300 \times 300$ pixels.
Lastly, the heart's motion and deformation is large, with the heart potentially moving out of the field of view. 
Together, these factors make optical mapping of contracting hearts or tissues a uniquely challenging problem.
By default and without further customization, stereoscopic imaging or photogrammetry techniques, which are frequently used for shape reconstruction, do not work out-of-the-box and lack the precision required for physiological measurements. 
Accordingly, attempts to reconstruct and optically map the 3D surface of beating hearts were demonstrated in only a few studies thus far \cite{Zhang2016,Christoph2017,Zhang2023}.

\begin{figure*}[htb]
  \centering
  \includegraphics[clip, trim=0.0cm 0.0cm 0.0cm 0.0cm, width=0.99\textwidth]{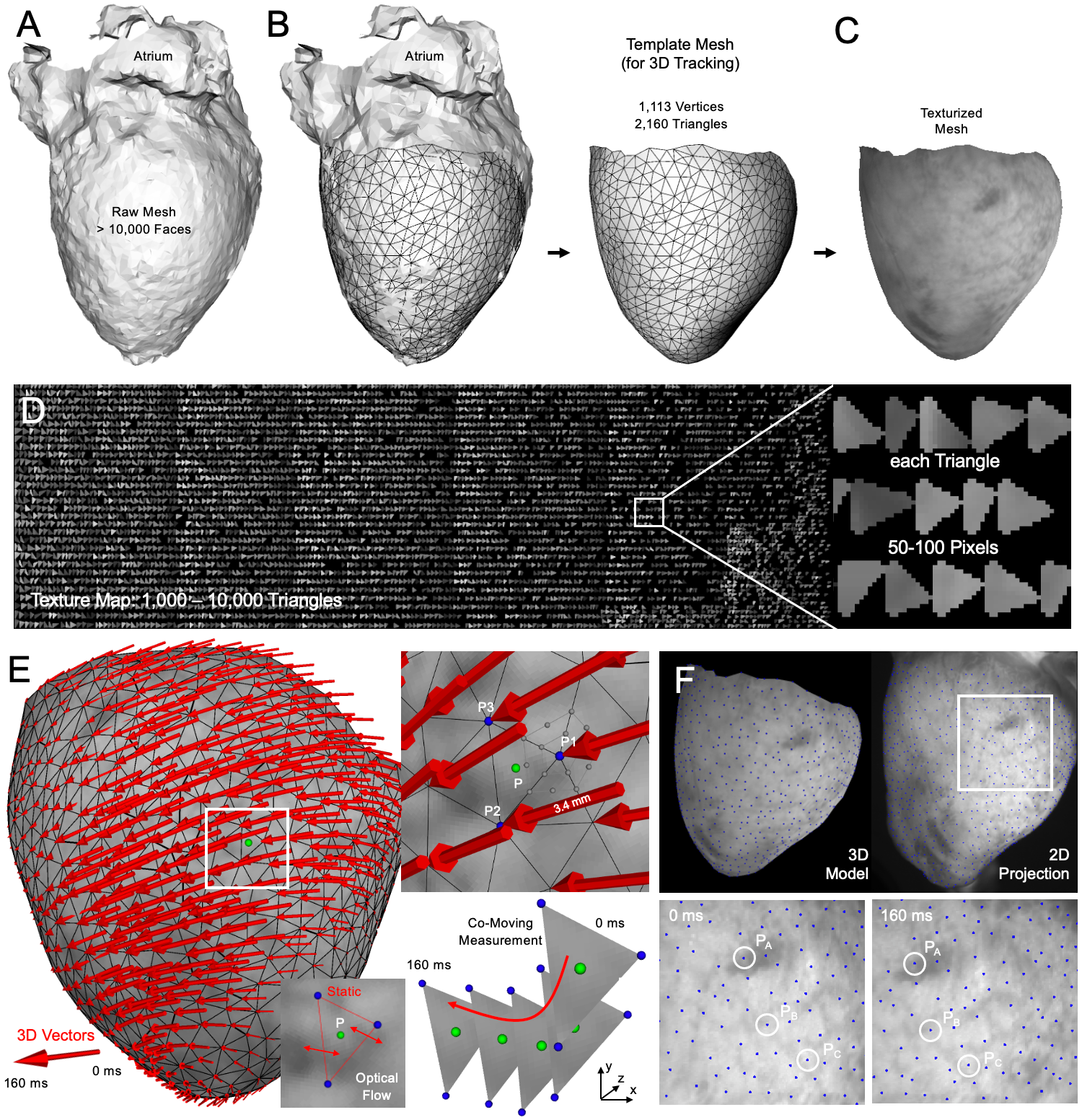}
  \caption{
Panoramic 3D reconstruction, tracking, and voltage-sensitive optical mapping of moving heart using multi-view cooperative numerical motion tracking with 12 calibrated cameras.
\textbf{A} High-resolution raw mesh resulting from static 3D reconstruction (photogrammetry-like technique or patch-matching with COLMAP) showing 3D rabbit heart surface during systole (\SI{160}{\milli \second}). The motion of the heart is initially represented by a sequence of independent raw meshes which each consist of an uncorrelated number of vertices and faces, see Supplementary Video \ref{video:SV3}.
\textbf{B} The motion is tracked over time using a template mesh (black wireframe mesh), which typically consists of approx. 2,000-10,000 (triangular) polygon faces. 3D Tracking is performed using a multi-view vertex-based mesh tracking approach developed by Klaudiny et al. \cite{Klaudiny2011}, see also panel E.
\textbf{C} After tracking, the moving mesh is texturized with the grayscale video data averaging data from multiple cameras, see also Supplementary Video \ref{video:SV2}.
\textbf{D} Grayscale texture map containing thousands of triangles used to texturize the mesh in C. The ventricular surface is resolved by about 0.5 - 1.0 million pixels, as each triangle contains approx. 50-150 pixels, at a spatial resolution of about \SI{120}{\micro\meter} per pixel. The pixels are superpositions of pixels from multiple cameras, see also Fig.~\ref{fig:SupplementCoverage}.
\textbf{E} Displacement vectors (red) indicating shifts of epicardial surface from diastole (0 ms, shortly before depolarization) to systole (160 ms).
Close-up: Tracking is performed per vertex (here shown for vertex P1): a mesh-based kernel consisting of a set of sample points (gray, illustration) surrounding each vertex is projected into and used to identify tissue movements in each camera, see \cite{Klaudiny2011} for details and Figs.~\ref{fig:SupplementTrackingElectrode} and \ref{fig:SupplementTrackingKernel}. 
\textbf{F} Mesh vertices (blue points) superimposed on texturized mesh (3D model) and projected into one of the camera images (2D). The projected vertices follow the motion of the heart in each individual video (e.g. points $P_A, P_B, P_C$), see also Supplementary Video \ref{video:SV5}.
  }
  \label{fig:Fig2}
\end{figure*}

Zhang et al.\ \cite{Zhang2016} demonstrated for the first time that action potential waves can be imaged on the contracting 3D left ventricular surface using multi-camera optical mapping.
The imaging, motion tracking, and 3D surface reconstruction was performed with 4 cameras and required fiducial markers glued to the epicardial surface, which were manually identified and then tracked through the video images in a semi-automatic fashion, as described in \cite{Bourgeois2011}.
Subsequently, the reconstructed 3D heart surface consisted of a low-resolution polygon surface comprising about $20$-$30$ polygons with vertices defined by the fiducial markers used for tracking.
Action potential measurements were averaged within each of the polygons without exploiting the full 128 $\times$ 128 pixel resolution of the video images.
The cameras were positioned at an angle of \SI{45}{\degree} to allow sufficient overlap between their fields of view.
This methodology was subsequently used to image beating hearts in vivo \cite{Zhang2023}.

Christoph and Schr\"oder-Schetelig et al.\ \cite{Christoph2017} demonstrated that action potential waves can be imaged on the contracting 3D left ventricular surface at higher spatial resolutions and without markers attached to the heart surface.
The imaging and 3D reconstruction was also performed with 4 cameras at \SI{45}{\degree}, but the heart's motion was tracked automatically in every pixel in each of the 4 videos, subsequently yielding a high-resolution polygon surface (4,000-5,000 polygons) with a spatial resolution comparable to the resolution of the video images ($128 \times 128$ pixels).
With the 4 cameras, it was possible to reconstruct nearly one half of the ventricular surface of the beating heart.
However, the approach required a static shape reconstruction of the pharmacologically contraction-inhibited heart following each experiment, which is a major limitation. 
After the optical mapping of the contracting heart was completed, the heart was uncoupled with the uncoupling agent Blebbistatin.
Once the heart was completely quiescent (15-30 minutes), it was rotated once around its long axis using a stepping motor while 72 images of its silhouette were taken every \SI{5}{\degree} throughout the \SI{360}{\degree} rotation. 
The silhouettes were then used to reconstruct the 3D geometry of the quiescent heart using a shape-from-contour algorithm.
Finally, the 3D mesh showing the static heart was deformed using the two-dimensional (2D) motion tracking data obtained with each of the video cameras showing the beating heart.
While this approach provided 3D reconstructions of the shape of the deforming heart over time, it could not provide these reconstructions immediately during imaging. 
The required additional experimental part at the end with Blebbistatin was difficult to execute, and added post-processing steps and potential complications to each experiment that could prohibit further analysis.
Most critically, the time available for imaging the beating heart was limited to a few minutes.

In this study, we address the previous shortcomings by introducing a stereoscopic multi-camera optical mapping approach, which uses multi-view imaging and 3D cooperative motion tracking of the heart surface with many low-cost cameras.
Our approach neither requires fiducial markers \cite{Zhang2016} nor a static shape reconstruction obtained with Blebbistatin \cite{Christoph2017}, and can be used to immediately and automatically reconstruct the heart surface while imaging action potential waves at high resolutions.

\begin{figure*}[htb]
    \centering
    \includegraphics[clip, trim=0.0cm 0.0cm 0.0cm 0.0cm, width=0.95\textwidth]{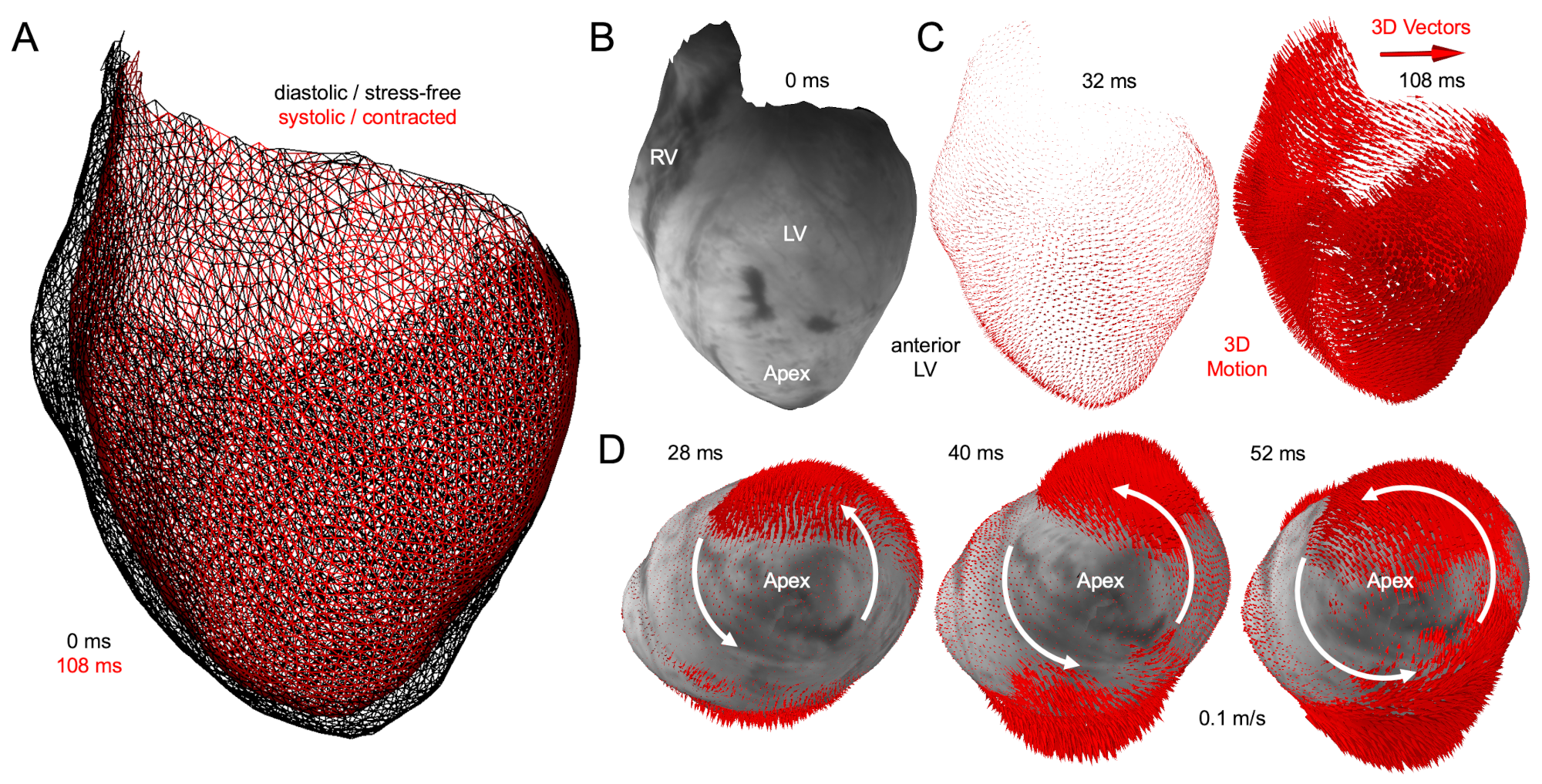}
    \caption{
    Ventricular wall motion imaged with voltage-sensitive 3D optical mapping during sinus rhythm, see also Figs.~\ref{fig:Sinus2} and \ref{fig:Sinus3}. Wall motion is tracked in 3D space using video data obtained with 12 calibrated high-speed cameras, see Figs.~\ref{fig:Setup} and \ref{fig:Fig2}.
   \textbf{A} Wireframe meshes of epicardial surface in diastolic, stress-free (black) and deformed, contracted (red) mechanical configuration. 
  The mesh consists of $3,325$ vertices or $6,349$ polygon triangles, respectively. Average nearest-neighbour distance: \textcolor{blue}{\SI{0.92 \pm 0.21}{\milli \meter}} (black mesh).
   \textbf{B} Raw video images of left ventricle (LV) projected onto mesh (same heart as in Fig.~\ref{fig:Sinus1}A,B, anterior wall). 
   \textbf{C} 3D displacement vector field (red) indicating motion at 32 ms and 108 ms, respectively, with respect to diastolic, stress-free mechanical configuration at 0 ms.
    \textbf{D} Apical view with displacement vectors (red) indicating motion with respect to the previous frame. 
    The fastest contractile motion occurs between 30 - 50 ms, and the ventricles exhibit rotational, torsional motion around the apex, see also Supplementary Videos \ref{video:SV2} and \ref{video:sinus}.
    }
    \label{fig:Sinus1}
\end{figure*}

\section*{Methods}

\subsection*{Soccer Ball-shaped Imaging Chamber}
We designed a spherical soccer ball-shaped imaging chamber, which facilitates panoramic optical mapping of contracting hearts with up to $24$ cameras simultaneously, see Fig.~\ref{fig:Setup}.
The chamber allows a dense coverage of the heart during imaging with overlapping fields of views, see Fig.~\ref{fig:SupplementCoverage}, while also providing an evenly distributed, unobstructed illumination of the heart with up to 48 light-emitting diodes (LEDs), see Figs.~\ref{fig:Setup}A-C,E) and \ref{fig:SupplementRatiometry}A-C).
The chamber's geometry is based on a truncated icosahedron or soccer ball geometry.
The soccer ball is open at the top, has 24 penta- or hexagonal surfaces and an outflow at the bottom.
Imaging can be performed through round windows in each of the 24 surfaces.
The angle of the optical axes between two adjacent windows is $37.4^{\circ}$ or $41.8^{\circ}$, respectively. 
Illumination can be provided through 48 small round windows (LED-ports) located at the vertices between the penta- or hexagonal surfaces.
The volume of the chamber is about $5$ liters.
The size of the chamber is determined by the working distance of the cameras and lenses, the size of the heart, and the size of the windows through which imaging is performed, respectively.
Hearts are positioned at the center of the imaging chamber and can be imaged equally through all windows without obstructions.
The plexiglas windows are glued into the window openings using silicone.
The chamber shown in Figs. ~\ref{fig:Setup}, \ref{fig:SupplementSoccerball}, \ref{fig:SupplementCoverage}C) and \ref{fig:SupplementRatiometry}A,B) was 3D-printed (3D printer: Creality CR-10 V2, material: white PLA+ by Overture).

\subsection*{Langendorff-Perfusion of isolated Hearts}
New Zealand White rabbits ($N=6$) were anaesthesized and hearts were quickly excised and transferred into ice-cold cardioplegic solution. 
All procedures were in accordance with animal welfare law and were approved by the Instutitional Animal Care and Use Committe (IACUC) at the University of California, San Francisco.
Hearts were prepared and subsequently placed inside the imaging chamber filled with warm, oxygenated Tyrode solution (\SI{5}{\liter}) and attached to retrograde Langendorff-perfusion within $15$-$20$ minutes after excision.
Hearts were perfused in constant-pressure mode.
The Langendorff-perfusion system comprised a \SI{5}{\liter} water-jacketed reservoir (Radnoti, model no. 120142-5, USA) that was positioned about $0.6$-$0.7\,\text{m}$ above the heart and connected via tubing to a heating coil with integrated bubble trap (Radnoti, model no. 158840, 10ml, USA) positioned directly above the heart.
The perfusion pressure was regulated by adjusting the height of the water reservoir to create perfusion pressures of $50 \pm 5 \,\text{mm Hg}$.
Perfusion pressure was monitored using a digital manometer (Hti-Xintai Instrument, HT-1890, China).
The Tyrode was oxygenated and its pH-level was kept constant at $7.4$ by bubbling Carbogen into the reservoir throughout the experiment.
The pH-level was monitored using a digital pH meter (Extech PH220, Taiwan).
An immersion circulator-heater (Polystat, Cole-Palmer, USA) was used to pre-heat water to $37^{\circ}\text{C} \pm 0.5^{\circ}\text{C}$, which was then pumped through the water-jacketed reservoir and heating coil to pre-heat the Tyrode solution. 
The resulting temperature of the Tyrode flowing out of the heating coil into the heart was $36.75^{\circ}\text{C} \pm 0.25^{\circ}\text{C}$.

\subsection*{Multi-Camera Optical Mapping}\label{sec:methods:stereoscopy}
Multi-camera optical mapping was performed using 12 high-speed CMOS cameras (acA720-520um, Basler, Germany), which were positioned around the imaging chamber on an optical table (TMC, USA), see Figs.~\ref{fig:Setup}A,D) and \ref{fig:SupplementDepthMaps}A) and Supplementary Video \ref{video:SV2}.
The cameras were aligned and fixed in place using articulator arms (Noga, Japan).
The average angle between neighboring cameras was approx. \SI{40 \pm 10}{\degree}, as shown in Figs.~\ref{fig:Setup} and \ref{fig:SupplementCoverage}C).
With 12 cameras, it is possible to achieve a sufficiently large overlap between the fields of view of the cameras, see Fig.~\ref{fig:SupplementCoverage}A), and dense coverage of the ventricular surface with at least 3 cameras observing one point simultaneously across the surface, see Fig.~\ref{fig:SupplementCoverage}B).
The cameras produce 12-bit (dynamic range: 0-4,096 intensity counts) greyscale video images with a maximal spatial resolution of $720\times540$ pixels.
Imaging was performed at speeds of $500\,\text{fps}$ with a spatial resolution of $440 \times 320$ pixels.
The cameras were triggered using a signal generator (FY8300, FeelElec, China) and custom software to synchronize the video acquisition.
Emission light was collected through machine vision lenses (3x CF8ZA-1s from below, 9x CF12ZA-1s horizontal, Fujinon, Japan) mounted onto each of the cameras together with emission filters, see next section.
With the magnification of the lenses, the video images covered a field of view of about $4 \times 5 \,\text{cm}$ at working distances of about 10-15cm, see Fig.~\ref{fig:Setup}A,D).
The videos from the cameras were streamed via USB onto a computer (AMD Ryzen Threadripper 3960X CPU, 2TB SSD working memory, Gigabyte Technology Co. TRX40 mainboard) using the 9 integrated USB ports and additional external USB hubs (10 Gbps, USB 3.2 Gen 2).
\

\subsection*{Ratiometric Voltage-sensitive Optical Mapping}
Hearts were stained with two \SI{250}{\micro \liter} bolus injections of voltage-sensitive fluorescent dye (Di-4-ANEPPS, Biotium, Germany) into the bubble trap. 
The second bolus was injected $5$ minutes after the first bolus and imaging was started about $10$-$15$ minutes after the first injection.
In this study, all hearts were imaged without pharmacological uncoupling agents, such as Blebbistatin.
Optical mapping was performed either in continuous mode with continuous green excitation light or in ratiometric mode with pulsed green and blue excitation light \cite{Bachtel2011, Bourgeois2011, Kappadan2020}.
We used up to 48 high-power, multi-color RGBW (red-green-blue-white) light-emitting diodes (LEDs) to illuminate the heart from all sides (Cree XML RGBW Star LED, USA), see Figs.~\ref{fig:Setup}A,F) and \ref{fig:SupplementRatiometry}A-C).
The LEDs were positioned in 48 ports between the imaging windows and directed at the center of the imaging chamber.
Each LED comprises 4 diodes, a red, green, blue and white diode, which are positioned next to each other in close proximity and can be powered individually, see Fig.~\ref{fig:SupplementRatiometry}C).
In continuous mode, we powered only the green diodes continuously.
In ratiometric mode, we powered the blue and green diodes alternatingly with two power sources, rapidly switching the power on and off between the blue and green sets ($2 \times 48 = 96$ total) using a custom-made LED driver.
We used two programmable direct current power supplies (KWR103, 60V/15A, Korad Technology, China) as power sources for the LEDs.
The triggering of the LED driver was synchronized with the camera acquisition, such that odd and even video frames were illuminated with blue or green excitation light for \SI{2}{\milli \second}, respectively, see also Figs.~\ref{fig:Setup}F) and \ref{fig:SupplementRatiometry}D,E) and Supplementary Video \ref{video:green-blue-video}.
Correspondingly, the effective acquisition speed of the blue and green videos was 250fps.
In continuous mode, we used red longpass emission filters (Dark Red \#29, Heliopan, Germany), which block green excitation light and transmit red and near-infrared fluorescent light.
In ratiometric or pulsed mode, we used orange-red bandpass filters (605-70, Chroma, USA) to create a negative fractional change in fluorescence $\Delta F / F$ under green illumination and no fractional change under blue illumination, respectively.
In this configuration, the blue channel can be used to measure changes in the illumination on the heart surface as it moves through an inhomogeneously illuminated scene.
During ratiometric imaging, the blue and green excitation light from the LEDs was filtered using shortpass filters (\SI{550}{\nano \meter} cut-off wavelength, model FESH0550, Thorlabs Inc, USA).

\subsection*{Rhythm Control}
Paced rhythms were induced using a custom-made micro-electrode (FHC Inc., USA), see Figs.~\ref{fig:SupplementDepthMaps}A) and \ref{fig:SupplementTrackingElectrode} connected to an isolated pulse stimulator (A-M Systems, model 2100, USA), which was externally triggered using custom-made software.
Regular ventricular pacing was induced using 8 ms long biphasic pulses applied at 3-8 Hz.
Ventricular fibrillation was induced using burst pacing at frequencies ranging from 15-50 Hz and terminated using a custom-made defibrillator.
Action potential prolongation was caused by switching to low-potassium (50\% concentration) Tyrode solution containing barium-chloride, as described in \cite{Myles2015}.
The electrocardiogram and camera triggers were recorded during the experiment using an analog-to-digital data acquisition system (PowerLab 16/35, ADInstruments, Australia).

\subsection*{Camera Calibration}\label{sec:calibration}
Camera calibration refers to the process of determining extrinsic and intrinsic camera parameters with the goal of undistorting the cameras and establishing the relative positioning and alignment of the cameras with respect to each other.
We used multical to perform the camera calibrations \cite{multical}.
Camera calibration was performed after each imaging experiment using an $8\times10$ ChArUco calibration target with a 4 mm checkerboard pattern and ArUco fiducial markers in each of the black squares \cite{GarridoJurado2014}, see also Fig.~\ref{fig:SupplementCalibration}.
We manually positioned and rotated the calibration target within the imaging chamber and took a full set of images per pose of the target with all cameras simultaneously.
The target was typically observable by at least 2-3 cameras.
The calibration procedure required typically 30-100 of image sets in total, while at least 2-3 sets needed to be taken for each camera.
With both the extrinsic and intrinsic camera parameters, we are able to transform a 3D world coordinate to an image coordinate, but for the transformation to be accurate we must account for camera distortion.
In addition to the camera's inherent distortion, the refraction from the light passing from water to glass to air introduces additional distortion. 
Kwon et al. showed that the refraction error can be accounted for using standard camera distortion calibration \cite{Kwon1999}.

\subsection*{Three-Dimensional Motion Tracking and Surface Reconstruction using Cooperative Multi-View Tracking}
Using the camera calibration, we used COLMAP \cite{Schoenberger2016mvs} to produce raw, static 3D reconstructions of the heart for each frame over time, see Fig.~\ref{fig:Fig2}A).
A frame corresponds to a set of video images acquired at the same time by the different cameras, see Fig.~\ref{fig:SupplementDepthMaps}A).
Using the 3D surface reconstructions, it is possible to create depth-maps, as shown in Fig.~\ref{fig:SupplementDepthMaps}B).
The static reconstructions were performed independently from each other frame-by-frame over the sequence of frames, producing a sequence of unique, uncorrelated raw meshes with different numbers of vertices and polygon faces, see also Supplementary Video \ref{video:SV3}.
Visually, these meshes reflect the motion of the heart when texturized, but they cannot be used to measure action potential waves or strain as the numbers of vertices and faces fluctuate over time. 
The topology of each mesh is different and the vertices and faces are uncorrelated over time across meshes.

To compute a single moving mesh that represents a dynamic reconstruction of the moving heart surface, we used a 3D mesh tracking technique by Klaudiny et al. \cite{Klaudiny2011}, see also Fig.~\ref{fig:Fig2}F).
This cooperative patch-based mesh tracking technique produces a single moving mesh with fixed topology and the same number of vertices over time from the individual meshes from the static reconstructions.
It does so by tracking the vertices of a template mesh, see Fig.~\ref{fig:Fig2}B,F), searching for the corresponding tissue segments in both the other already pre-computed raw meshes and in the different camera images of another frame.
In our case, the template mesh is derived from one of the raw meshes obtained during the static reconstructions which then serves as the reference frame (typically the first frame). 
The template meshes were created manually by cutting off the atria around the base of the heart from the selected raw mesh using Blender  \cite{Blender}. Next, a sequence of remeshing, decimating, and smoothing (Taubin filtering) was applied to the template mesh using Blender \cite{Blender}, VTK \cite{VTK}, and Open3D \cite{open3D}.
The remeshing and decimating determine the spatial resolution of the mesh and subsequently of the tracking and displacement vector fields. 
The resulting template meshes consist of approx. 1,000-5,000 vertices and about twice as many triangular faces, respectively, as shown in Fig.~\ref{fig:Fig2}B) and Supplementary Video \ref{video:SV3}.
The number of vertices or faces of the raw static meshes is, depending on the input parameters of the raw mesh reconstruction, in the order of tens to hundreds of thousands of vertices / faces ($10^4-10^5$).
The exact number is not critical as long as the resolution is about 2-5 times higher than the number of vertices / faces of the tracking mesh.
The reference frame shows either the heart shortly before the depolarization of the ventricles during sinus rhythm or is an arbitrary frame in case of arrhythmias.
To produce the next mesh in the sequence of frames, the algorithm uses first the reference mesh and then each previous mesh as a starting point and then computes and minimizes an error function for each vertex $v_e$ using a patch-match inspired approach \cite{Barnes2009}.
This results in a new time-varying mesh with vertices moving in accordance with the video data from the different cameras.

In more detail, the vertex error function is comprised of an image-based component and a mesh-based component. 
For the image component, we created a set of sample points surrounding the vertex on the mesh's surface, as shown in Fig.~1 in \cite{Klaudiny2011}, Fig.~\ref{fig:Fig2}F) (illustration) and Fig.~\ref{fig:SupplementTrackingElectrode}.
This is done for the vertex in the previous mesh and vertex in the new mesh, which is initially a replication of the previous mesh. 
These sample points are projected into the camera images in which they are visible in. 
The pixel values of the previous mesh's vertex sample points are compared with the new mesh's vertex sample points using a normalized cross-correlation similarity measure. 
This ensures that the area around the vertex in the new mesh is similar to the corresponding vertex in the previous mesh. 
For the mesh component, it computes the distance of the vertex from the surface of the raw, static mesh. 
This distance is passed through the Tukey biweight function to reduce the effect of outliers \cite{Klaudiny2011}. 
The image error $i_e$ and mesh error $m_e$ components are summed with a weighting term $w$ for the mesh error: $v_e = i_e + w \cdot m_e$.
The mesh error reduces the search space and drift over time. 
We optimized the vertex error using a cooperative optimization algorithm that optimizes all vertices together \cite{Barnes2009}. 
It begins by randomly translating each vertex multiple times and then keeping track of the displacement that produces the lowest error.
In the next step, each vertex applies its neighboring vertices' best displacements to itself and checks if it improves the error. 
The random search and neighbor searching stages are repeated in succession 4-5 times to find the vertex positions for the new mesh that minimize the error.
Subsequently, we take the displacements of the new mesh's vertices with respect to the previous mesh and apply a weighted smoothing to the displacements based on the vertex error \cite{Klaudiny2011}. 
This results in a completed mesh tracking step for one frame, which is then repeated sequentially for all frames to produce the time-varying mesh over the entire sequence of frames.

In continuous or single channel mode with green-only illumination, we performed tracking from frame to frame (\SI{2}{\milli \second} temporal resolution) using a template mesh created from the first raw mesh. 
In ratiometric or pulsed mode, we created a template mesh from the first raw mesh in the blue channel, and then used this mesh to track the motion separately in the blue and green channels from frame to frame (\SI{4}{\milli \second} temporal resolution).
Tracking fluorescent videos with substantial temporal fractional change in intensity (e.g. $ \Delta F/F \approx 10\%$) may yield tracking artifacts in that the tracking algorithm inadvertently tracks intensity changes caused by the action potential rather than motion, as described in \cite{Christoph2018Frontiers,Lebert2022}. 
To avoid tracking artifacts, we tracked the blue channel in ratiometric mode and contrast-enhanced videos, as described in \cite{Christoph2018Frontiers}, in continuous green mode.
The data shown in Supplementary Fig.~\ref{fig:SupplementCrosstalk} demonstrates that our tracking method is not sensitive to or compromised by the fluorescent signal in the green channel.
The insensitivity of our tracking method to the signal in the green channel is presumably facilitated by the normalized cross-correlation similarity measure in the image-based part of the error function, which suppresses temporal changes in the image intensity.
Ultimately, in ratiometric mode, we used the blue mesh resulting from tracking the blue videos to proceed with all post-processing, such as texturizing the meshes and extracting signals, and applied the different post-processing steps equally to both the video data from the blue and green channels, respectively.
The two time-varying meshes resulting from tracking the blue and green channels equally reproduce the motion of the ventricular surface, see Supplementary Videos \ref{video:green-blue-video-3D} and \ref{video:green-blue-video-3D-texture}.

\begin{figure*}[htb]
  \centering
  \includegraphics[clip, trim=0.0cm 0.0cm 0.0cm 0.0cm, width=0.95\textwidth]{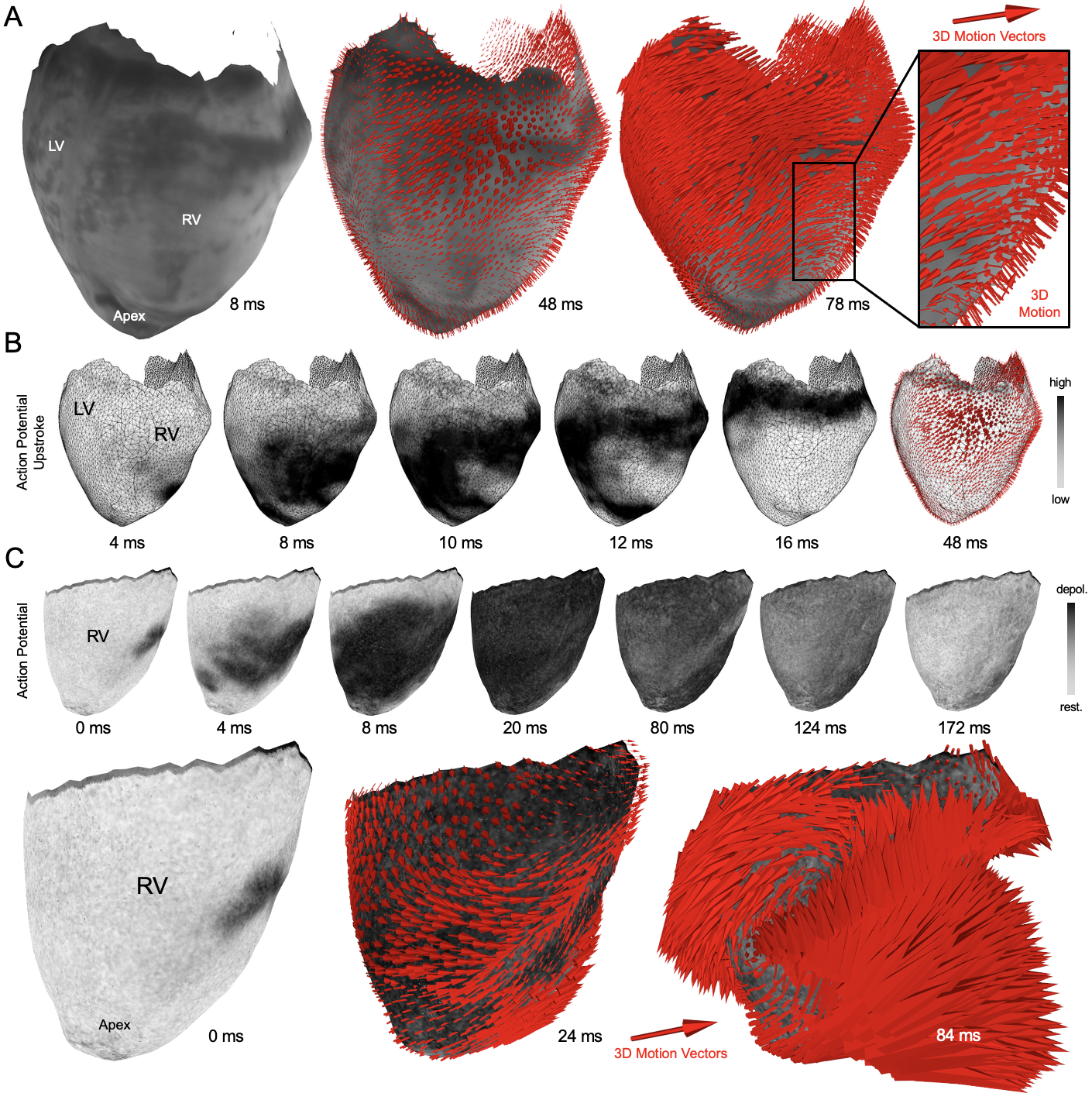}
  \caption{
  Sinus rhythm imaged in a contracting rabbit heart with voltage-sensitive multi-camera optical mapping with 12 cameras, see also Figs.~\ref{fig:Setup}, \ref{fig:Sinus1}, \ref{fig:Sinus3}, \ref{fig:SupplementSinusAPTraces}-\ref{fig:SupplementBaselineCorrection2} and Supplementary Video \ref{video:sinus}.
  Using 3D motion tracking, we imaged action potential waves and wavefronts across the entire contracting and strongly deforming ventricular surface (LV/RV: left/right ventricle).
 \textbf{A} Raw grayscale video data mapped onto 3D reconstructed ventricular surface together with motion vectors (red) indicating motion and deformation of the ventricular surface with respect to the mechanical configuration at 0 ms (beginning of depolarization).
  \textbf{B} Action potential wave front (dark: upstroke, high negative rate of change in fluorescence intensity) imaged with continuous green excitation propagating across ventricular surface triggering the contraction and deformation of the heart, see also Fig.~\ref{fig:Sinus3}A,B).
Noticeable contraction sets in about 30 ms after the beginning of the depolarization of the ventricles, c.f. Fig.~\ref{fig:Sinus2}C,D).
\textbf{C} Action potential wave (bright: resting tissue, black: depolarized tissue / plateau phase, gray: repolarizing tissue) measured with pulsed green-blue ratiometric excitation across ventricular surface in strongly contracting heart. Corresponding optical traces shown in Figs.~\ref{fig:Setup}G), \ref{fig:Sinus3}D) and \ref{fig:SupplementSinusAPTraces}. Motion vectors (red) as in A) (but with different scaling).
  }
  \label{fig:Sinus2}
\end{figure*}

\subsection*{Post-Processing} \label{sec:methods:post-processing}

The vertices of the time-varying mesh describe the motion of the heart surface. 
With that information, signal analysis can be performed in a co-moving frame of reference, see Fig.~\ref{fig:Fig2}G).
Accordingly, we warped the video images using the 2D projections of the movements of the 3D mesh in the camera images to be able to further process the data in an image-based co-moving frame of reference, see also Supplementary Video \ref{video:2Dvs3D}.
This step significantly reduced motion artifacts.
Because the resolution of the mesh is lower than the resolution of the video images, we further performed 2D optical flow-based motion tracking as described in Lebert et al. \cite{Lebert2022} using a GPU-accelerated version of the Farneb\"ack algorithm \cite{Farneback2003} on the already warped video images and warped them further with respect to the reference video images, see Supplementary Video \ref{video:2Dvs3Dwarping}. The 2D tracking and warping was performed using optimap \cite{optimap}.
This last step significantly reduced residual motion and motion artifacts in the triangles between the vertices, see Fig.~\ref{fig:SupplementTracking3D2D}.
Accordingly, all motion-compensated optical maps shown throughout this study were compensated first using the 3D tracking and then 2D tracking.
Otherwise, we texture-mapped the raw grayscale video images showing the moving heart onto the moving mesh to produce texturized meshes as shown in Supplementary Videos \ref{video:SV2}, \ref{video:SV3}, \ref{video:green-blue-video-3D-texture}, \ref{video:sinus} and \ref{video:VF1} and Figs.~\ref{fig:Fig2}C,E), \ref{fig:Sinus1}A), \ref{fig:Sinus2}B,D), \ref{fig:VF1}A), and \ref{fig:SupplementTracking}.
Similarly, we projected the warped, processed and then unwarped videos onto the moving mesh.
Projections were performed using a weighted average based on the angle between the mesh's surface normal and the camera's direction.
The texture data was saved in 8-bit format as a PNG file, so the textures could be processed by general mesh viewing tools.
The average angle between cameras was $54\pm18^{\circ}$, which includes nearest- and second-nearest-neighbouring cameras, as well as cameras further away (as some windows were not used for imaging).
Each camera overlaps with 2-4 other cameras, see Fig.\ref{fig:SupplementCoverage}.

All post-processing was performed in a co-moving frame of reference and in an image-based fashion in the individual warped video images using optimap \cite{optimap}.
The 3D post-processed (e.g. pixel-wise normalized) optical maps are the average of multiple post-processed videos and are combined using a weighted average with the weights being the angle between the triangle normals and camera's view direction.
To visualize action potential waves, as shown in Supplementary Videos \ref{video:sinus}, \ref{video:pacing1} and \ref{video:VF1}, we normalized the optical signals representing transmembrane potential changes in the warped (ratiometric) video data per pixel over time \cite{optimap}.
We used a temporal pixel-wise normalization with a sliding-window normalization (30 - 50 frames length), or a frame-wise difference computation with subsequent temporal sliding-window normalization to amplify the optical signal, see sections \ref{sec:results:sinus}-\ref{sec:results:vf} and Figs.~\ref{fig:Sinus1}-\ref{fig:VF2}.
Spatial or temporal smoothing of the optical maps was not applied throughout this study.
Lastly, we texture-mapped the post-processed pixel-wise normalized optical maps onto the 3D heart surface after rewarping them back into the static laboratory frame.

\subsection*{Action Potential Duration Measurement}\label{sec:methods:APD}
To calculate action potential durations (APD) and APD maps, we extracted raw ratiometric optical traces from the center of each triangle (averaging from $7 \times 7$ pixels), see also Supplementary Video \ref{video:SV4}, and further post-processed the traces using a custom detrending and baseline removal algorithm, see Figs.~\ref{fig:SupplementBaselineCorrection1} and \ref{fig:SupplementBaselineCorrection2}.
Action potential upstrokes were computed by detecting peaks in the temporal derivative of each optical trace.
Next, smoothed versions of the first and second temporal derivatives were calculated using Savitzky-Golay filtering, and negative peaks in the first and positive peaks in the second derivative were calculated as estimators for the repolarization phase and end of the repolarization, respectively, see Fig.~\ref{fig:SupplementBaselineCorrection1}.
The peak detections were only performed between \SI{100}{\milli \second} and \SI{300}{\milli \second} after each upstroke.
Detections yielding fewer or more than anticipated peaks were excluded from further analysis.
Subsequently, the parts of the trace between the upstrokes and repolarizations were removed and missing values were linealry interpolated.
The remaining baseline was smoothed using Savitzky-Golay filtering and then subtracted from the original trace yielding traces as shown in Fig.~\ref{fig:SupplementBaselineCorrection2}.
APDs were calculated for each action potential and averaged per trace, see Fig.~\ref{fig:SupplementAPD}.
Traces with substantial mismatches in APD beteeen two subsequent action potentials were excluded from further analysis. 
APD maps were obtained by repeating the APD measurement for every triangle.

\subsection*{Phase Maps and Phase Singularity Calculation}
Phase maps were computed using optimap \cite{optimap} from the motion-stabilized and pixel-wise normalized optical mapping videos in each camera individually. 
The phase angle was computed for each time-series using the Hilbert transform as described in \cite{IyerGray2001}.
The resulting phase maps were smoothed using a complex order parameter filter as described in \cite{Lebert2021}.
The smoothed 2D phase maps were then projected onto the 3D mesh and averaged from multiple camera perspectives if the respective pixels overlapped.
Phase singularities were computed on the 3D mesh surface by computing the circular integral of the gradient of the phase, as described in \cite{Li2020}.

\section*{Results}\label{sec:results}

\begin{figure*}[htb]
  \centering
  \includegraphics[clip, trim=0.0cm 0.0cm 0.0cm 0.0cm, width=0.95\textwidth]{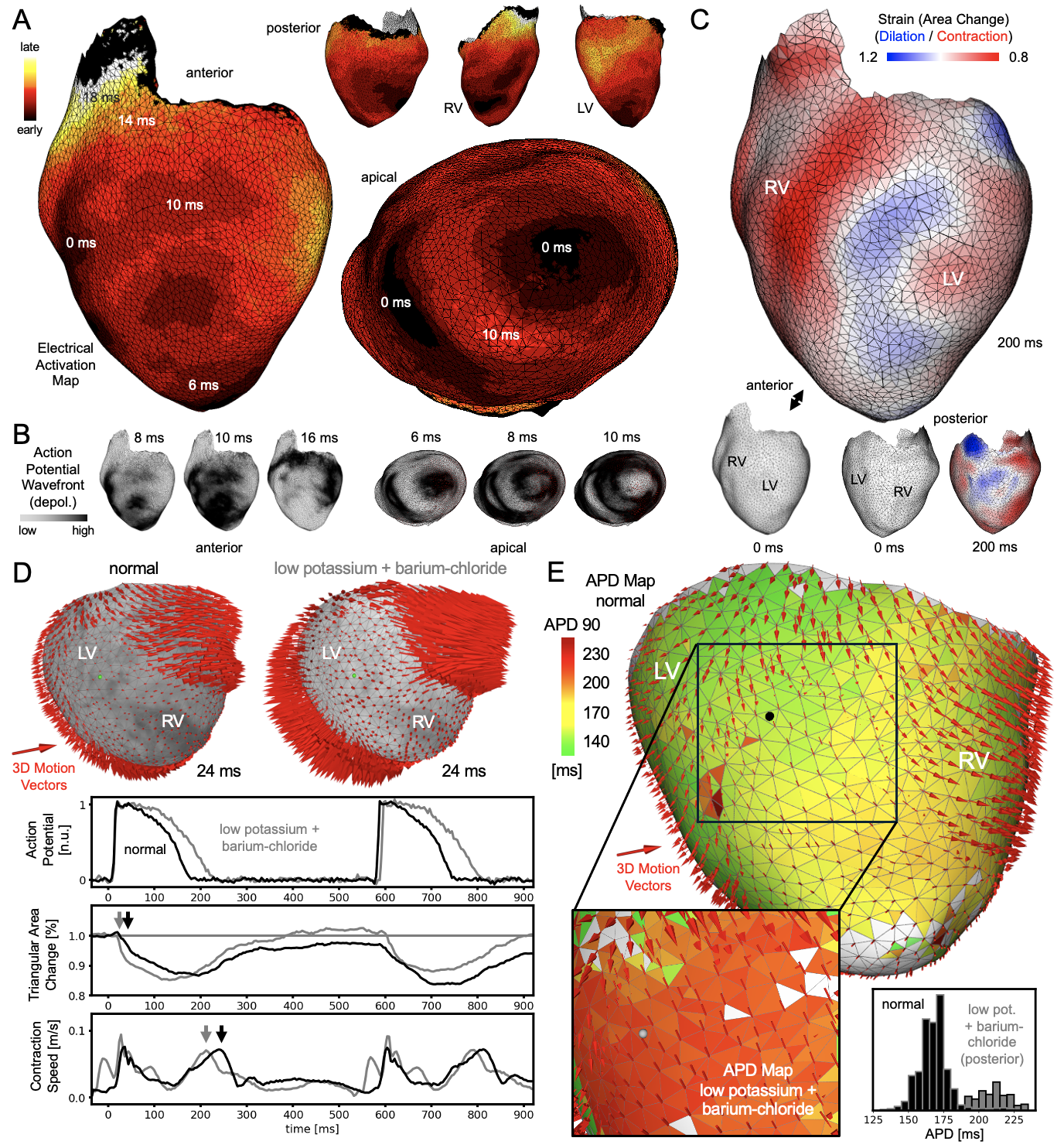}
  \caption{
  Electrical activation, strain, and action potential duration (APD) mapped across entire ventricular surface of isolated rabbit heart during sinus rhythm.
 \textbf{A} Electrical activation map with local activation times (black: early/0 ms, yellow/white: late/20ms; anterior, posterior, apical view, etc.) computed form the action potential upstroke shown in panel B) and Fig.~\ref{fig:Sinus1}B). The activation map indicates multiple early-activation or epicardial break-through sites (presumably correlated with the Purkinje system). The entire ventricles get electrically activated within 20 ms.
 \textbf{B} Action potential wavefront (black: upstroke, high negative rate of change of signal; 0 ms: beginning depolarization, continuous green excitation) as seen from anterior and apical walls, respectively, c.f. Fig.~\ref{fig:Sinus2}B).
 \textbf{C} Strain measured as local triangular area change across epicardial surface (blue: dilated, red: contracted, 1.2/0.8 = 20\% area increase/decrease, anterior wall).
 Bottom left: Undeformed, stress-free epicardial surface (0 ms: beginning depolarization, anterior wall).
 Bottom center and right: Stress-free and deformed epicardial surface (0 vs. 200 ms, posterior perspective), c.f. Fig.~\ref{fig:Sinus2}.
 \textbf{D} Top: Comparison of contraction (red vectors, frame-to-frame displacements) with normal and low potassium barium-chloride Tyrode. 
 Upper plot: Action potential measured on posterior wall (black dot in panel E) during normal (black) and abnormal sinus rhythm with low-potassium barium-chloride Tyrode (gray) causing delayed repolarization and prolonged action potential duration (APD). The optical traces were measured at the center of the same triangle (averaged from 7 $\times$ 7 pixels, ratiometric imaging with pulsed green-blue excitation).
 Center plot: Strain measured as local trianglular area change (normalized units with respect to intial triangle area) is comparable but rate of contraction is faster/stronger with delayed repolarization (arrows).
 Bottom plot: Contractile speed (magnitude of frame-to-frame displacements) shows that the heart contracts more strongly and relaxes earlier with low potassium barium-chloride Tyrode (arrows).
 \textbf{E} APD maps computed for normal APD across entire ventricular surface and for prolonged APD on posterior wall during sinus rhythm (APD-90 at 10\% height of the action potential). 
 Histogram showing peaks at \SI{170}{\milli \second} $\pm$ \SI{5}{\milli \second} with normal Tyrode and \SI{210}{\milli \second} $\pm$ \SI{5}{\milli \second} with low-potassium barium-chloride Tyrode (only posterior wall).
 Traces in D) sampled from black and gray dots (same location, same triangle).
 Heart in panels D,E) different than in panels A-C).
  }
  \label{fig:Sinus3}
\end{figure*}

\begin{figure*}[htb]
    \centering
    \includegraphics[clip, trim=0.0cm 0.0cm 0.0cm 0.0cm, width=0.99\textwidth]{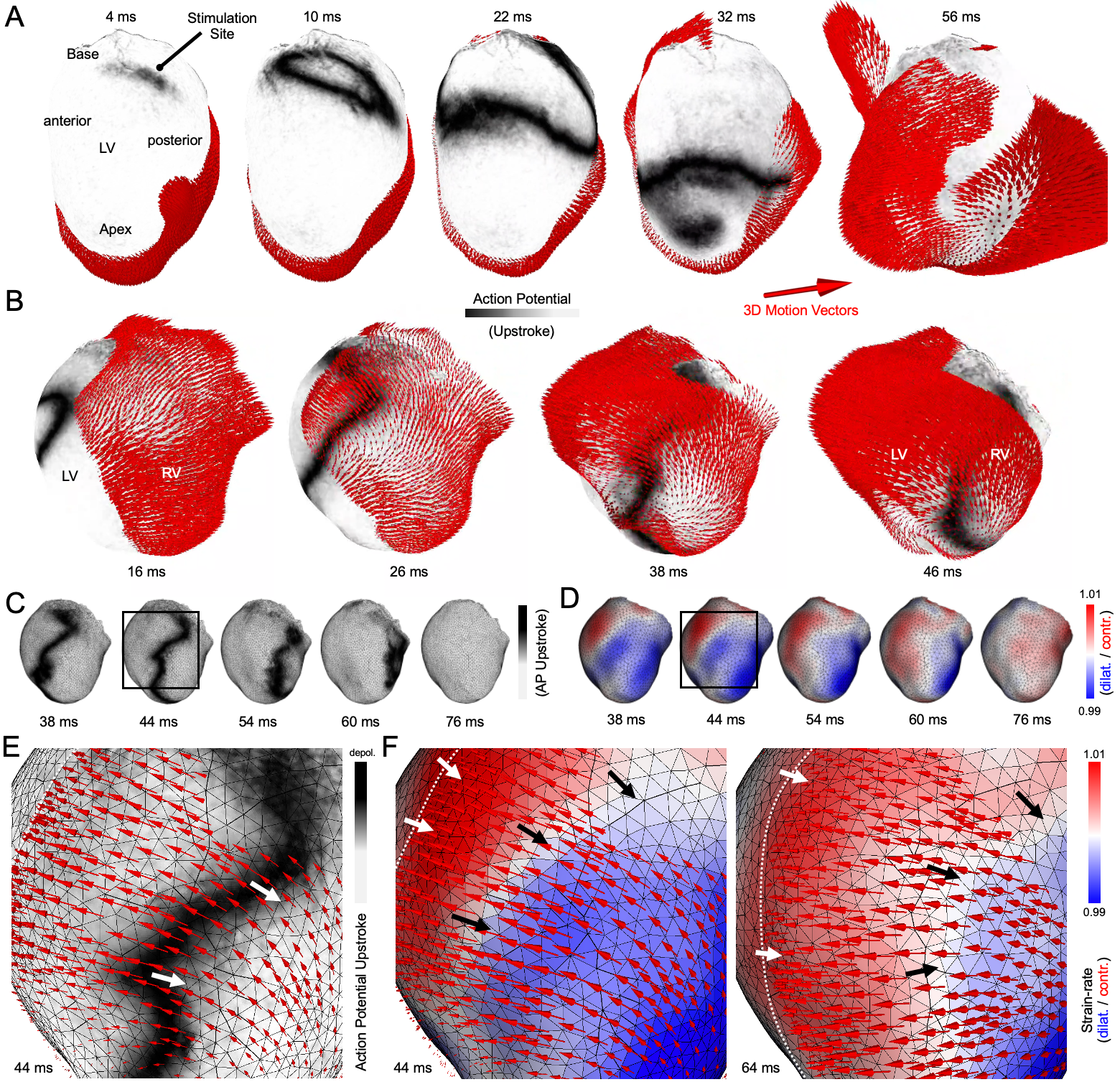}
    \caption{
    Electromechanical wave during pacing (4.5 Hz) in contracting isolated rabbit heart, see also Supplementary Videos \ref{video:pacing1} and \ref{video:pacing1vectors}.
    \textbf{A} Action potential (AP) wavefront propagating away from stimulation site across the deforming ventricular surface, see also Fig.~\ref{fig:Pacing2}. Black: Depolarizing tissue / action potential upstroke (temporal derivative of optical trace, continuous green excitation). Red: 3D displacement vectors indicating motion with respect to previous frame acquired 4 ms earlier (vectors pointing into the heart not visible). Frontal view showing focal wave propagating across left ventricular (LV) free wall away from stimulation site. The ventricles get electrically activated within about \SI{50}{\milli \second}.
    \textbf{B} Side view showing action potential wave propagating away from stimulation site across posterior wall and right ventricle (RV) towards opposite side of the heart.
    The tissue starts moving towards the approaching wave and contracts noticeably when the AP wavefront has propagated across and activated the whole muscle.
    \textbf{C} Electrical activation wavefront (black: action potential upstroke) propagating across posterior wall, c.f. panel A.
    \textbf{D} Mechanical activation wavefront consisting of a transition from dilating (blue) to contracting (red) strain-rates measured as the rate of triangular area change of each mesh triangle. The mechanical wave follows the electrical wave with a short delay of about \SI{5}{\milli \second}, see also Fig.~\ref{fig:Pacing2}G). The phenomenon occurs on both the posterior and anterior walls, see Figs.~\ref{fig:Pacing2} and \ref{fig:SupplementElectromechanicalWave}.
    \textbf{E} Close-up of electrical wave (white arrows) together with tissue displacement vectors (red, indicate motion with respect to the tissue configuration 4 ms earlier). The tissue is being pulled towards the approaching electrical wave.
    \textbf{F} Close-up of mechanical wave (black vectors) with dilating tissue (blue) in front of the electrical wave and contracting tissue (red) shortly after the electrical activation behind the wavefront, see also Fig.~\ref{fig:SupplementPacing2}B). The vectors undergo a rapid change in direction within the contracting region (white dotted line). The data was obtained with continuous green illumination at 500 fps.
    The electromechanical wave consists of two closely coupled electrical and mechanical activation waves, see also Fig.~\ref{fig:Pacing2}.
    }
    \label{fig:Pacing1}
\end{figure*}

We performed fully panoramic voltage-sensitive optical mapping with contracting isolated rabbit hearts, imaging action potential waves across the entire 3D contracting ventricular surface during sinus rhythm, ventricular pacing, and ventricular fibrillation (VF), see Figs.~\ref{fig:Sinus1}-\ref{fig:VF2} and Supplementary Videos \ref{video:SV2} and \ref{video:sinus}-\ref{video:VF2}.
Using our technique, we provide, for the first time, high-resolution, 3D measurements of electromechanical waves, including electromechanical vortices during VF, as well as measurements of action potential duration (APD) and contractile changes during pharmacological blockage of potassium ion channels in an isolated intact contracting rabbit heart.
The videos as well as interactive 3D renderings are available at \href{https://cardiacvision.ucsf.edu/videos/3d-optical-mapping/}{https://cardiacvision.ucsf.edu/videos/3d-optical-mapping/}.

\subsection*{Sinus Rhythm}\label{sec:results:sinus}
Sinus rhythm poses the most challenging condition in optical mapping studies with contracting hearts.
During sinus rhythm, the heart usually contracts and deforms more strongly, which makes the analysis more challenging in two specific ways: firstly, tracking the tissue and, secondly, measuring artifact-free optical traces of the action potential.
Tracking is difficult particularly when the heart exhibits large translational or torsional movements.
For instance, when tracking with a single camera, 2D tracking algorithms fail to detect shifts of the tissue if the heart rotates or moves out of the field of view, see Supplementary Video \ref{video:2Dvs3D}.
The second challenge lies in the strong back and forth movements of the heart relative to the excitation light, which can lead to local changes in the excitation of the fluorescent dye.
This may, depending on the strength of the motion and the inhomogeneity of the excitation light, produce strong measurement artifacts and prohibit measuring the time-course of the action potential. 

We overcome these challenges with our 3D multi-view mesh-based tracking approach combined with ratiometric voltage-sensitive imaging.
Figs.~\ref{fig:Sinus1}, \ref{fig:Sinus2}, and \ref{fig:SupplementTracking} as well as Supplementary Videos \ref{video:SV3}, \ref{video:SV5}, \ref{video:green-blue-video-3D}, and \ref{video:green-blue-video-3D-texture} show that we can track the entire strongly deforming ventricular surface during sinus rhythm, and Figs.~\ref{fig:Sinus2}B,C), \ref{fig:Sinus3}B) show that we can subsequently map action potential waves and action potential wavefronts propagating across the entire, strongly deforming 3D ventricular surface with ratiometric and non-ratiometric imaging, respectively. 
In particular, with ratiometric imaging, we can measure the time-course of the action potential, see Figs.~\ref{fig:Setup}F), \ref{fig:Sinus3}D), \ref{fig:SupplementSinusAPTraces} and measure its action potential duration (APD), see Figs.~\ref{fig:Sinus3}D,E) and \ref{fig:SupplementAPD}.

Figs.~\ref{fig:Setup}F), \ref{fig:Fig2}E), \ref{fig:Sinus1}-\ref{fig:Sinus3}, and \ref{fig:SupplementTracking} show the reconstructed 3D ventricular surface together with displacement vectors (red) indicating the motion and deformation of the heart as it contracts in response to the depolarization of the ventricles.
The vectors in Figs.~\ref{fig:Fig2}E), \ref{fig:SupplementTracking}, \ref{fig:Sinus1}C) and \ref{fig:Sinus2}A,C) indicate displacements of the epicardial surface with respect to its diastolic mechanical configuration just before the onset of the depolarization (at 0 ms), while the vectors in Figs.~\ref{fig:Setup}F), \ref{fig:Sinus1}D), and \ref{fig:Sinus3}D,E) indicate instantaneous displacements between subsequent frames.
The instantaneous displacements reflect contractile speed, see also Fig.~\ref{fig:Sinus3}D) (absolute values). 
The data in Fig.~\ref{fig:Sinus1}D) shows that the heart exhibits torsional motion around the apex. 
It moves upwards as it contracts, while its base remains roughly in its vertical position, see Fig.~\ref{fig:Sinus1}A).
The 3D tracking allows us to compensate motion artifacts and measure optical traces across the heart surface in a co-moving frame of reference, as illustrated in Figs.~\ref{fig:Fig2}E) and Supplementary Video \ref{video:SV4}.
Further, we can compute optical maps showing the spatio-temporal electrical activity per pixel (on each triangle) and correlate the electrical activity with the mechanical contraction of the heart.
The action potential wavefront (black) shown in Figs.~\ref{fig:Sinus2}B) and \ref{fig:Sinus3}B) (measured with continuous green excitation) corresponds to the action potential upstroke, which was computed in the co-moving frame as the difference between pixel intensities in subsequent frames (pixel-wise normalized temporal derivative of the optical maps).
The action potential wave originates in several locations activating the entire ventricular muscle within less than \SI{20}{\milli \second}, see also corresponding electrical activation map, see Fig.~\ref{fig:Sinus3}A).
The ventricles begin to contract and deform shortly thereafter, see Figs.~\ref{fig:Sinus1}C,D) and \ref{fig:Sinus2}B).
The epicardial surface experiences mostly contractile strain with a local area decrease of about 10-20\% (triangular area change) and tensile strain in some areas with a local area increase of about 5-15\%, see Fig.~\ref{fig:Sinus3}C).
Noticeable contraction sets in only about \SI{30}{\milli \second} after the onset of the depolarization of the ventricles.
The ventricles exhibit maximal contraction velocities of about 0.1 m/s about 40-50 ms after the onset of the depolarization, see Fig.~\ref{fig:Sinus1}D). 
These high contraction speeds then lead to the large deformations depicted by the wireframe mesh in Fig.~\ref{fig:Sinus1}A) and the corresponding displacements shown in Fig.~\ref{fig:Sinus1}C). 
Our data shows that the depolarization phase of the action potential and the onset of contraction are clearly separated in time.
By contrast, the action potential wave (black) shown in Fig.~\ref{fig:Sinus2}C) (measured with pulsed green-blue excitation) corresponds to the time-course of the action potential (pixel-wise normalized optical maps), not just its upstroke. 
Correspondingly, the depicted wave covers a larger timespan (about \SI{200}{\milli \second} instead of \SI{20}{\milli \second}) and the depicted displacement vectors indicate the overall displacement of the ventricular surface during the plateau phase of the action potential.
Figs.~\ref{fig:Setup}G), \ref{fig:Sinus3}D), \ref{fig:SupplementSinusAPTraces}, and \ref{fig:SupplementBaselineCorrection1}-\ref{fig:SupplementAPD} show measurements of the time-course of the action potential for various locations (from center of a triangle), and Figs.~\ref{fig:Sinus3}E) and \ref{fig:SupplementAPD} shows APD measurements across the entire ventricular surface (per triangle).
With low potassium barium-chloride Tyrode, the action potential becomes prolonged (APD increases) and the contraction shortens compared to with normal Tyrode, see arrows in two lower graph in Fig.~\ref{fig:Sinus3}D). 
We observe a positive and a negative electromechanical window with normal and low potassium barium-chloride Tyrode, respectively. 
A negative electromechanical window is associated with the action potential repolarization orccuring before the heart muscle's relaxation, see \cite{terBekke2015}.

Our data demonstrates that we can measure both the electrical and mechanical dynamics of the heartbeat simultaneously at high spatial and temporal resolutions.
The resolution of the electrical measurements, except the APD measurements, is higher than the mechanical measurements.
We measured electrical activation and optical traces per pixel (superimposed data from multiple cameras) and APD and strain per triangle.
In the example in Fig.~\ref{fig:Sinus1} and \ref{fig:Sinus2}A), both the wireframe mesh and the vectorfield consist of $3,325$ vertices or displacement vectors, respectively, the vertices forming $6,349$ polygon triangles. 
The average distance between two neighbouring vertices is \SI{0.68 \pm 0.30}{\milli \meter} and the average area of a triangle is about \SI{0.34 \pm 0.11}{\milli \meter^2} (at 0ms).
The entire ventricular surface is resolved by 668,793 pixels, where each triangle comprises about 50-150 pixels, see Fig.~\ref{fig:Fig2}D).
The spatial resolution of the texture map is so high that it is possible to identify individual small vessels and other anatomical features across the ventricular surface, see Figs.~\ref{fig:Sinus1}B) and \ref{fig:Sinus2}A).
Per pixel, the spatial resolution is about \SI{120}{\micro\metre}.
The horizontal (short-axis) diameters of the heart are \SI{2.69}{\centi \meter} and \SI{3.02}{\centi \meter}, respectively (along the shorter and longer axes of the elliptical shape of the heart).

\subsection*{Electromechanical Waves during Paced Rhythms}\label{sec:results:pacing}
\begin{figure*}[htb]
  \centering
  \includegraphics[clip, trim=0.0cm 0.0cm 0.0cm 0.0cm, width=0.98\textwidth]{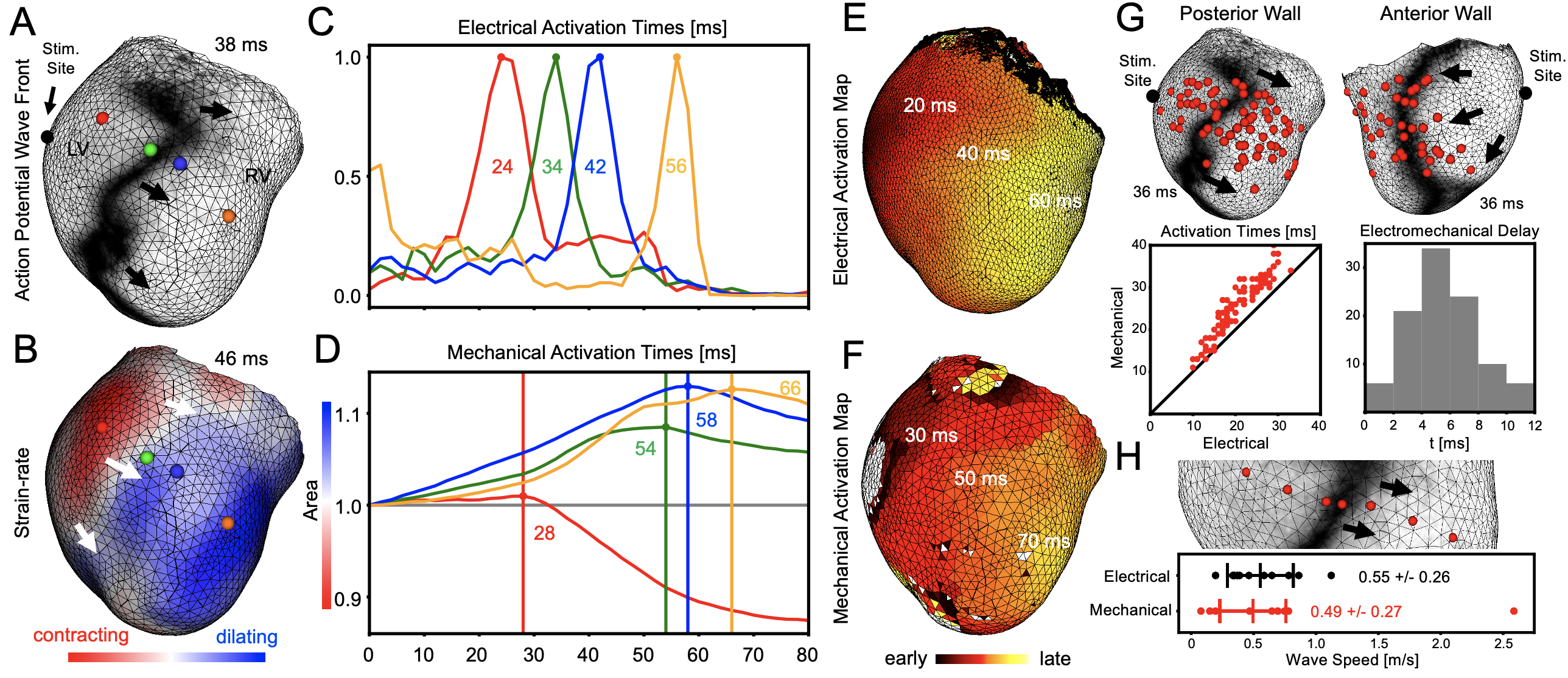}
  \caption{
  Electrical and mechanical activation during ventricular pacing in intact isolated rabbit heart.
  \textbf{A} Action potential wavefront (black) propagating across posterior wall activating subsequent measurement points (red, green, blue, orange).
  \textbf{B} Strain-rate measured as rate of triangular area change (blue: increasing, red: decreasing) across 3D ventricular epicardial surface following electrical wave front.
  \textbf{C} Electrical activation times (in milliseconds) computed from temporal derivative of voltage-sensitive optical signals highlighting depolarization phase or upstroke of the action potential.
  \textbf{D} Mechanical activation times measured as the transition from positive to negative strain-rates (dilating to contracting, slope of strain or area change) calculated from triangular area change. The triangular area is normalized to 1.0 with respect to the area at the time of the pacing pulse at 0 ms. The mechanical activation times are delayed with respect to the electrical activation times (4, 20, 16, 10 ms).
  \textbf{E} Electrical activation map computed from activation times.
  \textbf{F} Mechanical activation map computed from mechanical activation times.
  \textbf{G} Mechanical over electrical activation times measured from various points scattered across posterior and anterior walls, and distribution of electromechanical delays between electrical and mechanical activation times (mean: $5.0 \pm 2.4 ms$).
  \textbf{H} Speed of electrical and mechanical activation fronts measured along linear path (red dots). Electrical ($0.55 \pm 0.26 \, m/s$, CV) and mechanical ($0.49 \pm 0.27 \, m/s$) wave speeds.
  }
  \label{fig:Pacing2}
\end{figure*}

To study the relationship between electrical activation and mechanical deformation, we imaged focal action potential waves propagating across the ventricles away from a stimulation site, see Figs.~\ref{fig:Pacing1}-\ref{fig:Pacing2} as well as Supplementary Videos \ref{video:pacing1}-\ref{video:pacing2}.
The pacing pulses were applied using a thin, flexible stimulation electrode which was positioned on the ventricular epicardial surface, as shown in Figs.~\ref{fig:SupplementDepthMaps}, \ref{fig:SupplementTrackingElectrode}, and \ref{fig:SupplementCoverage}.
Using our multi-view reconstruction methodology, we were able to track the tissue underneath the electrode as it is always visible in at least 2-3 of the other camera images, see Fig.~\ref{fig:SupplementCoverage}A,B).

Fig.~\ref{fig:Pacing1}A,B) shows a focal action potential wavefront (black) propagating across the ventricles after the application of a pacing pulse during continuous pacing at \SI{4.5}{\hertz}.
The electrical wavefront corresponds to the upstroke or depolarization phase of the action potential (non-ratiometric imaging with continuous green excitation) and, therefore, to the moment when the tissue becomes electrically activated.
The electrical activation wavefront propagates away from the stimulation site towards the opposite side of the heart. 
We computed an electrical activation map from this data, see Figs.~\ref{fig:Pacing2}E) and \ref{fig:SupplementElectricalActivationMap} and Supplementary Video \ref{video:SV13}.
The electrical activation map exhibits an elliptical pattern around the pacing location, reflecting the anisotropic conduction of the action potential wave.
At the same time, the heart contracts and deforms in response to the electrical activation. 
Displacement vectors (red) indicate the motion of the epicardial surface (with respect to the mechanical configuration in a previous frame, acquired 4 ms earlier), see also Supplementary Video \ref{video:pacing1vectors}.
Initially, the heart wall contracts close to where the pacing pulse is applied, but the contractile motion sets in following the electrical activation.
Shortly thereafter (30-40 ms) the wall on the opposite side moves towards the origin of the wave.
Note that only vectors pointing outwards, away from the epicardial surface are visible, while vectors pointing beneath the surface are hidden behind the texturized surface.
The data demontrates a strong correlation between the ventricle's electrical activation and motion.
Here, motion is represented as kinematic data or as a displacement vector field.

Fig.~\ref{fig:Pacing1}C-F) shows that the focal action potential wave causes the ventricles to deform in a very characteristic way: 
as the muscle contracts close to the pacing location, tissue further away is being pulled toward the contracting part of the heart. 
Subsequently, the tissue that is being pulled experiences dilation or tensile stretch.
Moreover, as the electrical wavefront spreads across the ventricles, the contraction closely follows the electrical wave, see also Figs.~\ref{fig:SupplementElectromechanicalWave}-\ref{fig:SupplementCrosstalk} and Supplementary Videos \ref{video:pacing1} and \ref{video:pacing2}.
Plotting the epicardial strain-rate (measured as the rate of triangular area change), we see that the electrical wavefront is immediately followed by a mechanical wavefront that is defined as the transition from dilating (blue) to contracting (red) strain-rates, see Fig.~\ref{fig:Pacing2}C,D).
As the tissue is pulled towards the approaching electrical wave (red vectors) it dilates (blue) and then suddenly contracts (red) as the electrical wave passes through.
This mechanical activation front follows the electrical activation front with a short delay of about \SI{5.0 \pm 2.4}{\milli \second} as it propagates across the posterior and anterior ventricular walls, see Fig.~\ref{fig:Pacing2}C,D,G).
Based on this observation, we computed the mechanical activation times as the transition from positive or dilating to negative or contracting strain-rates, see Fig.~\ref{fig:Pacing2}D), and subsequently computed a mechanical activation map, see Fig.~\ref{fig:Pacing2}F), which exhibits similarities to the corresponding electrical activation map, see also Supplementary Video \ref{video:SV13}.
The delays between electrical and mechanical activation times were calculated for the points (red) shown in Fig.~\ref{fig:Pacing2}G).
Additionally, we computed the propagation speeds of the electrical and mechanical wavefronts across the posterior wall along the path shown in Fig.~\ref{fig:Pacing2}H) (between one to the next as well as to the second-next point), and the speeds are comparable (electrical: $0.55 \pm 0.26$ m/s, mechanical: $0.49 \pm 0.27$ m/s).
Our data is further evidence and the first direct high-resolution observation of electromechanical waves in the heart.
Previous studies described both electromechanical waves or correlated electrical and mechanical activation \cite{Wyman1999,Provost2011, Christoph2018Nature, Lebert2019, Molavi2022, Maffessanti2020}, but never obtained all data at once and at high resolutions (only mechanical data \cite{Wyman1999,Provost2011}, partly 2D electrical and 3D mechanical data \cite{Christoph2018Nature}, low resolution data of both electrical and mechanical activation \cite{Maffessanti2020}) or were {\it in silico} studies \cite{Lebert2019, Molavi2022}.
For instance, Provost et al. \cite{Provost2011} imaged mechanical waves in patients {\it in vivo}, and compared them to simulated electrical and mechanical waves obtained with computer models.
Until now, simultaneous high-resolution 3D observations of both the electrical {\it and} mechanical wave phenomena have been missing.

It is important to note that, while the electrical and mechanical activation times correlate well across a wide area of the ventricles, they do not correlate well everywhere, particularly in the vicinity to the pacing electrode, see also Supplementary Video \ref{video:SV13}.
This is not surprising given that the electrical and mechanical waves are fundamentally different physical phenomena emerging in an excitable and elastic system, respectively.
In general, we presume that there are many counter-examples where electrical and mechanical activation and wave speed do not match.
The mechanical wave shown in Fig.~\ref{fig:Pacing2}B,D) emerges while propagating along the fiber orientation. 
In this case, the contractile motion is aligned with the wavefront propagation direction. 
As previously described in Extended Data Fig. 8 in \cite{Christoph2018Nature}, Fig. 3 in \cite{Lebert2019}, and Figs. 5 and 6 in \cite{Molavi2022}, this configuration facilitates the mechanical wave phenomenon.
The observation is also determined by how we define the mechanical activation time: the sudden transition from dilating to contracting strains is dependent on the alignment between the wave normal and local muscle fiber direction and can only be measured sufficiently far away from the pacing location, as it requires the tissue to first dilate before it contracts.
The tissue surrounding the origin of the pulse starts contracting almost immediately.
However, the situation might be different in a different experimental setting (e.g. optical stimulation).
Our data shows that, while coupled electromechanical wave phenomena exist, mechanical wavefronts are unreliable approximations of electrical wavefronts and more systematic measurements are necessary. 
While often in good agreement, their relationship is more complicated. 
The degree of correlation likely depends on various factors, such as fiber anisotropy, elastic inhomogeneities, and mechanical boundary conditions, as also described in a numerical study by Molavi et al. \cite{Molavi2022}. 
Further research is needed to establish a better understanding of the relationship between electrics and strain mechanics and to determine whether there is a mapping between the two.

\subsection*{Ventricular Fibrillation}\label{sec:results:vf}
\begin{figure*}[ht]
  \centering
  \includegraphics[clip, trim=0.0cm 0.0cm 0.0cm 0.0cm, width=0.98\textwidth]{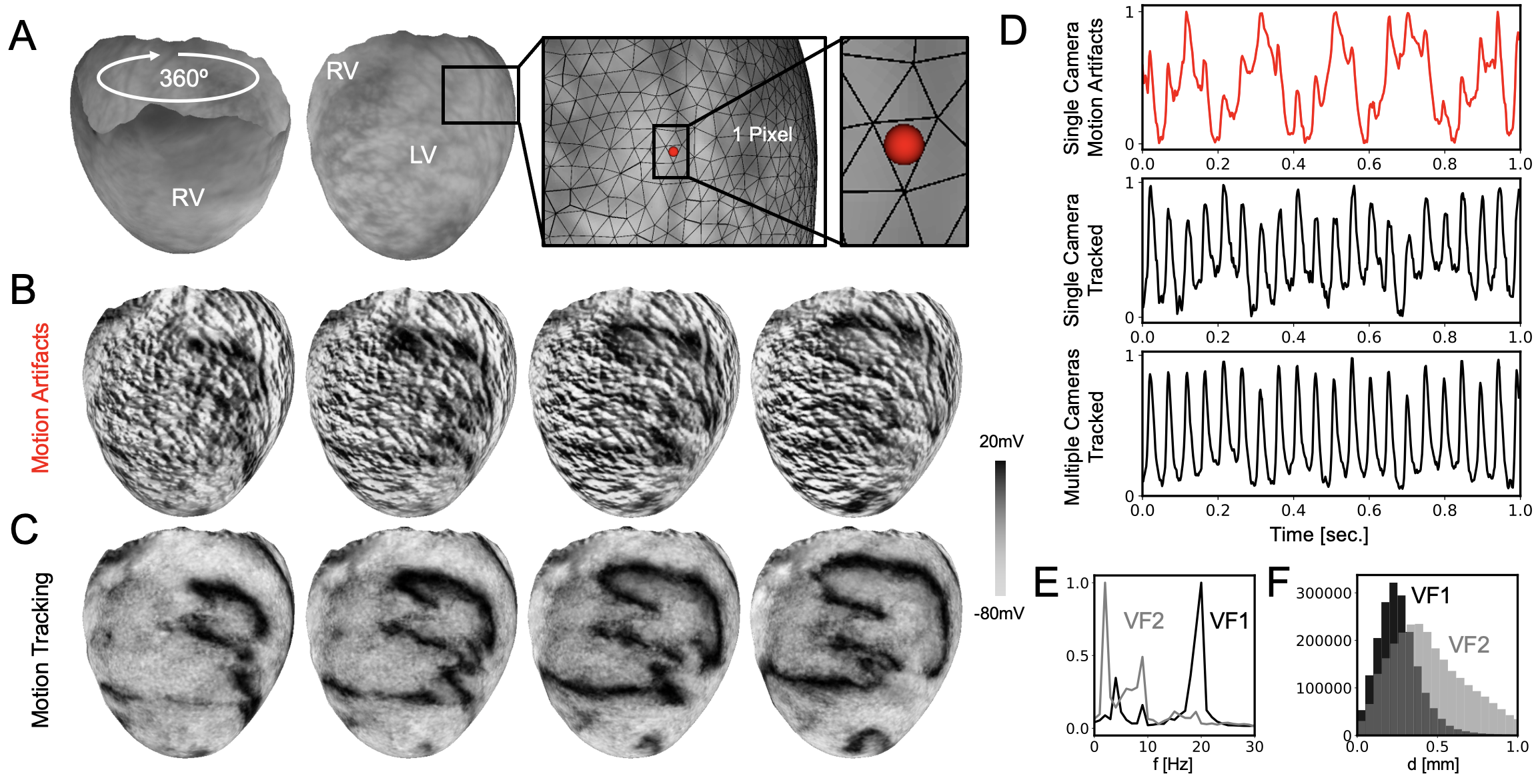}
  \caption{
Ventricular fibrillation (VF) imaged across the entire surface of a moving isolated rabbit heart using voltage-sensitive panoramic optical mapping with 12 cameras, see also Supplementary Video \ref{video:VF1}.
\textbf{A} Reconstructed 3D ventricular surface and location (red dot) in which optical traces in D were measured.
\textbf{B} Optical maps (3D) before motion tracking containing motion artifacts.
\textbf{C} Tracked and motion-stabilized optical maps (3D) without motion artifacts showing action potential vortex waves on left ventricular (LV) surface (black: depolarized tissue, light gray: repolarized tissue, pixel-wise normalized video, 30-50 frames sliding window as in B).
\textbf{D} Optical traces before (top) and after motion tracking with a single (center) and multiple (bottom) cameras (measured in one pixel, location in center of triangle shown in A).
\textbf{E} Frequency spectra (cumulative, 1 sec. long video, computed from all motion-compensated optical traces) of fast VF1 episode (black, dominant peak at \SI{20}{\hertz}) shown in A-D and F, and 'slow' VF2 episode (light gray, dominant peak at \SI{10}{\hertz}) shown in Fig.~\ref{fig:VF2} (secondary peaks caused by residual motion artifacts).
\textbf{F} Episode VF1 (black) exhibits minimal motion, while episode VF2 (light gray) exhibits larger contractile motion: distribution of displacement magnitudes $d$ [mm] with respect to first frame (1 sec. long video). Motion artifacts in B are substantial even though the motion is less than \SI{0.5}{\milli \meter}.
  }
  \label{fig:VF1}
\end{figure*}

\begin{figure*}[htb]
  \centering
  \includegraphics[clip, trim=0.0cm 0.0cm 0.0cm 0.0cm, width=0.98\textwidth]{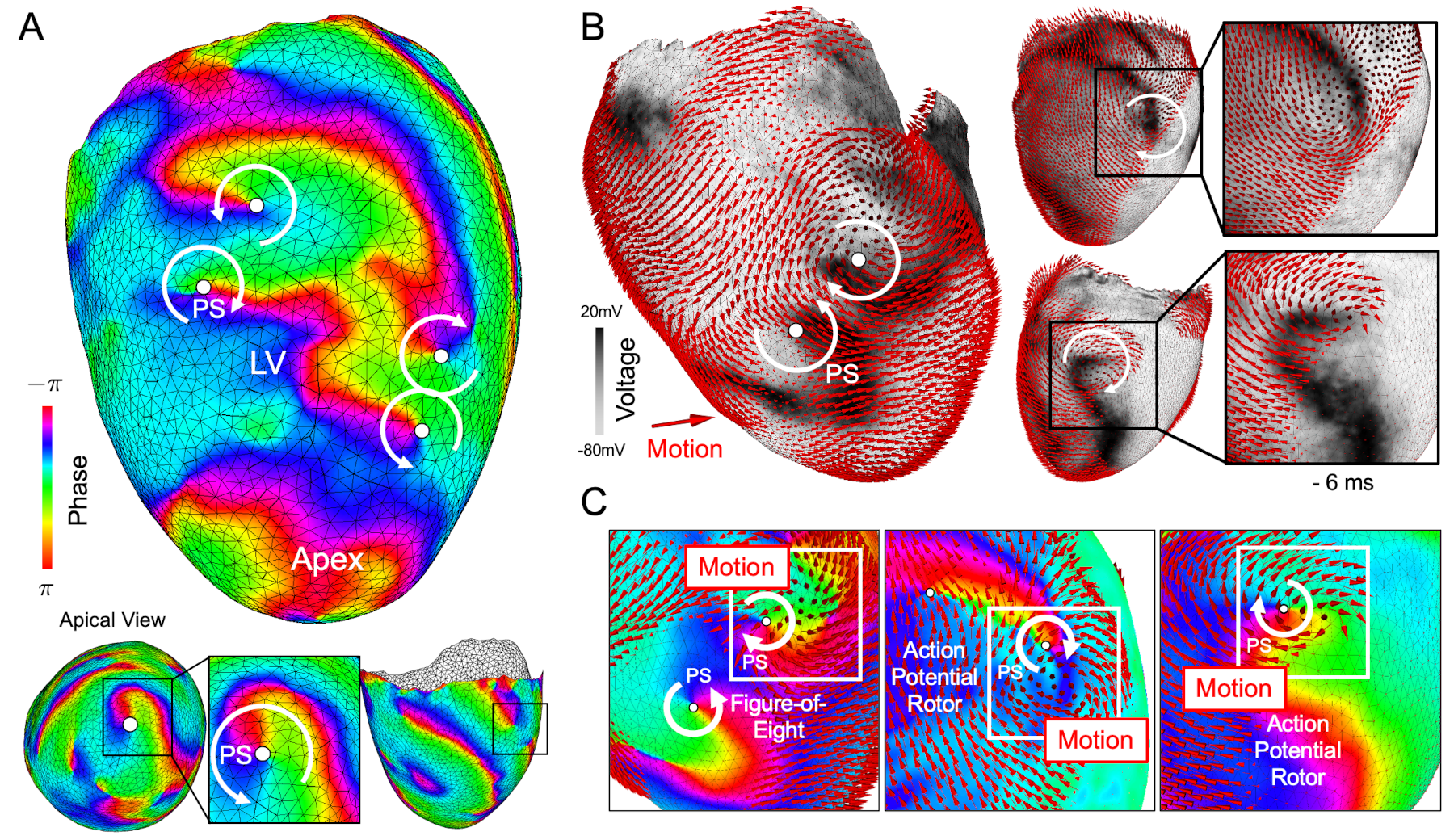}
  \caption{
Vortex-like rotating action potential waves on the ventricular surface of an isolated fibrillating and contracting rabbit heart, see also Supplementary Videos \ref{video:VF1} and \ref{video:VF2}.
\textbf{A} Phase maps showing action potential vortex waves including rotors with core regions marked by phase singularities (PS) in a rapidly fibrillating heart with little residual motion (VF1, same data as shown in Fig.~\ref{fig:VF1}).
\textbf{B} Rotating action potential vortex waves (black) cause curl-like rotating mechanical deformation patterns (red vectors).
In order to highlight curls in the displacement field, the vectors are normalized and, therefore, do not encode magnitude but only direction. 
Tissue displacements calculated with respect to tissue's mechanical configuration $20$ ms earlier. 
Inserts show the same area $6$ ms earlier. 
In some regions, the vectors point into the heart muscle and are therefore not visible.
This episode (VF2) shows a more strongly contracting heart than in panel A.
\textbf{C} Phase maps of action potential vortex wave activity displaying phase singularities (PS) indicating rotational core region together with mechanical curls in the displacement vector fields, which emerge in close vicinity to electrical PS.
  }
  \label{fig:VF2}
\end{figure*}

Hearts contract only minimally during ventricular fibrillation (VF).
However, the residual contractile motion during VF is yet strong enough to produce substantial motion artifacts, as shown in Fig.~\ref{fig:VF1}B). 
In this study, we discuss two episodes of VF: one with very little to almost no motion (VF1), shown in Figs.~\ref{fig:VF1}, \ref{fig:VF2}A) and Supplementary Video \ref{video:VF1}, and one with moderate contractile motion (VF2), shown in Figs.~\ref{fig:VF2}B,C) and Supplementary Video \ref{video:VF2}. 
The different contractile strengths between the two VF episodes can be associated with a much shorter action potential duration and accelerated fibrillatory dynamics (\SI{20}{\hertz}) which leads to very minimal motion in episode VF1, versus a longer action potential duration with slower fibrillatory dynamics (\SI{10}{\hertz}) resulting in moderate contractile motion in episode VF2, see Fig.~\ref{fig:VF1}E,F).
While episode VF1 is an interesting example to discuss motion artifacts, episode VF2 is interesting in terms of its ventricular mechanics.
Both episodes were recorded in the same heart.

The VF1 episode in Fig.~\ref{fig:VF1} shows that even with very little to almost no motion it is necessary to track and compensate motion in optical mapping studies, as otherwise motion artifacts become so strong that they prohibit further analysis of the data.
Fig.~\ref{fig:VF1}B) shows pixel-wise normalized optical maps, which were not tracked and subsequently not analyzed in a co-moving frame. 
Motion artifacts manifest as the high-frequency dark-bright pattern, which superimposes the action potential waves, see also Fig. 3C) in \cite{Christoph2018Nature} and Figs. 3, 5B) and 7B,D) in \cite{Christoph2018Frontiers}.
By contrast, panel C) shows pixel-wise normalized optical maps computed using co-moving signal analysis. 
They do not exhibit any significant motion artifacts, but instead multiple reentrant vortex-like action potential wave patterns.
Note that in Figs.~\ref{fig:VF1} and \ref{fig:VF2}, we display the action potential wave and not the action potential wavefront as in Figs.~\ref{fig:Sinus2}B), \ref{fig:Sinus3}B) and \ref{fig:Pacing1}.
The dark area corresponds to depolarized tissue, the bright area to repolarized tissue (normalized units [0,1]), as in Fig.~\ref{fig:Sinus2}C).
In both Figs.~\ref{fig:VF1}B,C) and \ref{fig:VF2}B), we amplified intensity fluctuations numerically using pixel-wise normalization with a sliding window (with a length of 50 frames or \SI{100}{\milli \second}).
The data was generated with continuous green excitation without ratiometric motion artifact compensation. 
Therefore, changes in the excitation were not compensated and are accordingly visible as slight baseline changes, see Supplementary Videos \ref{video:VF1} and \ref{video:VF2}.
In episode VF1, the action potential duration is extremely short in the order of 30-40 ms and the wavelength is in the order of 1-3 mm, respectively.
The tissue only moves on average by about 0.2-0.3 mm, see Fig.~\ref{fig:VF1}F).
The fact that the strong motion artifacts in Fig.~\ref{fig:VF1}A) occur at all with this little motion highlights the necessity to track the tissue's motion to compensate motion artifacts.
Fig.~\ref{fig:VF1}D) shows optical traces before and after motion tracking with and without motion artifacts, respectively. 
The top trace shows a measurement without tracking obtained from a single pixel of one of the cameras (close to the red dot shown in panel A). 
The time-series shows an unrecognizable, distorted sequence of action potential waves superimposed by strong intensity modulations, which are caused by the back and forth movement of the tissue (the darker nearby vessel moves across the pixel).
The trace below shows the corresponding measurement obtained in a co-moving frame (red dot in panel A), again from a single pixel of a single camera, but after 3D and 2D tracking. 
Due to the tracking and co-moving signal analysis, the motion artifacts are inhibited and the trace shows instead a sequence of action potentials.
The bottom trace shows the averaged signal obtained with multiple cameras.
While ratiometry can effectively reduce measurement artifacts in general, it is not necessarily required during VF, see also \cite{Kappadan2020}.
We computed phase maps from the motion-stabilized traces, see Fig.~\ref{fig:VF2}A). 
The phase maps emphasize the vortex-like structure and topological organization of the action potential wave pattern, and indicate the rotational centers of the electrical vortex waves as phase singular points (PS, white dots).
In this example, there are several rotating vortices including a figure-of-eight pattern in the anterior left ventricle (LV) and a persistent rotating vortex close to the apex.
The phase singularities allow us to automatically track electrical vortex waves across the moving 3D ventricular surface.

With the larger contractions in episode VF2, it is possible to observe reentrant action potential vortex waves inducing vortex-like 3D deformation patterns on the epicardial ventricular surface, see Fig.~\ref{fig:VF2}B,C) and Supplementary Video \ref{video:VF2}.
Panel B) in Fig.~\ref{fig:VF2} shows three examples of spiral wave-like action potential waves (black: depolarized tissue, light gray: repolarized tissue, normalized units [0,1]): a figure-of-eight pattern and two clockwise spiral waves in the left and right ventricles, respectively.
In this slower episode, the action potential waves induce noticeable contractions, which lead to macroscopic deformations of the ventricles in response to the electrical activation.
The dominant VF frequency is about 10Hz, see Fig.~\ref{fig:VF1}E).
At these slower activation frequencies, the contractions are stronger than at the higher frequencies in episode VF1 in Fig.~\ref{fig:VF1}, which is an observation that was also made in \cite{Christoph2018Nature}.
The tissue moves on average by about \SI{0.5}{\milli \meter}, see Fig.~\ref{fig:VF1}F). 
The larger deformations are associated with a spatio-temporal displacement vectorfield (red).
Note that the displacement vectors are scaled differently than in Figs.~\ref{fig:Sinus1}-\ref{fig:Pacing2}.
Here they display only the direction of the movements of the vertices and not the magnitude of these movements.
We chose this visualization to highlight vortices or curls in the displacement vectorfields, which emerge close to electrical phase singularities (PS).
Note that displacement vectors pointing into the heart muscle are not visible.
Fig.~\ref{fig:VF2}C) shows the corresponding phase maps and the mechanical curls in the vicinity of electrical PS.

Our data demonstrates a strong correlation between action potential waves and tissue deformation during VF, and hints at topological defects in the deformation of the heart wall which co-localize with electrical PS.
Our data confirms previous observations that were made when analyzing single-camera optical mapping data \cite{Christoph2018Nature}.
Here, we present, for the first time, corresponding 3D measurements of action potential vortex waves and co-localized vortex-like tissue deformation across the ventricular surface during VF.

\section*{Discussion}

In this study, we demonstrated that it is possible to image action potential waves at high resolutions across the entire deforming ventricular surface of contracting isolated hearts using ratiometric voltage-sensitive multi-camera optical mapping and 3D numerical motion tracking.
With our setup, we are able to measure electrical and mechanical phenomena simultaneously across the heart surface at high spatial and temporal resolutions.
For instance, we can measure electrical and mechanical activation times, action potential durations, tissue strain, and strain-rates at thousands of locations across the 3D heart surface.
The measurement data is invaluable as it can be used to study the heart's electromechanics in unprecedented detail.
In the future, it could be used to study disease mechanisms and fundamental properties of cardiac physiology, or to inform and calibrate electromechanical computer models of the heart.

\subsection{Practical Considerations}

While we used 12 cameras in the present study, our imaging chamber fits up to 24 cameras. 
We anticipate that, in the future, we will use up to 24 cameras to image the whole heart including the atria, and eventually use 2 sets of cameras to image action potential and calcium waves.
This would enable us to study the three main physiological players underlying the heartbeat: voltage, calcium, and contraction.
Our overall approach and design of the imaging setup utilizes the recent availability of industrial-grade low-cost cameras (\citenum{Lee2017}). 
These cameras produce lower quality videos, but averaging the overlapping video images across the 3D heart surface produces high quality optical signals.
The lower costs enabled us to integrate far more cameras into our setup than previously possible: 
in the past, multi-camera optical mapping systems consisted of usually not more than 4 scientific cameras \cite{Zhang2016,Christoph2017}, the main limiting factor being cost.
While low-cost cameras were already used for panoramic optical mapping of non-moving hearts \cite{Lee2017} and while 3-4 cameras are usually sufficient for mapping the entire surface of non-moving hearts \cite{Bray2000,Kay2004,Kay2006,Rogers2007}, it was estimated in Christoph \& Schr\"oder-Schetelig \cite{Christoph2017} that at least 8 cameras would be necessary to image the entire ventricular surface of a contracting heart.

In this study, we showed that operating 12 or potentially more cameras is technically and practically feasible.
We streamed the video data from 12 cameras onto a single computer using USB, and we did not experience major issues with camera failures, the high data rates, or calibrating the cameras.
Occasionally, cameras failed or could not be included during the data analysis, but this could be tolerated to some extent as one camera could usually be compensated by surrounding cameras if there was sufficient coverage of that area of the heart.
Further, the high data rates and large amounts of data that need to be captured, stored and processed during and after each experiment do pose a challenge.
With 24 cameras and video image resolutions of $440 \times 320$ pixels, we would obtain data rates of 3 gigabytes per second.
Accordingly, a 10 second long recording with 24 cameras would create 30 gigabytes of data.
In our case with 12 cameras, a single experiment with 25-50 recordings produced on average about 0.5-1.0 terabytes of data.
While we crop the video images and discard unnecessary data during the experiment to save storage space, further work is needed to compress the raw data into a suitable format for long-term storage.

Despite the many cameras, our imaging setup is cost-effective with an overall approximate cost of \$20,000 for 24 cameras (\$400 each) including the respective lenses (\$500-800 each).
By comparison, a single scientific camera costs \$10,000-\$50,000 (e.g. Brainvision Scimedia MiCAM Ultima / N256, Teledyne Photometrics Evolve 128, Oxford Instruments Andor) and a setup of 4-10 scientific cameras would correspondingly cost \$100,000-\$500,000, which is prohibitively expensive for most laboratories.
Adopting our low-cost design with industrial-grade cameras could enable more researchers to venture into the world of panoramic 3D optical mapping.
Costs for emission or excitation filters vary widely depending on the application (\$30-50 per filter for standard non-ratiometric longpass emission filters vs. \$500-1000 per filter for custom-made ratiometric bandpass filters; no excitation filters for non-ratiometric imaging and \$100-200 per excitation filter for ratiometric imaging).

\subsection{Advantages}

The two main advantages of our panoramic optical mapping approach are that it does neither rely on 1) fiducial markers \cite{Zhang2016} nor on 2) any additional shape reconstruction methods \cite{Christoph2017}. 
It therefore provides dynamic high-resolution 3D reconstructions immediately, and it could, in principle, also provide these reconstructions automatically and in real-time.
As already mentioned in the introduction, the method described in Zhang et al. \cite{Zhang2016} required fiducial markers, which impose lower spatial resolutions and may affect the tissue's mechanics.
In Christoph and Schr\"oder-Schetelig et al. \cite{Christoph2017} it was necessary to perform a static shape reconstruction with the excitation-contraction uncoupling agent Blebbistatin.
This additional step made the approach impractical as it limited the time available for imaging to 5-15 minutes, because the motion tracking and shape reconstructions could become compromised due to swelling of the heart or photo-bleaching or wash-out of the fluorescent dye.
With our approach, there are no such constraints.
We can perform measurements over 1-2 hours, calibrate the cameras, track the tissue and generate 3D dynamic reconstructions in a fully automatic fashion. 
In principle, with further streamlining and optimization of the algorithms, the measurements could be done close to or in real-time.

Another advantage is that the soccerball-shaped imaging chamber allows us to image with up to 24 cameras and illuminate the heart from as many as  48 pre-defined locations, respectively. 
Both the cameras and light sources are evenly spaced and distributed around the heart.
This circumvents several issues with regard to space restrictions, uneven illumination and excitation of the heart surface, and other complications that arose with previous experimental setups \cite{Christoph2017,Christoph2018Nature}.
A crucial element in optical mapping studies with contracting hearts is to illuminate the heart as evenly as possible.
Homogeneous illumination minimizes illumination-related motion artifacts, which are caused by relative motion between the tissue and the light sources, see limitations section in the discussion in \cite{Lebert2022}.
Even though it seems impossible to illuminate the heart perfectly homogeneously, our illumination scheme produces smaller illumination gradients than many other approaches.
For instance, in octagonal baths with 8 glass walls, which were used in \cite{Christoph2017} and \cite{Christoph2018Nature}, the fewer windows restrict the positioning of both the cameras and light sources (e.g. LEDs).
The cameras need to image perpendicularly through the windows, and the light sources need to be positioned in a way that back-reflections into the camera images are avoided.
Additionally, objects in the way of the light-sources (e.g. silicone seams between glass wall parts) can create shadows on the heart surface.
Our chamber design allows the unobstructed, even illumination and imaging of the heart. 
A comparable approach using a spherical geometry for an optical fiber-based imaging setup was reported in \cite{Rieger2021}.

Lastly, we achieved high spatial resolutions thanks to the many cameras with their low-cost CMOS sensor, which produces images in the order of $300 \times 300$ pixels at 500 fps.
We measured electrical wave phenomena on the entire ventricular surface with about 0.5-1 million pixels at spatial resolutions of about \SI{120}{\micro\meter} per pixel.
At these resolutions, we can identify single small vessels across the ventricular surface.
Currently, optical mapping is typically performed only on parts of the ventricular surface, with video images in the order $100 \times 100$ pixels.
The sensor comprises a lower dynamic range and smaller pixels (\SI{6.9}{\micro\metre} $\times$ \SI{6.9}{\micro\metre}) associated with lower light-efficiency and signal-to-noise ratio (than e.g. Teledyne Photometrics Evolve 128), see also \citep{Lee2017}.
However, we demonstrated that the signal quality is very high when averaging the signals from multiple cameras.
In a previous study, we determined that the video quality of the low-cost cameras is sufficient for tracking \cite{Lebert2022}.

\subsection{Limitations}

Measuring action potential waves is still challenging compared to measuring action potential upstroke timings or electrical activation maps on the contracting heart surface.
Although ratiometric imaging reduces illumination-related motion artifacts, residual light field inhomogeneities and differences between the green and blue light fields, see traces in Fig.~\ref{fig:SupplementSinusAPTraces}, diminish the efficacy of artifact compensation.
Hence, it will be either necessary to further increase the homogeneity of the illumination or to measure or numerically estimate the inhomogeneity of the green and blue light fields during or after each experiment.

One challenging technical aspect during the experiments is associated with the low dynamic range of the cameras (12-bit: 4,096 counts): we had to manually adjust the apertures and brightness of each of the cameras before and throughout the experiment to ensure that it would match across cameras and that the full dynamic range of each sensor was utilized.
The final adjustments could only be made after staining and eventually needed to be repeated with each restaining.
In addition to the aperture adjustments, we also occasionally had to adjust the brightness of some of the videos during post-processing to match the average brightness across the videos.
This effectively minimizes artifacts during texture mapping, but a more automated approach would be desirable in the future.

The accuracy of the 3D reconstructions is demonstrated by the negligible residual motion artifacts and sharpness of the projected, averaged video images.
However, currently, we do not have means to directly validate the reconstructions, e.g. with synthetic data or a co-registered imaging modality serving as ground-truth. 
For instance, one could co-register the reconstructions with high-resolution computerized tomography scans of the heart.

Lastly, even though we used GPU-accelerated motion tracking, processing a 1.0 second long recording with 12 cameras currently takes several hours, excluding the time that is necessary to manually create a template mesh for tracking.
We anticipate that the processing can be further streamlined and eventually be performed in real time in the future.

\section*{Conclusions}
For over 30 years, optical mapping of action potential or calcium waves has been performed in contraction-inhibited hearts.
As a result, studying the interplay of cardiac bioelectricity and biomechanics, two of the main players involved in the heartbeat, has been limited.
Here, we demonstrated, for the first time, panoramic optical mapping of action potential waves in beating hearts.
We imaged the electrical wave phenomena across the entire, 3D deforming ventricular surface during normal and abnormal rhythms.
Our optical mapping system provides high-resolution imaging data, which can be used to study both the electrical and mechanical heart muscle physiology in unprecedented detail, and could allow new insights into the mechanisms of mechano-electric feedback.

\section*{Funding}
This research was funded by the University of California, San Francisco, the National Institutes of Health (DP2HL168071), and the Sandler Program for Breakthrough Biomedical Research, which is partially funded by the Sandler Foundation (to J.C.). This research was also supported through the Academic Hardware Grant program by the NVIDIA Corporation (to J.L. and J.C.).

\section*{Conflict of Interest Statement}
The authors declare that the research was conducted in the absence of any commercial or financial relationships that could be construed as a potential conflict of interest.

\section*{Author Contributions}
SC developed the numerical tracking and reconstruction algorithms.
SD and JL assisted with developing and refining the algorithms.
SC, JL, and JC analyzed the data.
JL and JC performed the experiments.
SC assisted during the experiments.
JL and JC designed and built the imaging setup. 
JL developed hardware and software for imaging and data acquisition.
JL established the barium chloride protocol.
JC designed and built the imaging chamber, developed the ratiometric imaging and performed the corresponding data analysis.
CG developed the LED driver.
JC designed the figures and wrote the manuscript.
JL and JC conceived the research.

\section*{Acknowledgements}
We would like to thank Ilija Uzelac and Vineesh Kappadan for fruitful discussions about cameras, lenses, filters, and optical mapping in general.

\section*{Bibliography}
\bibliographystyle{unsrtnat}
\bibliography{refs.bib}

\begin{thebibliography}{50}
\providecommand{\natexlab}[1]{#1}
\providecommand{\url}[1]{\texttt{#1}}
\expandafter\ifx\csname urlstyle\endcsname\relax
  \providecommand{\doi}[1]{doi: #1}\else
  \providecommand{\doi}{doi: \begingroup \urlstyle{rm}\Url}\fi

\bibitem[Salama et~al.(1987)Salama, Lombardi, and Elson]{Salama1987}
G.~Salama, R.~Lombardi, and J.~Elson.
\newblock Maps of optical action potentials and {NADH} fluorescence in intact working hearts.
\newblock \emph{American Journal of Physiology-Heart and Circulatory Physiology}, 252\penalty0 (2):\penalty0 H384--H394, February 1987.
\newblock \doi{10.1152/ajpheart.1987.252.2.h384}.
\newblock URL \url{https://doi.org/10.1152/ajpheart.1987.252.2.h384}.

\bibitem[Zhang et~al.(2016)Zhang, Iijima, Huang, Walcott, and Rogers]{Zhang2016}
Hanyu Zhang, Kenichi Iijima, Jian Huang, Gregory~P. Walcott, and Jack~M. Rogers.
\newblock Optical mapping of membrane potential and epicardial deformation in beating hearts.
\newblock \emph{Biophysical Journal}, 111\penalty0 (2):\penalty0 438--451, July 2016.
\newblock \doi{10.1016/j.bpj.2016.03.043}.

\bibitem[Christoph et~al.(2017)Christoph, Schr\"oder-Schetelig, and Luther]{Christoph2017}
J.~Christoph, J.~Schr\"oder-Schetelig, and S.~Luther.
\newblock Electromechanical optical mapping.
\newblock \emph{Progress in Biophysics and Molecular Biology}, 130:\penalty0 150--169, 2017.
\newblock \doi{10.1016/j.pbiomolbio.2017.09.015}.

\bibitem[Garrot et~al.(2017)Garrot, A., H., H., J., and M.]{Garrot2017}
K.~Garrot, Kuzmiak-Glancy A., Wengrowski H., Zhang H., Rogers J., and Kay M.
\newblock K(atp) channel inhibition blunts electromechanical decline during hypoxia in left ventricular working rabbit hearts.
\newblock \emph{The Journal of Physiology}, 595\penalty0 (12):\penalty0 3799--3813, 2017.

\bibitem[Christoph et~al.(2018)Christoph, Chebbok, Richter, Schr\"oder-Schetelig, Bittihn, Stein, Uzelac, Fenton, Hasenfuss, Gilmour, and Luther]{Christoph2018Nature}
J.~Christoph, M.~Chebbok, C.~Richter, J.~Schr\"oder-Schetelig, P.~Bittihn, S.~Stein, I.~Uzelac, F.~H. Fenton, G.~Hasenfuss, R.~Jr. Gilmour, and S.~Luther.
\newblock Electromechanical vortex filaments during cardiac fibrillation.
\newblock \emph{Nature}, 555:\penalty0 667 -- 672, 2018.
\newblock \doi{10.1038/nature26001}.

\bibitem[Christoph and Luther(2018)]{Christoph2018Frontiers}
J.~Christoph and S.~Luther.
\newblock Marker-free tracking for motion artifact compensation and deformation measurements in optical mapping videos of contracting hearts.
\newblock \emph{Frontiers in Physiology}, 9:\penalty0 1483, 2018.
\newblock \doi{10.3389/fphys.2018.01483}.

\bibitem[Kappadan et~al.(2020)Kappadan, Telele, Uzelac, Fenton, Parlitz, Luther, and Christoph]{Kappadan2020}
Vineesh Kappadan, Saba Telele, Ilija Uzelac, Flavio Fenton, Ulrich Parlitz, Stefan Luther, and Jan Christoph.
\newblock High-resolution optical measurement of cardiac restitution, contraction, and fibrillation dynamics in beating vs. {Blebbistatin}-uncoupled isolated rabbit hearts.
\newblock \emph{Frontiers in Physiology}, 11, 2020.
\newblock \doi{10.3389/fphys.2020.00464}.
\newblock URL \url{https://doi.org/10.3389/fphys.2020.00464}.

\bibitem[Lebert et~al.(2022)Lebert, Ravi, Kensah, and Christoph]{Lebert2022}
Jan Lebert, Namita Ravi, George Kensah, and Jan Christoph.
\newblock Real-time optical mapping of contracting cardiac tissues with gpu-accelerated numerical motion tracking.
\newblock \emph{Frontiers in Cardiovascular Medicine}, 9, 2022.
\newblock \doi{10.3389/fcvm.2022.787627}.

\bibitem[Zhang et~al.(2023)Zhang, Patton, Wood, Yan, Loew, Acker, Walcott, and Rogers]{Zhang2023}
Hanyu Zhang, Haley~N Patton, Garrett~A Wood, Ping Yan, Leslie~M Loew, Corey~D Acker, Gregory~P Walcott, and Jack~M Rogers.
\newblock Optical mapping of cardiac electromechanics in beating in vivo hearts.
\newblock \emph{Biophys. J.}, 122\penalty0 (21):\penalty0 4207--4219, November 2023.

\bibitem[Kappadan et~al.(2023)Kappadan, Sohi, Parlitz, Luther, Uzelac, Fenton, Peters, Christoph, and Ng]{Kappadan2023}
Vineesh Kappadan, Anies Sohi, Ulrich Parlitz, Stefan Luther, Ilija Uzelac, Flavio Fenton, Nicholas~S Peters, Jan Christoph, and Fu~Siong Ng.
\newblock Optical mapping of contracting hearts.
\newblock \emph{The Journal of Physiology}, n/a\penalty0 (n/a), 2023.
\newblock \doi{https://doi.org/10.1113/JP283683}.
\newblock URL \url{https://physoc.onlinelibrary.wiley.com/doi/abs/10.1113/JP283683}.

\bibitem[Nash et~al.(2006)Nash, Mourad, Clayton, Sutton, Bradley, Hayward, Paterson, and Taggart]{Nash2006}
Martyn~P. Nash, Ayman Mourad, Richard~H. Clayton, Peter~M. Sutton, Chris~P. Bradley, Martin Hayward, David~J. Paterson, and Peter Taggart.
\newblock Evidence for multiple mechanisms in human ventricular fibrillation.
\newblock \emph{Circulation}, 114\penalty0 (6):\penalty0 536--542, 2006.
\newblock \doi{10.1161/CIRCULATIONAHA.105.602870}.
\newblock URL \url{https://www.ahajournals.org/doi/abs/10.1161/CIRCULATIONAHA.105.602870}.

\bibitem[Narayan and John(2024)]{Narayan2024}
S.~M. Narayan and R.~M. John.
\newblock Advanced electroanatomic mapping: Current and emerging approaches.
\newblock \emph{Curr Treat Options Cardio Med}, 26:\penalty0 69--91, 2024.
\newblock \doi{10.1007/s11936-024-01034-6}.
\newblock URL \url{https://link.springer.com/article/10.1007/s11936-024-01034-6}.

\bibitem[Chen et~al.(2023)Chen, Nguyen, Kowalik, Shi, Tian, Doshi, Alber, Guan, Liu, Ning, Kay, and Lu]{Chen2023}
Zhiyuan Chen, Khanh Nguyen, Grant Kowalik, Xinyu Shi, Jinbi Tian, Mitansh Doshi, Bridget~R. Alber, Xun Guan, Xitong Liu, Xin Ning, Matthew~W. Kay, and Luyao Lu.
\newblock Transparent and stretchable au–ag nanowire recording microelectrode arrays.
\newblock \emph{Advanced Materials Technologies}, 8\penalty0 (10):\penalty0 2201716, 2023.
\newblock \doi{https://doi.org/10.1002/admt.202201716}.

\bibitem[Fullenkamp et~al.(2024)Fullenkamp, Maeng, Oh, Luan, Kim, Chychula, Kim, Yoo, Holgren, Demonbreun, George, Li, Hsu, Chung, Yoo, Koo, Park, Efimov, McNally, and Rogers]{Fullenkamp2024}
Dominic~E. Fullenkamp, Woo-Youl Maeng, Seyong Oh, Haiwen Luan, Kyung~Su Kim, Ivana~A. Chychula, Jin-Tae Kim, Jae-Young Yoo, Cory~W. Holgren, Alexis~R. Demonbreun, Sharon George, Binjie Li, Yaching Hsu, Gooyoon Chung, Jeongmin Yoo, Jahyun Koo, Yoonseok Park, Igor~R. Efimov, Elizabeth~M. McNally, and John~A. Rogers.
\newblock Simultaneous electromechanical monitoring in engineered heart tissues using a mesoscale framework.
\newblock \emph{Science Advances}, 10\penalty0 (37):\penalty0 eado7089, 2024.
\newblock \doi{10.1126/sciadv.ado7089}.
\newblock URL \url{https://www.science.org/doi/abs/10.1126/sciadv.ado7089}.

\bibitem[Gray et~al.(1995)Gray, Jalife, Panfilov, Baxter, Cabo, Davidenko, and Pertsov]{Gray1995}
Richard~A. Gray, José Jalife, Alexandre Panfilov, William~T. Baxter, Cándido Cabo, Jorge~M. Davidenko, and Arkady~M. Pertsov.
\newblock Nonstationary vortexlike reentrant activity as a mechanism of polymorphic ventricular tachycardia in the isolated rabbit heart.
\newblock \emph{Circulation}, 91\penalty0 (9):\penalty0 2454--2469, 1995.
\newblock \doi{10.1161/01.CIR.91.9.2454}.

\bibitem[Swift et~al.(2021)Swift, Kay, Ripplinger, and Posnack]{Swift2021}
Luther~M. Swift, Matthew~W. Kay, Crystal~M. Ripplinger, and Nikki~Gillum Posnack.
\newblock Stop the beat to see the rhythm: excitation-contraction uncoupling in cardiac research.
\newblock \emph{American Journal of Physiology-Heart and Circulatory Physiology}, 321\penalty0 (6):\penalty0 H1005--H1013, 2021.
\newblock \doi{10.1152/ajpheart.00477.2021}.

\bibitem[Efimov et~al.(2004)Efimov, Nikolski, and Salama]{Efimov2004}
I.~R. Efimov, V.~P. Nikolski, and G.~Salama.
\newblock Optical imaging of the heart.
\newblock \emph{Circ. Res.}, 95:\penalty0 21--33, 2004.

\bibitem[Kay et~al.(2004)Kay, Amison, and Rogers]{Kay2004}
M.W. Kay, P.M. Amison, and J.M. Rogers.
\newblock Three-dimensional surface reconstruction and panoramic optical mapping of large hearts.
\newblock \emph{IEEE Transactions on Biomedical Engineering}, 51\penalty0 (7):\penalty0 1219--1229, 2004.
\newblock \doi{10.1109/TBME.2004.827261}.

\bibitem[Kay et~al.(2006)Kay, Walcott, Gladden, Melnick, and Rogers]{Kay2006}
Matthew~W. Kay, Gregory~P. Walcott, James~D. Gladden, Sharon~B. Melnick, and Jack~M. Rogers.
\newblock Lifetimes of epicardial rotors in panoramic optical maps of fibrillating swine ventricles.
\newblock \emph{American Journal of Physiology-Heart and Circulatory Physiology}, 291\penalty0 (4):\penalty0 H1935--H1941, 2006.
\newblock \doi{10.1152/ajpheart.00276.2006}.

\bibitem[Rogers et~al.(2007)Rogers, Walcott, Gladden, Melnick, and Kay]{Rogers2007}
J.~M. Rogers, P.~G. Walcott, J.~D. Gladden, S.~B. Melnick, and M.~W. Kay.
\newblock Panoramic optical mapping reveals continuous epicardial reentry during ventricular fibrillation in the isolated swine heart.
\newblock \emph{Biophysical Journal}, 92:\penalty0 1090--1095, 2007.

\bibitem[Bray et~al.(2000)Bray, Lin, and Wikswo]{Bray2000}
M.-A. Bray, Shien-Fong Lin, and J.P. Wikswo.
\newblock Three-dimensional surface reconstruction and fluorescent visualization of cardiac activation.
\newblock \emph{{IEEE} Transactions on Biomedical Engineering}, 47\penalty0 (10):\penalty0 1382--1391, 2000.
\newblock \doi{10.1109/10.871412}.
\newblock URL \url{https://doi.org/10.1109/10.871412}.

\bibitem[Rohde et~al.(2005)Rohde, Benoit, and Lin]{Rohde2005}
G.~K. Rohde, B.~M. Benoit, and S.~Lin.
\newblock Corrections of motion artifacts in cardiac optical mapping using image registration.
\newblock \emph{IEEE Trans. Biomed. Eng.}, 52:\penalty0 338--341, 2005.

\bibitem[Svrcek et~al.(2009)Svrcek, Rutherford, and Smaill]{Svrcek2009}
M.~Svrcek, S.~Rutherford, and B.~Smaill.
\newblock Characteristics of motion artifacts in cardiac optical mapping studies.
\newblock \emph{Conference Proceedings, 31st Annual International Conference of the IEEE EMBS, Minneapolis, USA}, 31:\penalty0 3240--3243, 2009.
\newblock \doi{10.1109/IEMBS.2009.5333531}.

\bibitem[Khwaounjoo et~al.(2015)Khwaounjoo, Rutherford, Svrcek, LeGriece, Trew, and Smaill]{Khwaounjoo2015}
P.~Khwaounjoo, S.~L. Rutherford, M.~Svrcek, I.~J. LeGriece, M.~L Trew, and B.~H. Smaill.
\newblock Image-based motion correction for optical mapping of cardiac electrical activity.
\newblock \emph{Ann. Biomed. Eng.}, 43:\penalty0 1235--1246, 2015.

\bibitem[Rodriguez and Nygren(2015)]{Rodriguez2015}
M.~P. Rodriguez and A.~Nygren.
\newblock Motion estimation in cardiac fluorescence imaging with scale-space landmarks and optical flow: a comparative study.
\newblock \emph{IEEE Trans. Biomed. Eng.}, 62:\penalty0 774--782, 2015.

\bibitem[Klaudiny and Hilton(2011)]{Klaudiny2011}
Martin Klaudiny and Adrian Hilton.
\newblock Cooperative patch-based 3d surface tracking.
\newblock In \emph{2011 Conference for Visual Media Production}. {IEEE}, November 2011.
\newblock \doi{10.1109/cvmp.2011.14}.
\newblock URL \url{https://doi.org/10.1109/cvmp.2011.14}.

\bibitem[Bourgeois et~al.(2011)Bourgeois, Bachtel, Huang, Walcott, and Rogers]{Bourgeois2011}
Elliot~B. Bourgeois, Andrew~D. Bachtel, Jian Huang, Gregory~P. Walcott, and Jack~M. Rogers.
\newblock Simultaneous optical mapping of transmembrane potential and wall motion in isolated, perfused whole hearts.
\newblock \emph{Journal of Biomedical Optics}, 16\penalty0 (9):\penalty0 096020, 2011.
\newblock \doi{10.1117/1.3630115}.
\newblock URL \url{https://doi.org/10.1117/1.3630115}.

\bibitem[Bachtel et~al.(2011)Bachtel, Gray, and Rogers]{Bachtel2011}
A.~D. Bachtel, R.~A. Gray, and J.~M. Rogers.
\newblock A novel approach to dual excitation ratiometric optical mapping of cardiac action potentials with di-4-anepps using pulsed led excitation.
\newblock \emph{IEEE Trans. Biomed. Eng.}, 58:\penalty0 2120--2126, 2011.

\bibitem[Myles et~al.(2015)Myles, Wang, Bers, and Ripplinger]{Myles2015}
Rachel~C. Myles, Lianguo Wang, Donald~M. Bers, and Crystal~M. Ripplinger.
\newblock Decreased inward rectifying k+ current and increased ryanodine receptor sensitivity synergistically contribute to sustained focal arrhythmia in the intact rabbit heart.
\newblock \emph{The Journal of Physiology}, 593\penalty0 (6):\penalty0 1479--1493, 2015.
\newblock \doi{https://doi.org/10.1113/jphysiol.2014.279638}.
\newblock URL \url{https://physoc.onlinelibrary.wiley.com/doi/abs/10.1113/jphysiol.2014.279638}.

\bibitem[{Oliver Batchelor}(2023)]{multical}
{Oliver Batchelor}.
\newblock multical, 2023.

\bibitem[Garrido-Jurado et~al.(2014)Garrido-Jurado, Mu{\~{n}}oz-Salinas, Madrid-Cuevas, and Mar{\'{\i}}n-Jim{\'{e}}nez]{GarridoJurado2014}
S.~Garrido-Jurado, R.~Mu{\~{n}}oz-Salinas, F.J. Madrid-Cuevas, and M.J. Mar{\'{\i}}n-Jim{\'{e}}nez.
\newblock Automatic generation and detection of highly reliable fiducial markers under occlusion.
\newblock \emph{Pattern Recognition}, 47\penalty0 (6):\penalty0 2280--2292, June 2014.
\newblock \doi{10.1016/j.patcog.2014.01.005}.

\bibitem[Kwon(1999)]{Kwon1999}
Young-Hoo Kwon.
\newblock Object plane deformation due to refraction in two-dimensional underwater motion analysis.
\newblock \emph{Journal of Applied Biomechanics}, 15\penalty0 (4):\penalty0 396--403, November 1999.
\newblock \doi{10.1123/jab.15.4.396}.
\newblock URL \url{https://doi.org/10.1123/jab.15.4.396}.

\bibitem[Sch\"{o}nberger et~al.(2016)Sch\"{o}nberger, Zheng, Pollefeys, and Frahm]{Schoenberger2016mvs}
Johannes~Lutz Sch\"{o}nberger, Enliang Zheng, Marc Pollefeys, and Jan-Michael Frahm.
\newblock Pixelwise view selection for unstructured multi-view stereo.
\newblock In \emph{European Conference on Computer Vision (ECCV)}, 2016.

\bibitem[Blender-Online-Community(2018)]{Blender}
Blender-Online-Community.
\newblock \emph{Blender - a 3D modelling and rendering package}.
\newblock Blender Foundation, Stichting Blender Foundation, Amsterdam, 2018.
\newblock URL \url{http://www.blender.org}.

\bibitem[Schroeder et~al.(2006)Schroeder, Martin, and Lorensen]{VTK}
Will Schroeder, Ken Martin, and Bill Lorensen.
\newblock \emph{The Visualization Toolkit (4th ed.)}.
\newblock Kitware, 2006.
\newblock ISBN 978-1-930934-19-1.

\bibitem[Zhou et~al.(2018)Zhou, Park, and Koltun]{open3D}
Qian-Yi Zhou, Jaesik Park, and Vladlen Koltun.
\newblock {Open3D}: {A} modern library for {3D} data processing.
\newblock \emph{arXiv:1801.09847}, 2018.

\bibitem[Barnes et~al.(2009)Barnes, Shechtman, Finkelstein, and Goldman]{Barnes2009}
Connelly Barnes, Eli Shechtman, Adam Finkelstein, and Dan~B Goldman.
\newblock {PatchMatch}: A randomized correspondence algorithm for structural image editing.
\newblock \emph{ACM Transactions on Graphics (Proc. SIGGRAPH)}, 28\penalty0 (3), August 2009.

\bibitem[Farnebäck(2003)]{Farneback2003}
Gunnar Farnebäck.
\newblock Two-{Frame} {Motion} {Estimation} {Based} on {Polynomial} {Expansion}.
\newblock In Gerhard Goos, Juris Hartmanis, Jan van Leeuwen, Josef Bigun, and Tomas Gustavsson, editors, \emph{Image {Analysis}}, volume 2749, pages 363--370. Springer Berlin Heidelberg, Berlin, Heidelberg, 2003.
\newblock ISBN 978-3-540-40601-3 978-3-540-45103-7.
\newblock \doi{10.1007/3-540-45103-X_50}.
\newblock URL \url{http://link.springer.com/10.1007/3-540-45103-X_50}.

\bibitem[{Jan Lebert and Jan Christoph}(2023)]{optimap}
{Jan Lebert and Jan Christoph}.
\newblock optimap: an open-source library for the processing of fluorescence video data.
\newblock 2023.
\newblock \doi{doi:10.5281/ZENODO.8336455}.
\newblock URL \url{https://github.com/cardiacvision/optimap}.

\bibitem[Iyer and Gray()]{IyerGray2001}
Anand~N. Iyer and Richard~A. Gray.
\newblock An experimentalist's approach to accurate localization of phase singularities during reentry.
\newblock \emph{Annals of Biomedical Engineering}, 29:\penalty0 47--59.
\newblock \doi{10.1114/1.1335538}.

\bibitem[Lebert et~al.(2021)Lebert, Ravi, Fenton, and Christoph]{Lebert2021}
Jan Lebert, Namita Ravi, Flavio~H. Fenton, and Jan Christoph.
\newblock Rotor localization and phase mapping of cardiac excitation waves using deep neural networks.
\newblock \emph{Frontiers in Physiology}, 12, 2021.
\newblock ISSN 1664-042X.
\newblock \doi{10.3389/fphys.2021.782176}.
\newblock URL \url{https://www.frontiersin.org/articles/10.3389/fphys.2021.782176}.

\bibitem[Li et~al.(2020)Li, Almeida, Dastagir, Guillem, Salinet, Chu, Stafford, Schlindwein, and Ng]{Li2020}
Xin Li, Tiago~P. Almeida, Nawshin Dastagir, Mar{\'{\i}}a~S. Guillem, Jo{\~{a}}o Salinet, Gavin~S. Chu, Peter~J. Stafford, Fernando~S. Schlindwein, and G.~Andr{\'{e}} Ng.
\newblock Standardizing single-frame phase singularity identification algorithms and parameters in phase mapping during human atrial fibrillation.
\newblock \emph{Frontiers in Physiology}, 11, July 2020.
\newblock \doi{10.3389/fphys.2020.00869}.
\newblock URL \url{https://doi.org/10.3389/fphys.2020.00869}.

\bibitem[ter Bekke et~al.(2015)ter Bekke, Haugaa, van~den Wijngaard, Bos, Ackerman, Edvardsen, and Volders]{terBekke2015}
R.~ter Bekke, K.~H. Haugaa, A.~van~den Wijngaard, M.~Bos, M.~J. Ackerman, T.~Edvardsen, and P.~G.~A. Volders.
\newblock Electromechanical window negativity in genotyped long-qt syndrome patients: relation to arrhythmia risk.
\newblock \emph{European Heart Journal}, 36:\penalty0 179--186, 2015.

\bibitem[Wyman et~al.(1999)Wyman, Hunter, Prinzen, and McVeigh]{Wyman1999}
Bradley~T. Wyman, William~C. Hunter, Frits~W. Prinzen, and Elliot~R. McVeigh.
\newblock Mapping propagation of mechanical activation in the paced heart with mri tagging.
\newblock \emph{American Journal of Physiology - Heart and Circulatory Physiology}, 276\penalty0 (3):\penalty0 H881--H891, 1999.
\newblock \doi{10.1152/ajpheart.1999.276.3.H881}.
\newblock URL \url{https://doi.org/10.1152/ajpheart.1999.276.3.H881}.
\newblock PMID: 10070071.

\bibitem[Provost et~al.(2011)Provost, Nguyen, Legrand, Okrasinski, Costet, Gambhir, Garan, and Konofagou]{Provost2011}
Jean Provost, Vu~Thanh-Hieu Nguyen, Di{\'{e}}go Legrand, Stan Okrasinski, Alexandre Costet, Alok Gambhir, Hasan Garan, and Elisa~E Konofagou.
\newblock Electromechanical wave imaging for arrhythmias.
\newblock \emph{Physics in Medicine and Biology}, 56\penalty0 (22):\penalty0 L1--L11, 2011.
\newblock \doi{10.1088/0031-9155/56/22/f01}.

\bibitem[Lebert and Christoph(2019)]{Lebert2019}
Jan Lebert and Jan Christoph.
\newblock Synchronization-based reconstruction of electromechanical wave dynamics in elastic excitable media.
\newblock \emph{Chaos: An Interdisciplinary Journal of Nonlinear Science}, 29, 2019.
\newblock \doi{10.1063/1.5101041}.

\bibitem[Molavi~Tabrizi et~al.(2022)Molavi~Tabrizi, Mesgarnejad, Bazzi, Luther, Christoph, and Karma]{Molavi2022}
A.~Molavi~Tabrizi, A.~Mesgarnejad, M.~Bazzi, S.~Luther, J.~Christoph, and A.~Karma.
\newblock Spatiotemporal organization of electromechanical phase singularities during high-frequency cardiac arrhythmias.
\newblock \emph{Phys. Rev. X}, 12:\penalty0 021052, Jun 2022.
\newblock \doi{10.1103/PhysRevX.12.021052}.
\newblock URL \url{https://link.aps.org/doi/10.1103/PhysRevX.12.021052}.

\bibitem[Maffessanti et~al.(2020)Maffessanti, Jadczyk, Kurzelowski, Regoli, Caputo, Conte, Gołba, Biernat, Wilczek, Dabrowska, Pezzuto, Moccetti, Krause, Wojakowski, Prinzen, and Auricchio]{Maffessanti2020}
Francesco Maffessanti, Tomasz Jadczyk, Radoslaw Kurzelowski, Francois Regoli, Maria~Luce Caputo, Giulio Conte, Krzysztof~S Gołba, Jolanta Biernat, Jacek Wilczek, Magdalena Dabrowska, Simone Pezzuto, Tiziano Moccetti, Rolf Krause, Wojciech Wojakowski, Frits~W Prinzen, and Angelo Auricchio.
\newblock The influence of scar on the spatio-temporal relationship between electrical and mechanical activation in heart failure patients.
\newblock \emph{EP Europace}, 22\penalty0 (5):\penalty0 777--786, 01 2020.
\newblock ISSN 1099-5129.
\newblock \doi{10.1093/europace/euz346}.
\newblock URL \url{https://doi.org/10.1093/europace/euz346}.

\bibitem[Lee et~al.(2017)Lee, Calvo, Alfonso-Almazin, Quintanilla, Chorro, Yan, Loew, Filgueiras-Rama, and Millet]{Lee2017}
P.~Lee, C.~J. Calvo, J.~M. Alfonso-Almazin, J.~G. Quintanilla, F.~J. Chorro, P.~Yan, L.~N. Loew, D.~Filgueiras-Rama, and J.~Millet.
\newblock Low-cost optical mapping systems for panoramic imaging of complex arrhythmias and drug-action in translational heart models.
\newblock \emph{Scientific Reports}, 7, 2017.
\newblock \doi{10.1038/srep43217}.

\bibitem[Rieger et~al.(2021)Rieger, Dellenbach, vom Berg, Beil-Wagner, Maguy, and Rohr]{Rieger2021}
Michael Rieger, Christian Dellenbach, Johannes vom Berg, Jane Beil-Wagner, Ange Maguy, and Stephan Rohr.
\newblock Enabling comprehensive optogenetic studies of mouse hearts by simultaneous opto-electrical panoramic mapping and stimulation.
\newblock \emph{Nature Communications}, 12\penalty0 (1), October 2021.
\newblock \doi{10.1038/s41467-021-26039-8}.
\newblock URL \url{https://doi.org/10.1038/s41467-021-26039-8}.

\end{thebibliography}

%  CONFIGURE NEW SINGLE-PAGE FORMAT 

\onecolumn % go back to one column
\fancyhead{} % make sure we get no headers
\renewcommand{\floatpagefraction}{0.1}
\lfoot[\bSupInf]{\dAuthor}
\rfoot[\dAuthor]{\cSupInf}
\newpage

\captionsetup*{format=largeformat} % make figure legend slightly larger than in the paper
\setcounter{figure}{0} % reset figure counter for Supp. Figures
\setcounter{equation}{0} % reset equation counter for Supp. Equations
\makeatletter 
\renewcommand{\thefigure}{S\@arabic\c@figure} % make Figure legend start with Figure S
\makeatother
\def\theequation{S\arabic{equation}}

%  MAIN TEXT 

\newpage
\section*{Supplementary Information}

\begin{figure*}[!ht]
\centering
\includegraphics[width=0.65\linewidth]{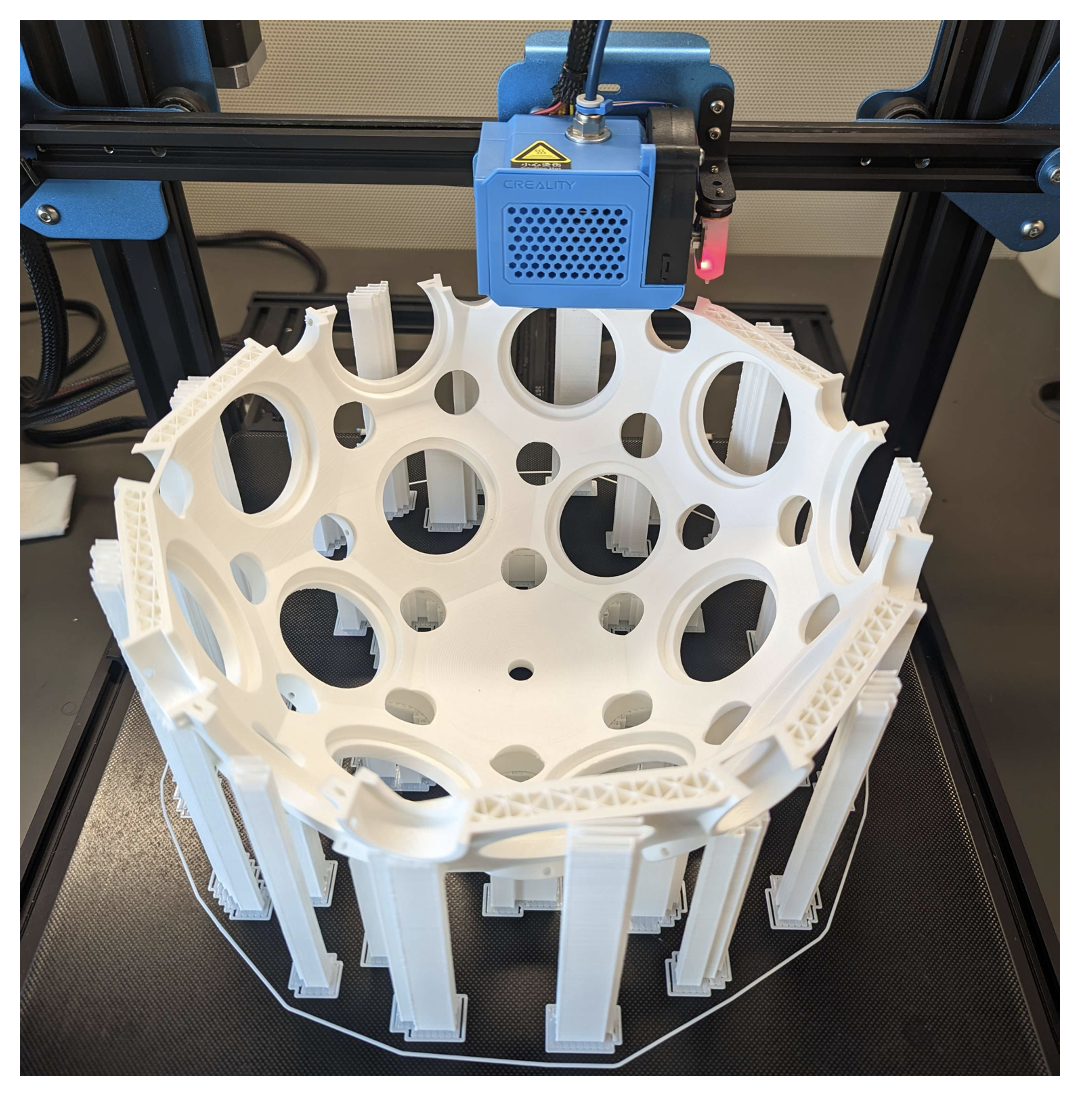}
\caption{\textbf{Soccer ball-shaped imaging chamber.}\\
3D printing of soccer ball-shaped imaging chamber for panoramic optical mapping of isolated hearts, see also Fig.~\ref{fig:Setup}. The chamber has 24 windows for imaging and 48 ports for LEDs.}
\label{fig:SupplementSoccerball}
\end{figure*}

\begin{figure*}[!ht]
    \centering
    \includegraphics[clip, trim=0.0cm 0.0cm 0.0cm 0.0cm, width=0.95\textwidth]{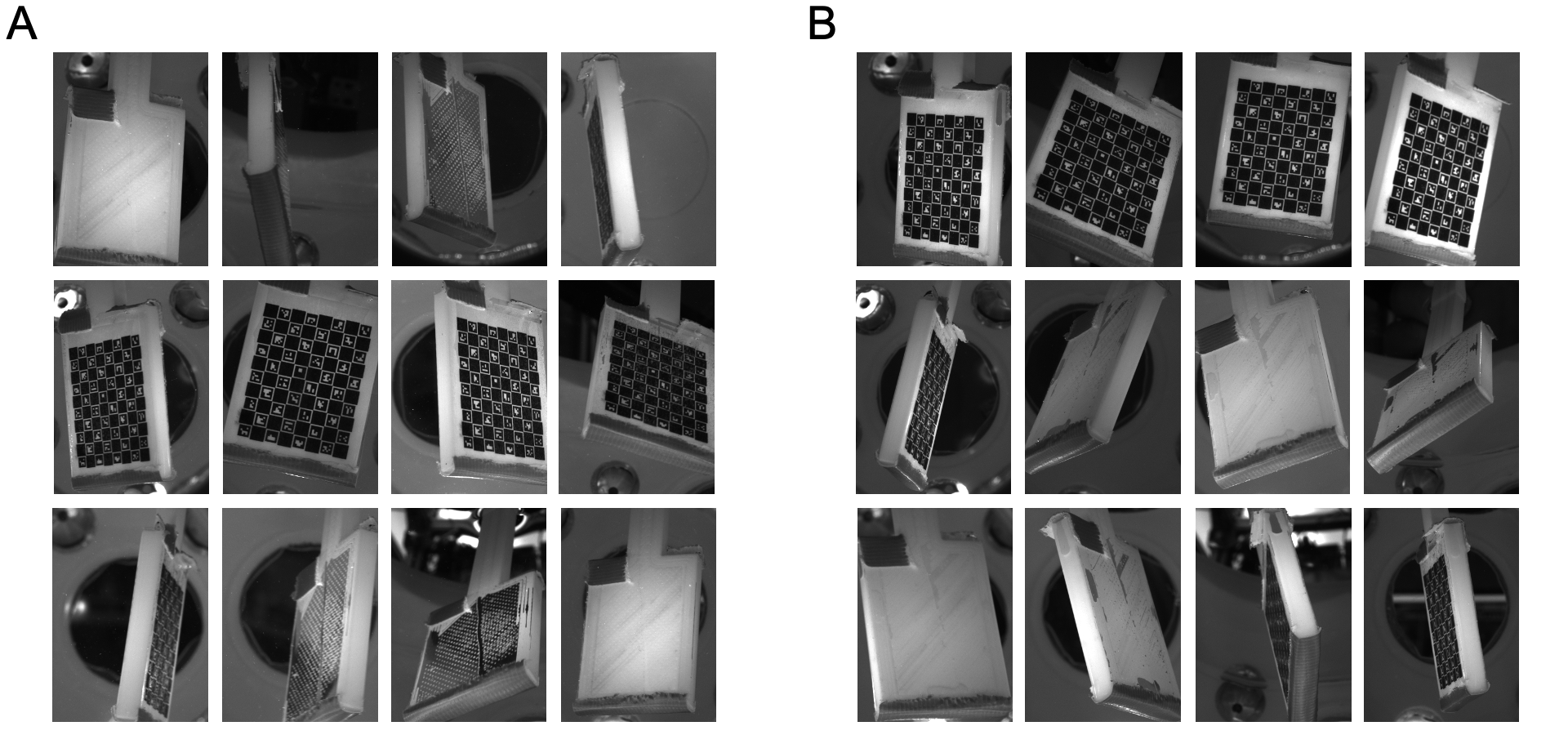}
    \caption{\textbf{Camera Calibration.}\\
    Calibration target with ChArUco pattern inside imaging chamber as seen by 12 cameras for two different poses of the target (A,B). At least 36 different sets of images were taken such that the target was visible at least 3 times in each camera while also being visible in 2-3 of the neighboring cameras.}
    \label{fig:SupplementCalibration}
\end{figure*}

\begin{figure*}[!ht]
    \centering
    \includegraphics[clip, trim=0.0cm 0.0cm 0.0cm 0.0cm, width=0.6\textwidth]{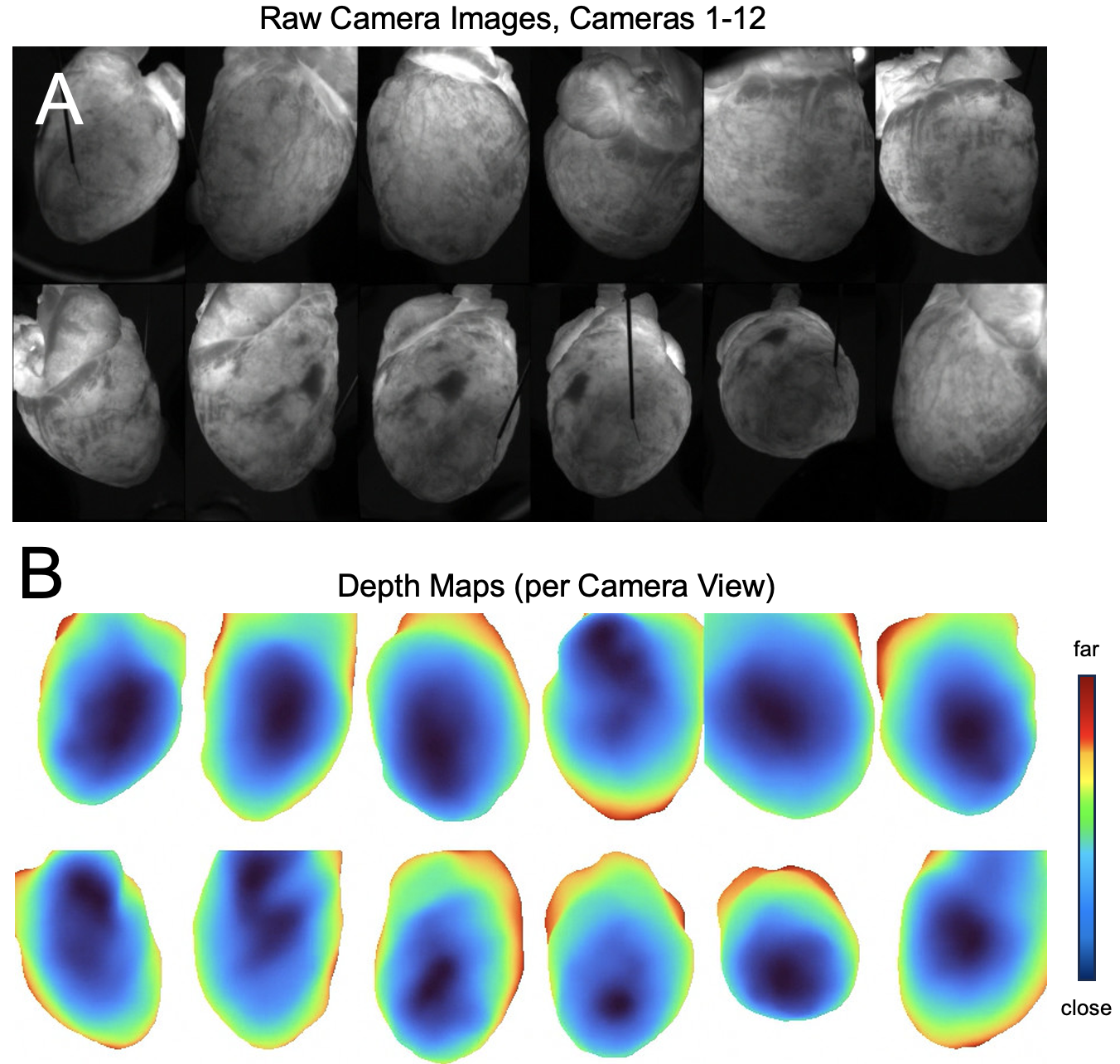}
    \caption{
  \textbf{A} Raw grayscale camera images from cameras 1-12 showing isolated rabbit heart during voltage-sensitive optical mapping with continuous green illumination. In this study, we refer to each set of (12) camera images as one frame. 
    \textbf{B} Corresponding depth maps indicating the distance of the heart surface to the camera (dark blue: close, red: far). The information can be obtained from the three-dimensional reconstruction and multi-camera tracking of the heart surface.
    }
    \label{fig:SupplementDepthMaps}
  \end{figure*}

\begin{figure*}[!ht]
    \centering
    \includegraphics[clip, trim=0.0cm 0.0cm 0.0cm 0.0cm, width=0.95\textwidth]{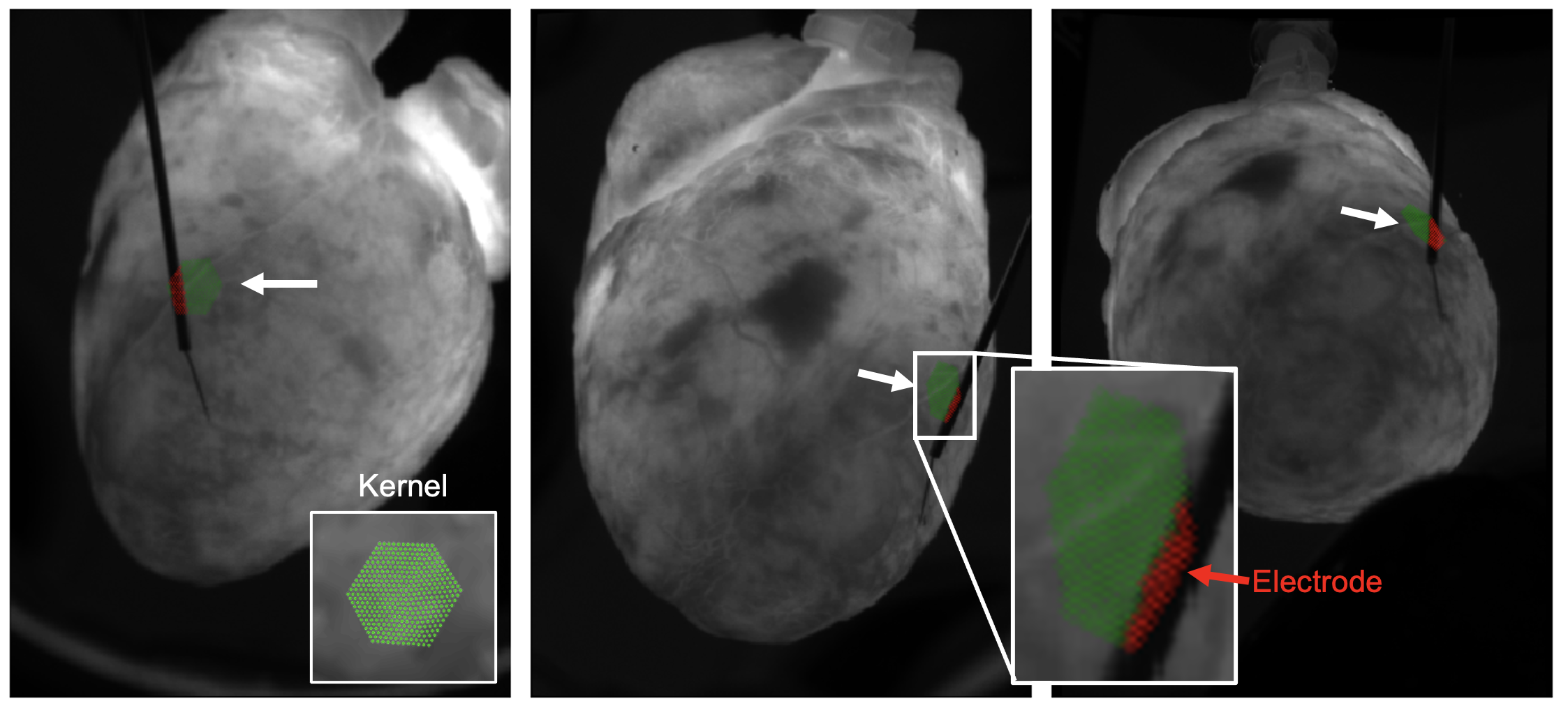}
    \caption{\textbf{Tracking with occlusions.}\\
    Tracking algorithm ignores pacing electrode. The tracking kernel from the mesh tracking method developed by Klaudiny \& Hilton, see Methods, samples image points from the ventricular surface (green). With our extensions, it ignores tissue that is occluded by the pacing electrode (red). The tissue behind the electrode can be tracked nevertheless, because it is in almost all cases visible in multiple other camera views, see also Fig.~\ref{fig:SupplementCoverage}A,B).}
    \label{fig:SupplementTrackingElectrode}
\end{figure*}

\begin{figure*}[!ht]
    \centering
    \includegraphics[clip, trim=0.0cm 0.0cm 0.0cm 0.0cm, width=0.95\textwidth]{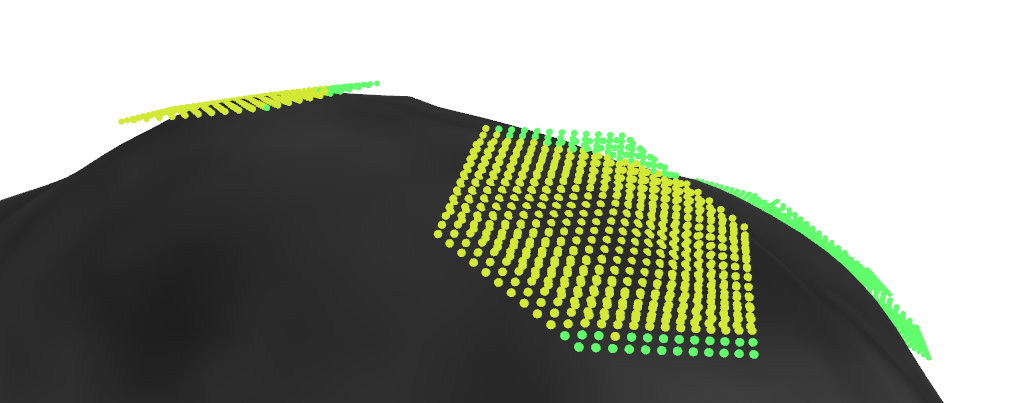}
    \caption{\textbf{Tracking kernel.}\\
    The tracking kernel from the mesh tracking method developed by Klaudiny \& Hilton, see Methods, samples image points across the surface using a 2D plane that is an approximation of the surface and is aligned with the center vertex normal, see also Fig.~\ref{fig:SupplementTrackingElectrode}.}
    \label{fig:SupplementTrackingKernel}
\end{figure*}

\begin{figure*}[!ht]
    \centering
    \includegraphics[clip, trim=0.0cm 0.0cm 0.0cm 0.0cm, width=0.9\textwidth]{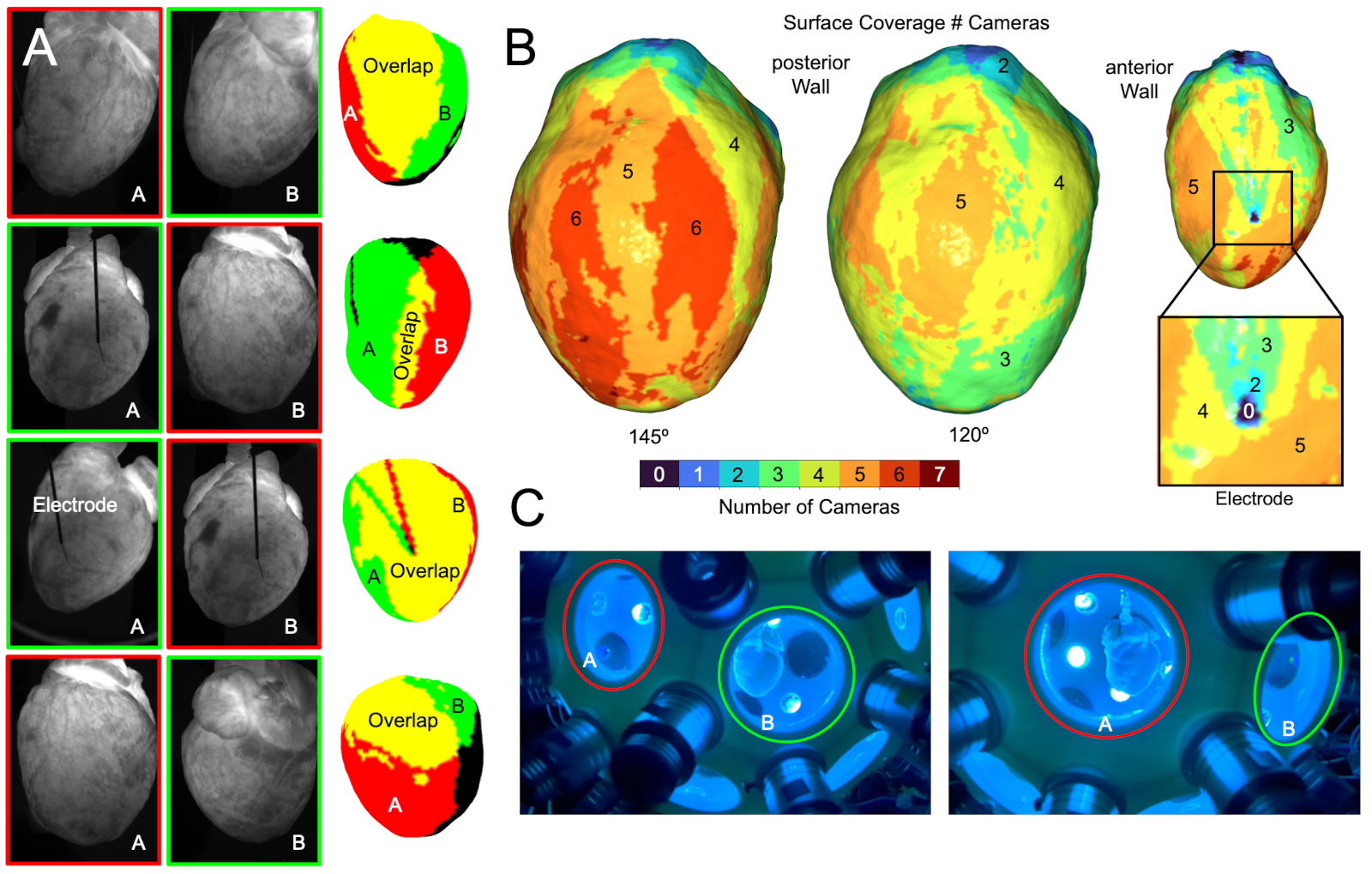}
    \caption{\textbf{Coverage.}\\
    Coverage of heart surface with 12 cameras configured as in Fig.~\ref{fig:Setup}E).
    \textbf{A} Camera images from two adjacent cameras (here two random pairs A\&B indicated in red/green) and overlap (yellow). Tissue covered by the pacing electrode in one camera is visible in the other camera. 
    \textbf{B} Coverage of tissue surface by number (\#) of cameras. The tissue is always covered by at least 4-6 (left) or 3-5 cameras (center, right) when allowing camera normals within angles of \SI{145}{\degree} or \SI{120}{\degree} around the local surface normal, respectively. The tissue behind the pacing electrode can always be seen by at least 3 cameras, except for a small area underneath the tip of the electrode (right).
    \textbf{C} View through two adjacent windows into imaging chamber. The average angle between cameras is about \SI{35}{\degree}-\SI{45}{\degree}.
    }
    \label{fig:SupplementCoverage}
\end{figure*}

\begin{figure*}[!ht]
    \centering
    \includegraphics[clip, trim=0.0cm 0.0cm 0.0cm 0.0cm, width=0.95\textwidth]{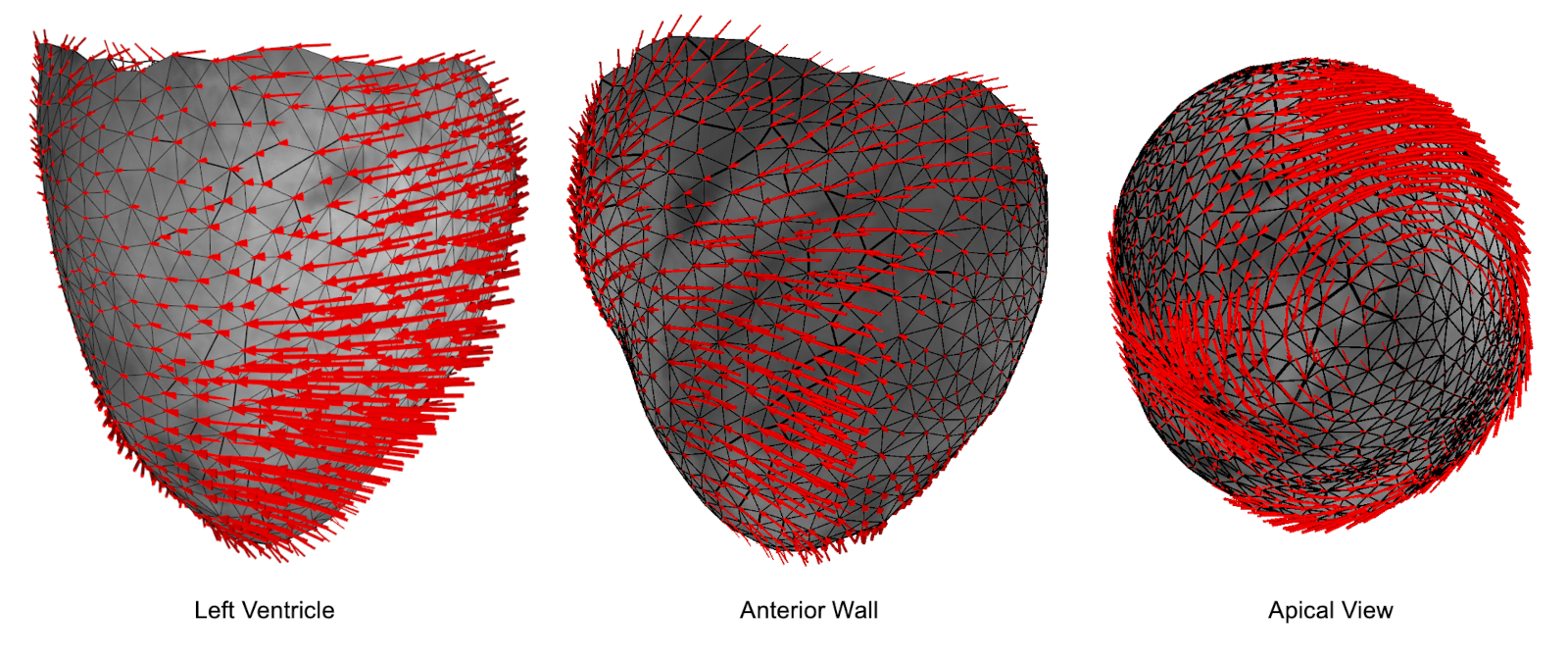}
    \caption{\textbf{Tracked ventricular surface.}\\
    Tracked ventricular surface of isolated rabbit heart during sinus rhythm, the visualizations showing the texturized triangular polygon mesh and displacement vectors (red) from different views (anterior wall, posterior wall, apical view), see also Fig.~\ref{fig:Fig2} and \ref{fig:SupplementTracking3D2D}. The displacement vectors indicate absolute displacements of the tissue with respect to the diastolic configuration of the tissue shortly before the depolarization of the ventricles.}
    \label{fig:SupplementTracking}
\end{figure*}

\begin{figure*}[!ht]
    \centering
    \includegraphics[clip, trim=0.0cm 0.0cm 0.0cm 0.0cm, width=0.9\textwidth]{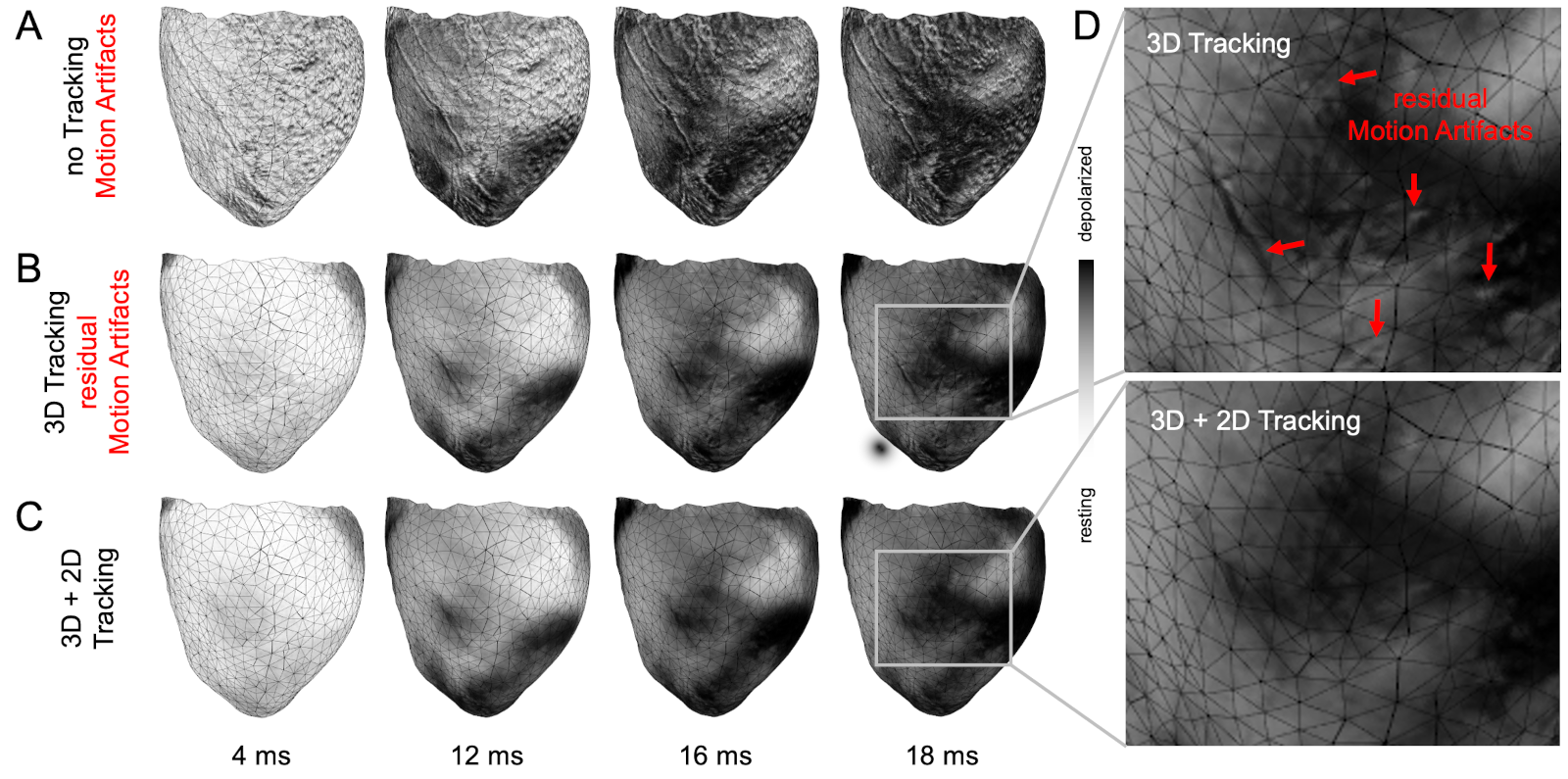}
    \caption{\textbf{Motion artifact compensation using a combination of 3D and 2D tracking.}\\
    Motion artifacts can be significantly reduced using a combination of 3D and 2D numerical motion tracking and co-moving signal analysis, see also Fig.~\ref{fig:Fig2}. First, the tissue is tracked in 3D space using our multi-view tracking methodology based on Klaudiny \& Hilton, see Methods. When using only the 3D tracking vector data to stabilize the motion in the video images, the motion-stabilized videos exhbit residual motion (within each triangle), because the spatial resolution of the 3D tracking is lower than the resolution of the video data. 
    \textbf{A} Pixel-wise normalized 3D optical maps without numerical motion tracking and motion artifact compensation (sliding-window normalization, \SI{100}{\milli \second} interval).
    \textbf{B} Pixel-wise normalized 3D optical maps (black: depolarized tissue, white: resting tissue) after 3D motion tracking. The maps exhibit residual motion artifacts (red arrows, see panel D).
    \textbf{C} Pixel-wise normalized 3D optical maps after 3D and subsequent 2D motion tracking. The residual motion artifacts are largely inhibited. The residual motion in each triangle was tracked and stabilized using 2D motion tracking applied to the already stabilized videos which were warped using only the 3D tracking vector data.
    The data was acquired during sinus rhythm and with continuous green illumination.
    To further suppress motion artifacts, it is required to employ ratiometric imaging to compensate artifacts associated with relative motion between LEDs and tissue, see Figs.~\ref{fig:Setup}F,G) and \ref{fig:SupplementRatiometry}-\ref{fig:SupplementComparisonGreenRatio}.}
    \label{fig:SupplementTracking3D2D}
\end{figure*}

\begin{figure*}[!ht]
    \centering
    \includegraphics[clip, trim=0.0cm 0.0cm 0.0cm 0.0cm, width=0.95\textwidth]{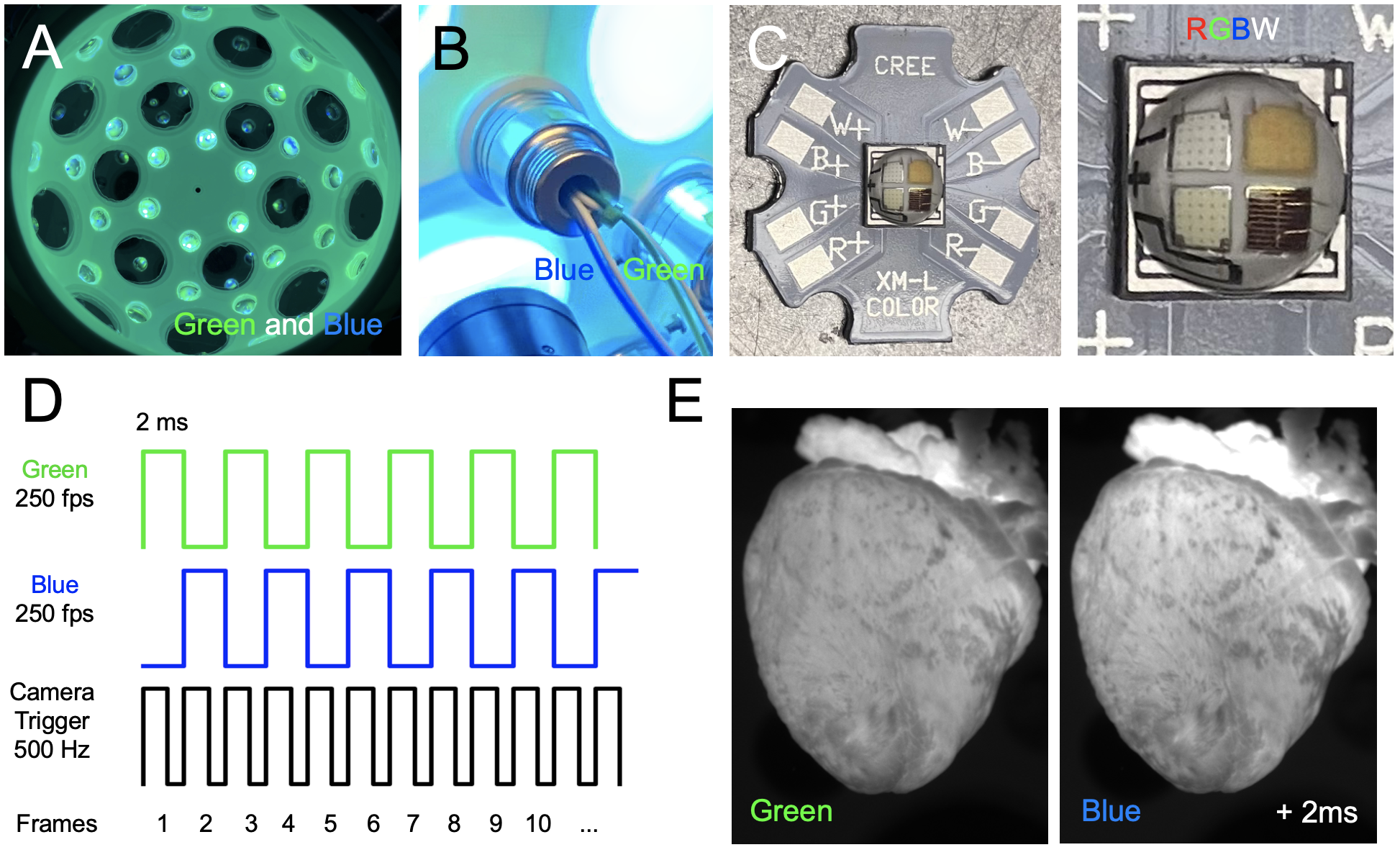}
    \caption{\textbf{Panoramic ratiometric voltage-sensitive optical mapping.}\\
    Panoramic ratiometric voltage-sensitive optical mapping with alternating blue and green excitation. 
    \textbf{A} Interior of imaging chamber under simultaneous green and blue illumination.
    \textbf{B} Single LED mounted onto heatsink and inserted into LED port of imaging chamber with green and blue wiring for the two green and blue diodes, respectively.
    \textbf{C} Red-green-blue-white (RGBW) LED (Cree XML RGBW Star LED, USA) with adjacent diodes which can be powered individually.
    \textbf{D} Triggering scheme for pulsed green-blue illumination and video frame acquisition. The switching back and forth, alternatingly powering the green and blue channel, is synchronized with the video acquisition (500f ps) yielding 2 videos at half the video acquisition rate (2 $\times$ 250f ps), see also Supplementary Video \textcolor{red}{\ref{video:green-blue-video}}.
    \textbf{E} Blue and green video frames acquired 2 ms apart using pulsed blue and green illumination, see also Fig.~\ref{fig:Setup}G). The pulsed illumination and acquisition is performed simultaneously with all 12 cameras. The signal reporting the action potential wave is only present in the green channel. It is common that the atria appear brighter and saturated in the video images. This did not affect the tracking or signal analysis in the ventricles.}
    \label{fig:SupplementRatiometry}
\end{figure*}

\begin{figure*}[!ht]
    \centering
    \includegraphics[clip, trim=0.0cm 0.0cm 0.0cm 0.0cm, width=0.95\textwidth]{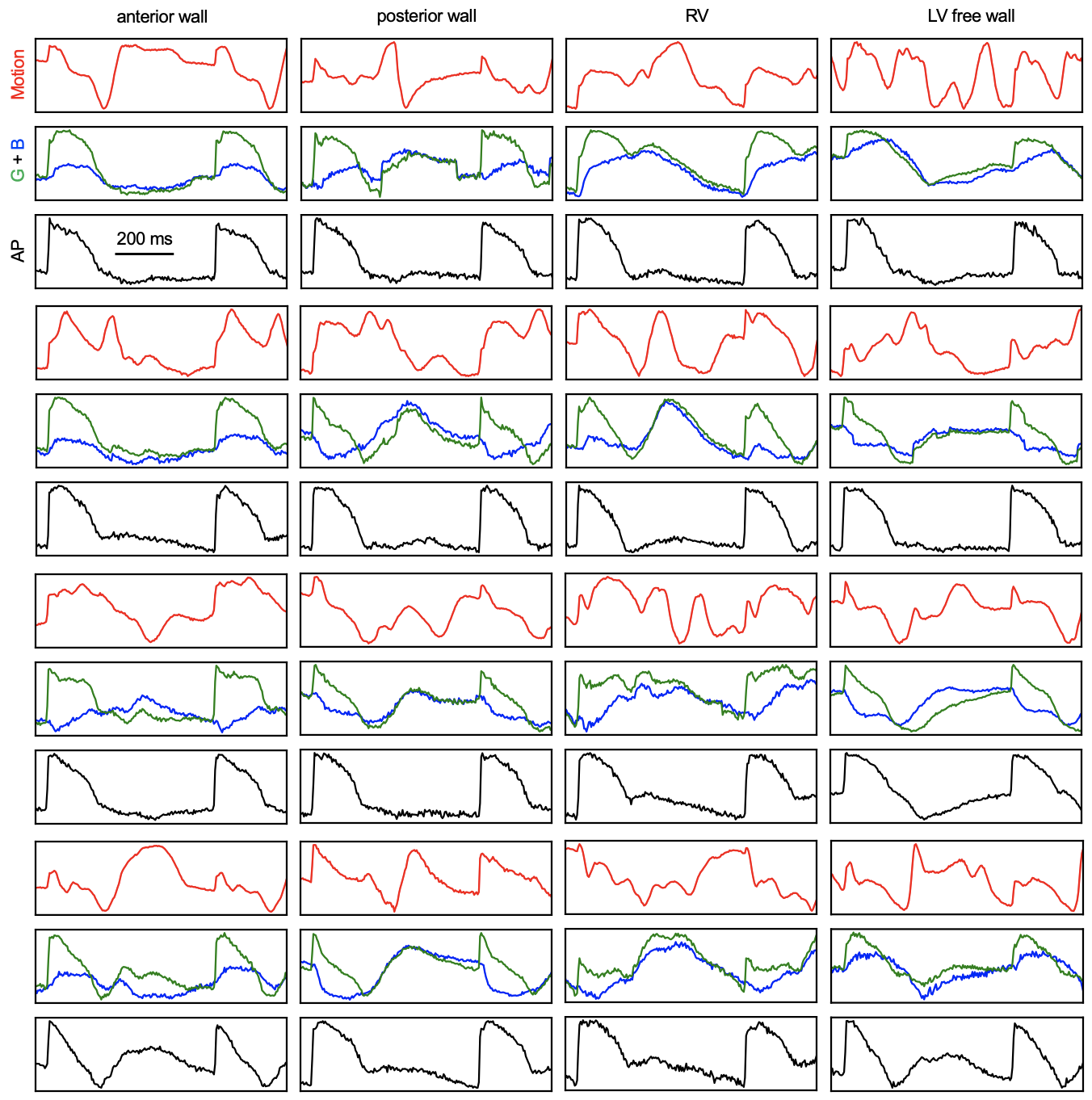}
    \caption{\textbf{Optical measurements of action potentials on the surface of strongly contracting isolated rabbit heart during sinus rhythm.}\\
    Optical measurements ($4 \times 4 = 16$ representative examples) of action potentials across the ventricular surface of a contracting isolated rabbit heart during sinus rhythm performed with voltage-sensitive panoramic 3D optical mapping. The optical traces (normalized dimensionless units) were obtained from a $7 \times 7$ pixel region located at the center of one of the triangles of the mesh representing the ventricular surface, c.f. Fig.~\ref{fig:Fig2}E). The triangles were selected from random locations on either the anterior wall (first column), the posterior wall (second column), the right ventricular wall (RV, third column), or the free left ventricular wall (LV, fourth column). The red traces correspond to motion artifacts obtained before tracking. The green and blue traces were obtained after motion tracking with green and blue excitation, respectively. The black trace is the ratiometric combination of the green and blue traces (one divided by the other) and corresponds to a measurement of the time-course of the transmembrane potential after numerical motion tracking and ratiometric compensation of motion artifacts, see also Fig.~\ref{fig:Setup}G). 
    While the motion artifact compensation works well in many cases, there are residual artifacts, which can be removed using the detrending and baseline correction method shown in Figs.~\ref{fig:SupplementBaselineCorrection1} and \ref{fig:SupplementBaselineCorrection2}.
    In all cases, numerical motion tracking and ratiometry reduces motion artifacts substantially.}
    \label{fig:SupplementSinusAPTraces}
\end{figure*}

\begin{figure*}[!ht]
    \centering
    \includegraphics[clip, trim=0.0cm 0.0cm 0.0cm 0.0cm, width=0.9\textwidth]{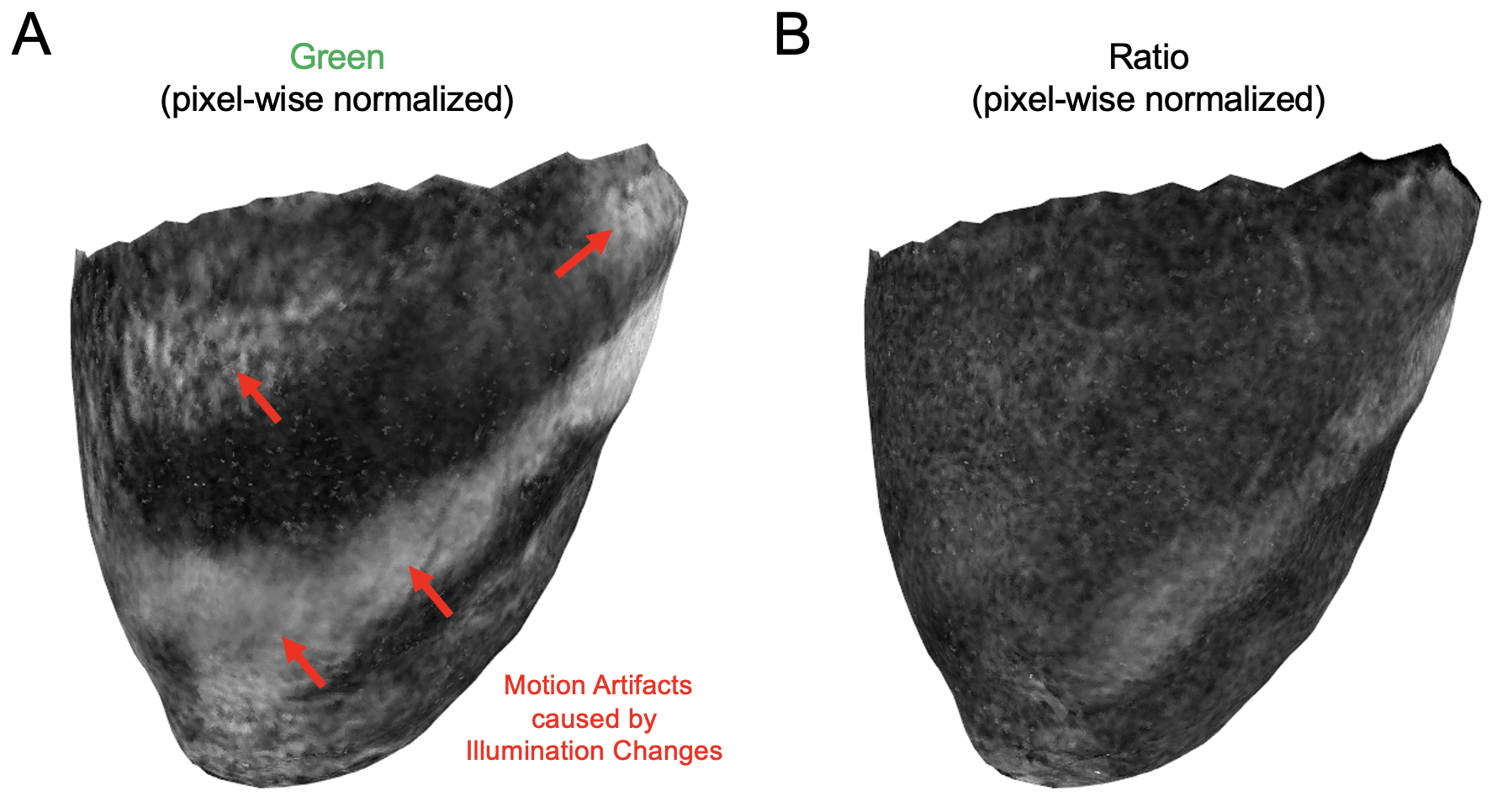}
    \caption{\textbf{Efficacy of ratiometric imaging combined with 3D motion tracking.}\\
    Comparison of 3D optical maps showing the right ventricular surface of a strongly contracting heart during the plateau phase of the action potential (dark: depolarized tissue), see also Fig.~\ref{fig:Sinus2}C). Both datasets were 3D tracked and the optical maps were equally processed in a co-moving frame of reference using the 3D tracking and mesh data.
    The heart was imaged during sinus rhythm.
    \textbf{A} Pixel-wise normalized optical map computed from only the green video data. Without ratiometry, the optical maps exhibit motion artifacts (bright areas) which stem from illumination changes in the excitation light as the heart moves back and forth between slightly differently illuminated areas.
    \textbf{B} Pixel-wise normalized optical map computed from the ratiometric (combined green and blue) video data. With ratiometry, the optical maps are more homogeneous indicating that illumination-related motion artifacts are reduced by the ratiometric artifact compensation.
    }
    \label{fig:SupplementComparisonGreenRatio}
\end{figure*}

\begin{figure*}[!ht]
    \centering
    \includegraphics[clip, trim=0.0cm 0.0cm 0.0cm 0.0cm, width=0.6\textwidth]{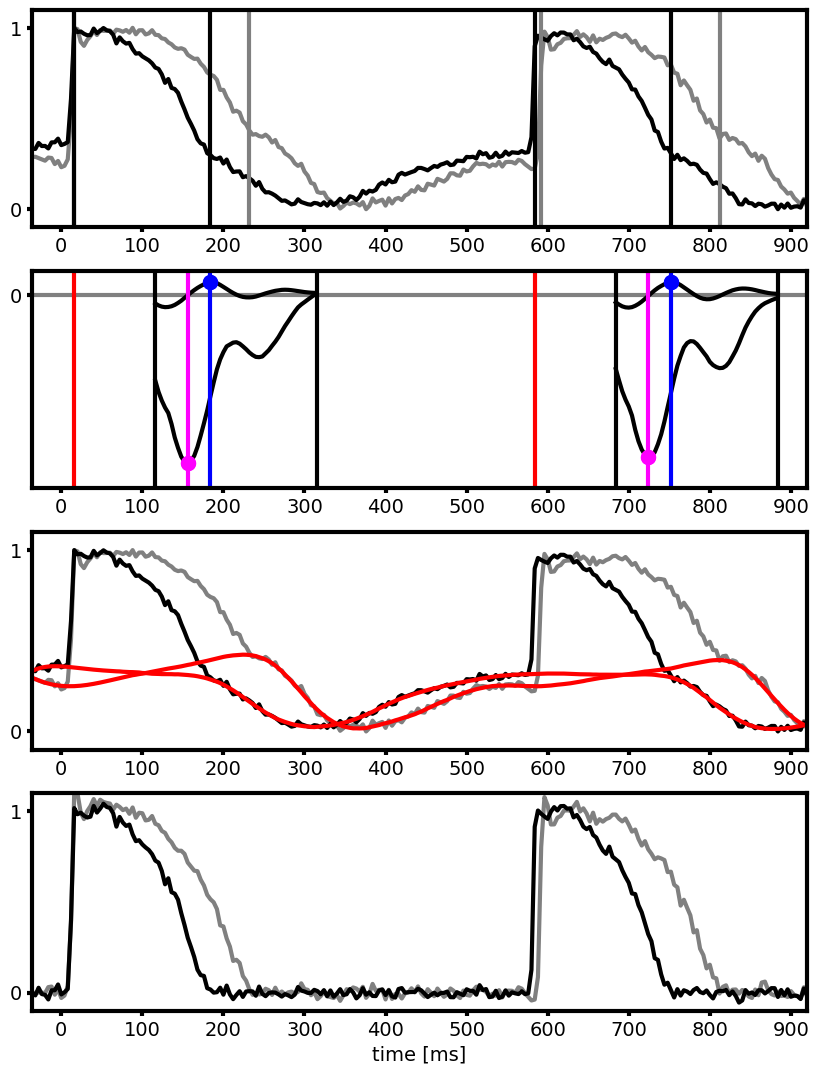}
    \caption{\textbf{Upstroke and repolarization detection for baseline correction of optical traces to compensate residual motion artifacts.}\\
    Residual artifacts which remain after numerical motion tracking and ratiometric compensation in optical traces measured across the surface of strongly moving hearts can be further inhibited using a custom baseline correction method. First, upstrokes and repolarization times were detected by calculating the first and second temporal derivatives of each optical trace (shown in second plot from the top). Each optical trace and derivative was smoothed using Savitzky-Golay filtering prior to each processing step. Upstrokes (red vertical bar) were detected as maximal positive peaks in the first derivative (not shown). Repolarizations were detected as a combination of a negative peak in the first derivative (pink vertical bar and dot) indicating the steepest part of the repolarization phase followed by a slight positive peak in the second derivative (blue vertical bar and dot) indicating a sudden change and return to baseline. The detections were performed between \SI{100}{\milli \second} and \SI{300}{\milli \second} after each upstroke. The two optical traces were measured in the same location with normal Tyrode (black) and low potassium Tyrode plus barium-chloride to prolong the action potential (gray). The derivatives (shown in the second plot from the top) correspond to the optical trace measured with normal Tyrode (black), but the detection performs the same with the optical trace showing the prolonged action potential.
    Using the upstroke and repolarization time, the action potential was removed and missing values were linearly interpolated between upstrokes and repolarizations. The resulting trace was smoothed using Savitzky-Golay filtering (red) and subsequently subtracted from the original trace yielding the traces shown in the bottom plot.
    The baseline correction yields consistent, uniform action potential duration measurements across the ventricular surface, see Fig.~\ref{fig:SupplementAPD}.
    }
    \label{fig:SupplementBaselineCorrection1}
\end{figure*}

\begin{figure*}[!ht]
    \centering
    \includegraphics[clip, trim=0.0cm 0.0cm 0.0cm 0.0cm, width=0.8\textwidth]{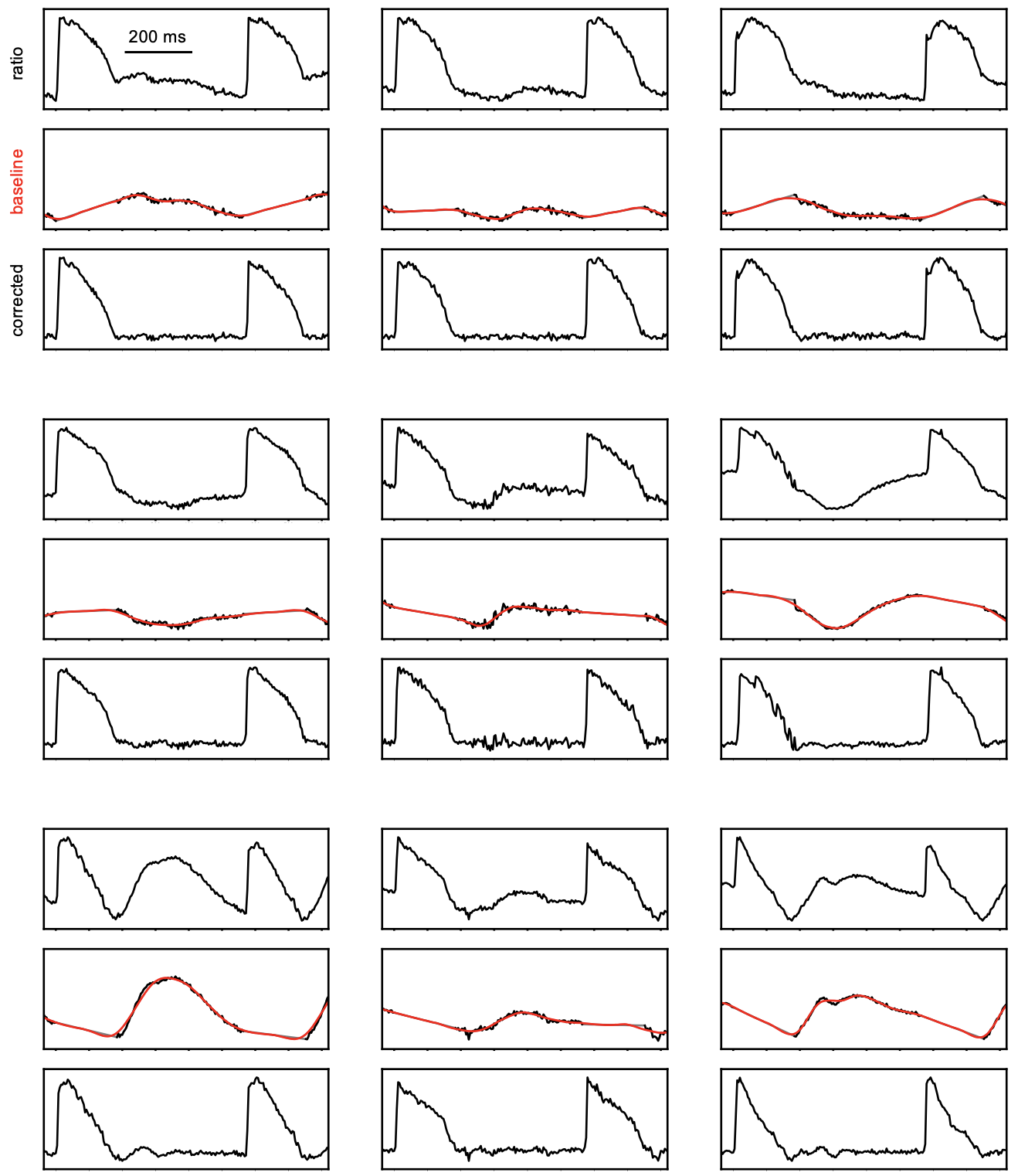}
    \caption{\textbf{Baseline correction for compensation of residual motion artifacts during post-processing.}\\
    Residual artifacts which remain after numerical motion tracking and ratiometric compensation in optical traces measured across the surface of strongly moving hearts can be further inhibited using a custom baseline correction method applied to each optical trace, see also Fig.~\ref{fig:SupplementBaselineCorrection1}. Shown are $3 \times 3$ representative examples. Each panel: optical trace after tracking and ratiometry (top), baseline fitting using Savitzky-Golay filtering (red) after removal of action potential (center), and corrected optical trace showing flat diastolic intervals for most optical traces. Residual artifacts remain in some of the optical traces (bottom panels).
    Scalebar \SI{200}{\milli \second}, y-axis: normalized dimensionless units.
    }
    \label{fig:SupplementBaselineCorrection2}
\end{figure*}

\begin{figure*}[!ht]
    \centering
    \includegraphics[clip, trim=0.0cm 0.0cm 0.0cm 0.0cm, width=0.8\textwidth]{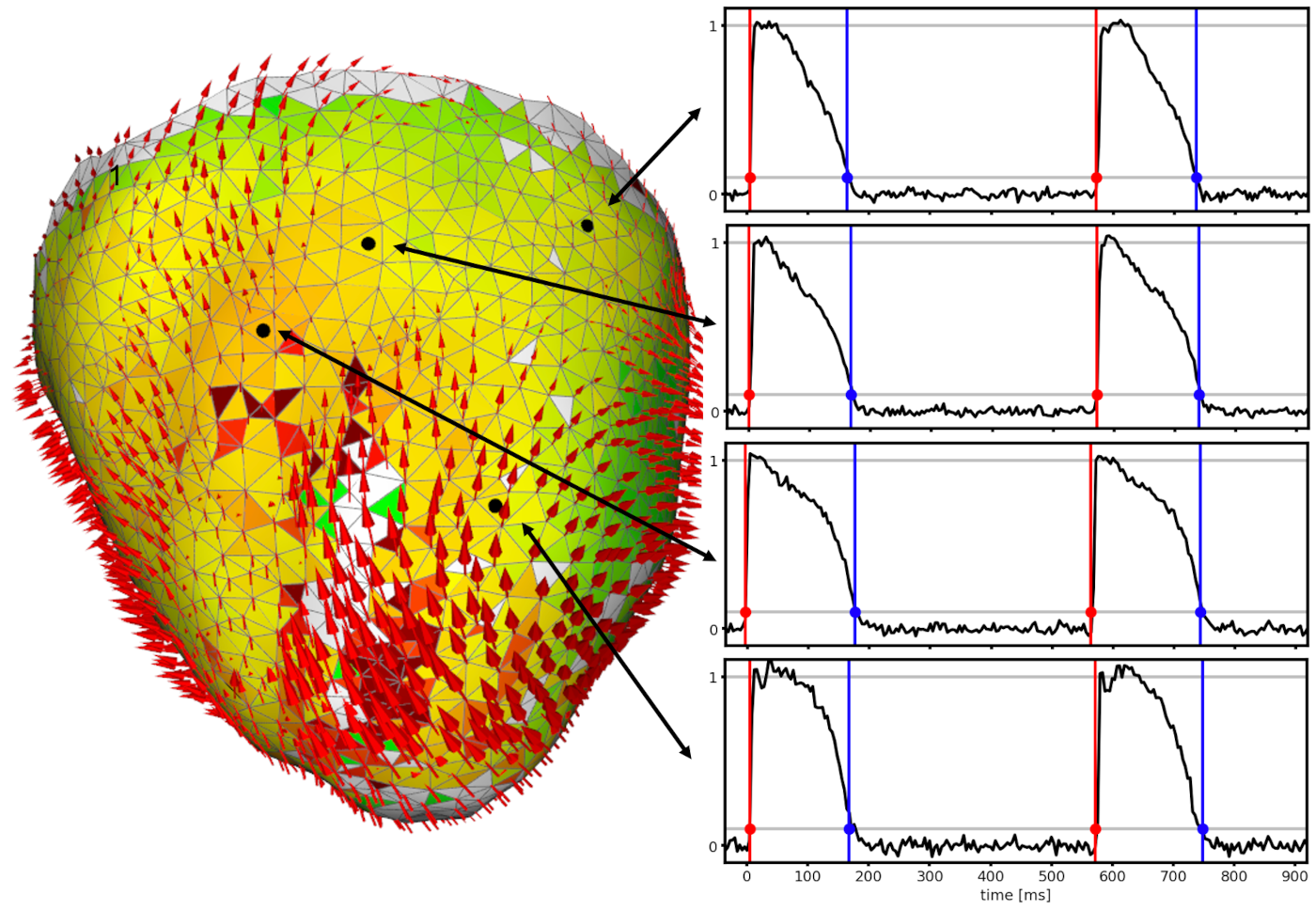}
    \caption{\textbf{Action potential duration (APD) map measured optically across contracting heart surface.}\\
    Left: Action potential duration (APD) map computed from optical traces measured at the center of each triangle across entire ventricular surface of contracting rabbit heart (here shown across anterior wall). Same colorcode as in Fig.~\ref{fig:Sinus3}E). Red displacement vectors indicate movements from one frame to the next (arbitrary scaling of magnitude). White triangles represent rejected/excluded APD measurements. Right: Representative optical traces obtained from 4 locations on anterior wall (black dots) after post-processing as shown in Figs.~\ref{fig:SupplementBaselineCorrection1}-\ref{fig:SupplementBaselineCorrection2}. An APD measurement was rejected/excluded either when the first and second APDs were significantly different or if there was an inconsistent number of upstrokes (red) or downstrokes (blue). APDs were computed at 10\% of the height of the action potential (gray line, APD 90).
    }
    \label{fig:SupplementAPD}
\end{figure*}

\begin{figure*}[!ht]
    \centering
    \includegraphics[clip, trim=0.0cm 0.0cm 0.0cm 0.0cm, width=0.8\textwidth]{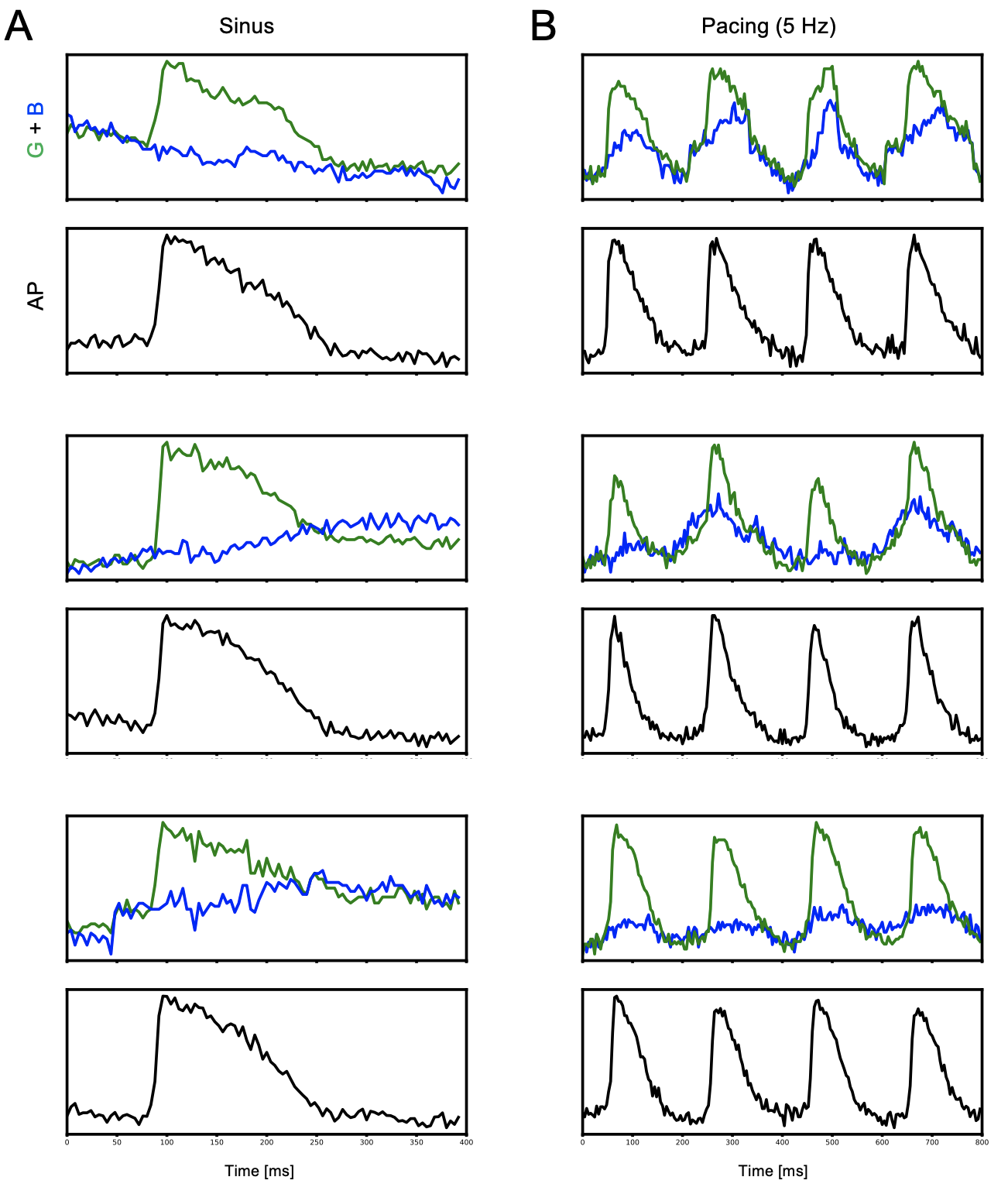}
    \caption{\textbf{Efficacy of ratiometric imaging combined with 3D motion tracking.}\\
    Baseline correction methods as shown in Figs.~\ref{fig:SupplementBaselineCorrection1}-\ref{fig:SupplementBaselineCorrection2} are not necessary with hearts exhibiting mild to moderate motion. 
    In this example, the heart moved only minimally during \textbf{A} sinus rhythm and \textbf{B} pacing (3 examples each). 
    The traces show that numerical motion tracking and ratiometry can be sufficient to suppress motion artifacts (all traces obtained after 3D motion tracking).
    }
    \label{fig:SupplementTracesRatio}
\end{figure*}

\begin{figure*}[!ht]
    \centering
    \includegraphics[clip, trim=0.0cm 0.0cm 0.0cm 0.0cm, width=0.9\textwidth]{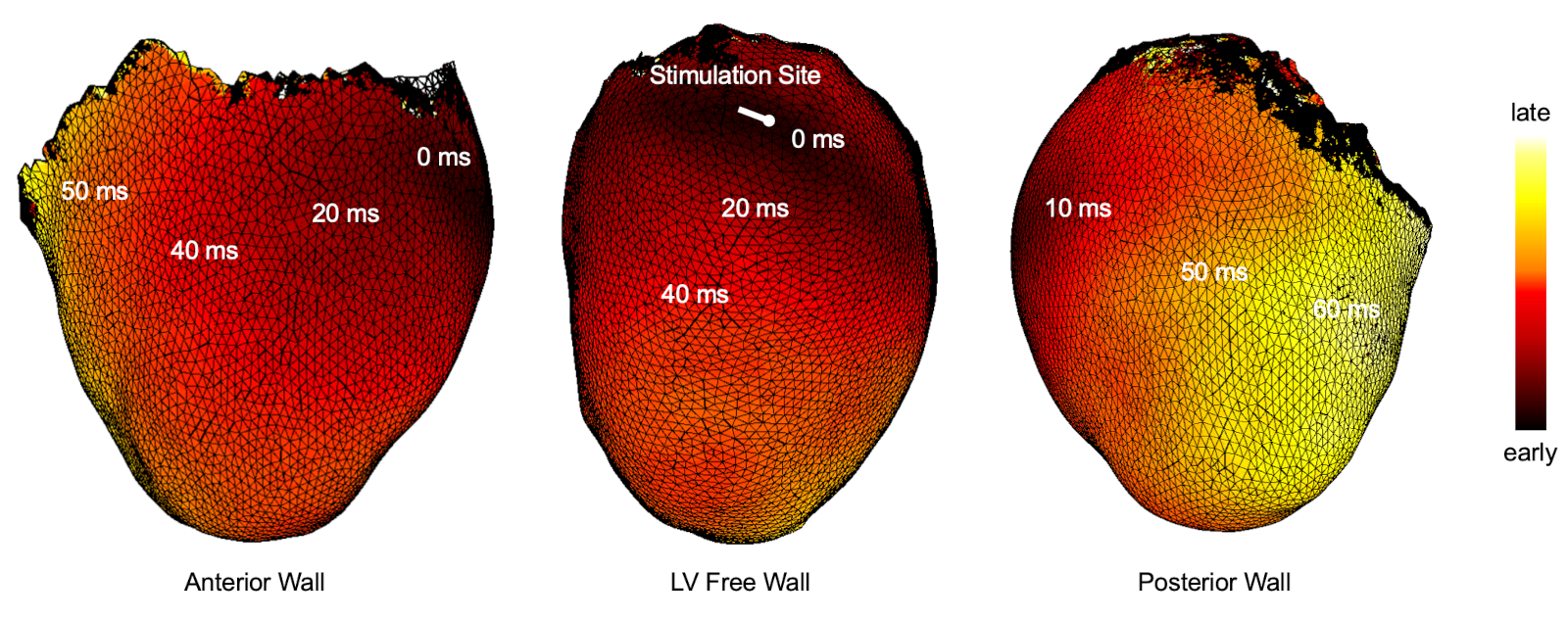}
    \caption{\textbf{Electrical activation map.}\\
    Panoramic electrical activation map showing anisotropic propagation of action potential wave across entire ventricular surface of a contracting isolated rabbit heart during ventricular pacing (dark: early activation, yellow: late activation), c.f. Fig.~\ref{fig:Pacing2}E). The ventricles become electrically activated within about \SI{60}{\milli \second}.
    The view onto the free left ventricular (LV) wall shows the typical elliptical shape of the electrical conduction and activation associated with the muscle fiber anisotropy of the heart.}
    \label{fig:SupplementElectricalActivationMap}
\end{figure*}

\begin{figure*}[!ht]
    \centering
    \includegraphics[clip, trim=0.0cm 0.0cm 0.0cm 0.0cm, width=0.9\textwidth]{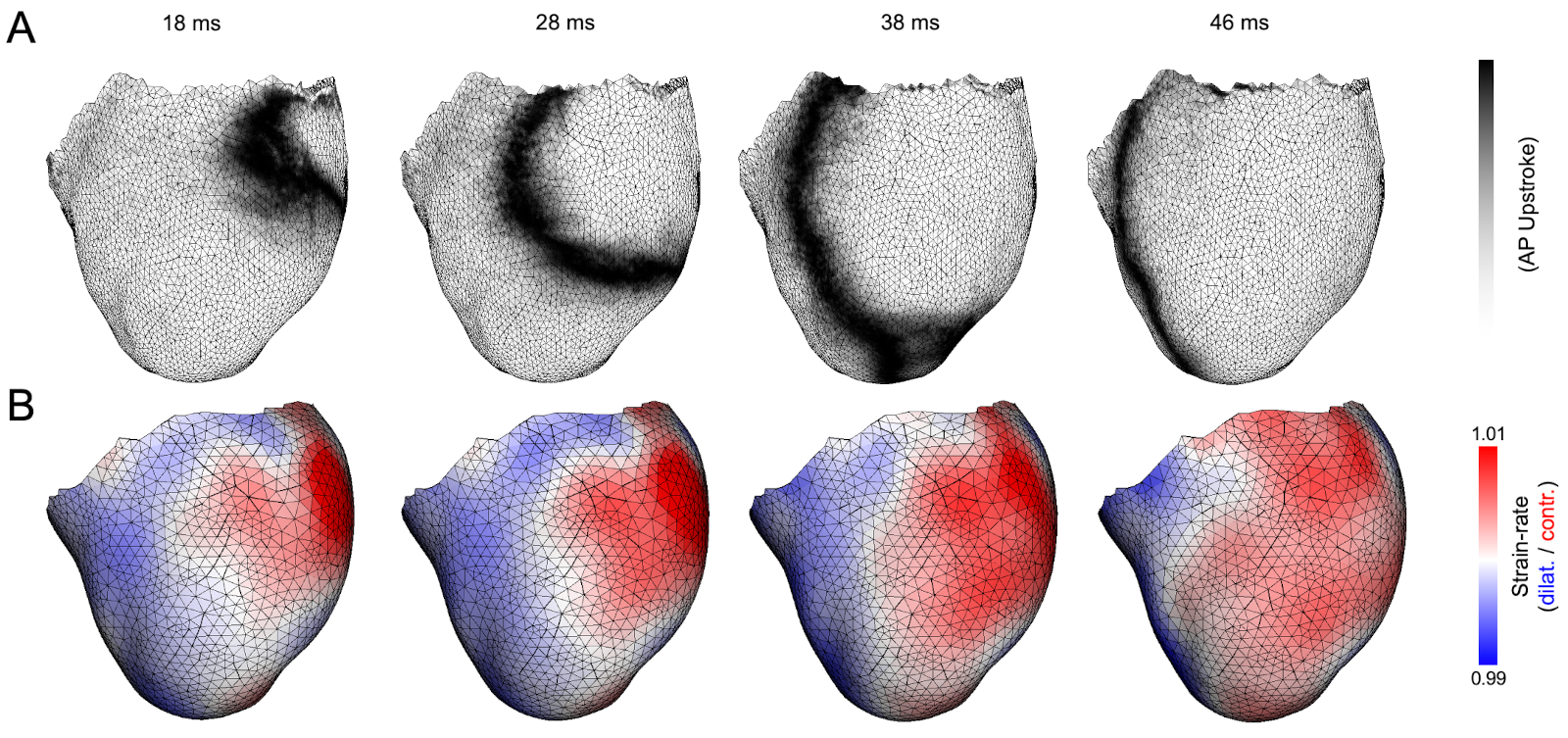}
    \caption{\textbf{Electromechanical wave propagating across ventricular wall.}\\
    Electromechanical wave propagating across the ventricles of a rabbit heart during pacing (4.5 Hz), here shown on the opposite side of the heart shown in Fig.~\ref{fig:Pacing2}.
    \textbf{A)} Action potential wavefront (black: upstroke) propagating across deforming ventricular surface.
    \textbf{B)} Mechanical activation wavefront consisting of a transition from dilating (blue) to
    contracting (red) strain-rates measured as the rate of triangular area change of each mesh triangle, see also Fig.~\ref{fig:Pacing2}B,D).
    The electrical and mechancial activation are correlated, see also Figs.~\ref{fig:Pacing2} and \ref{fig:SupplementElectricalActivationMap}.}
    \label{fig:SupplementElectromechanicalWave}
\end{figure*}

\begin{figure*}[!ht]
    \centering
    \includegraphics[clip, trim=0.0cm 0.0cm 0.0cm 0.0cm, width=0.95\textwidth]{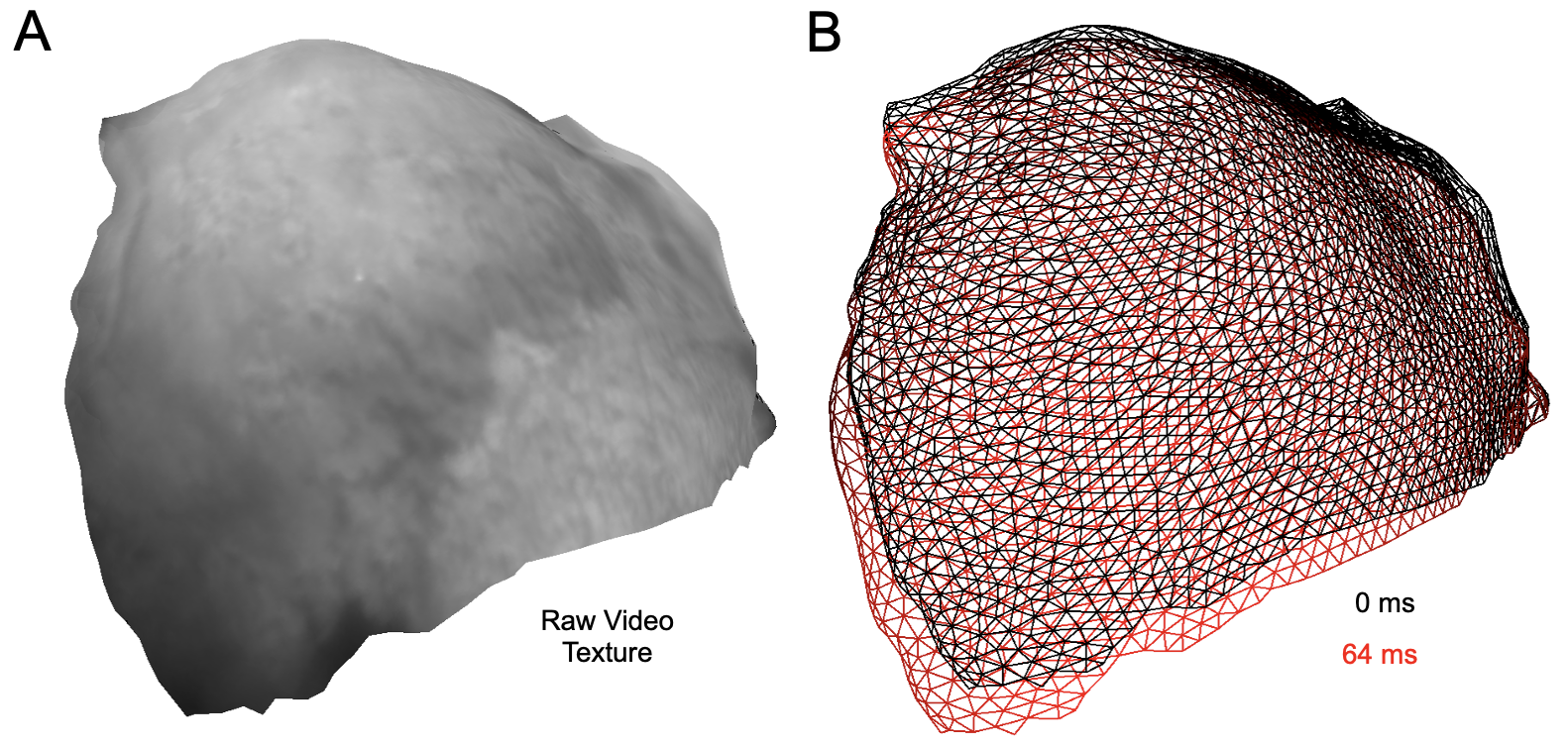}
    \caption{\textbf{Motion and deformation of ventricular wall during pacing.}\\
    Part of the ventricular surface reconstructed and tracked using 5 cameras, see also Supplementary Video \ref{video:pacing2}.
    \textbf{A)} Texturized mesh.
    \textbf{B)} Wireframe mesh before stimulation (black) and \SI{64}{\milli \second} after stimulation (red) during ventricular pacing, see also Fig.~\ref{fig:SupplementPacing2}.
    }
    \label{fig:SupplementPacing1}
\end{figure*}

\begin{figure*}[!ht]
    \centering
    \includegraphics[clip, trim=0.0cm 0.0cm 0.0cm 0.0cm, width=0.95\textwidth]{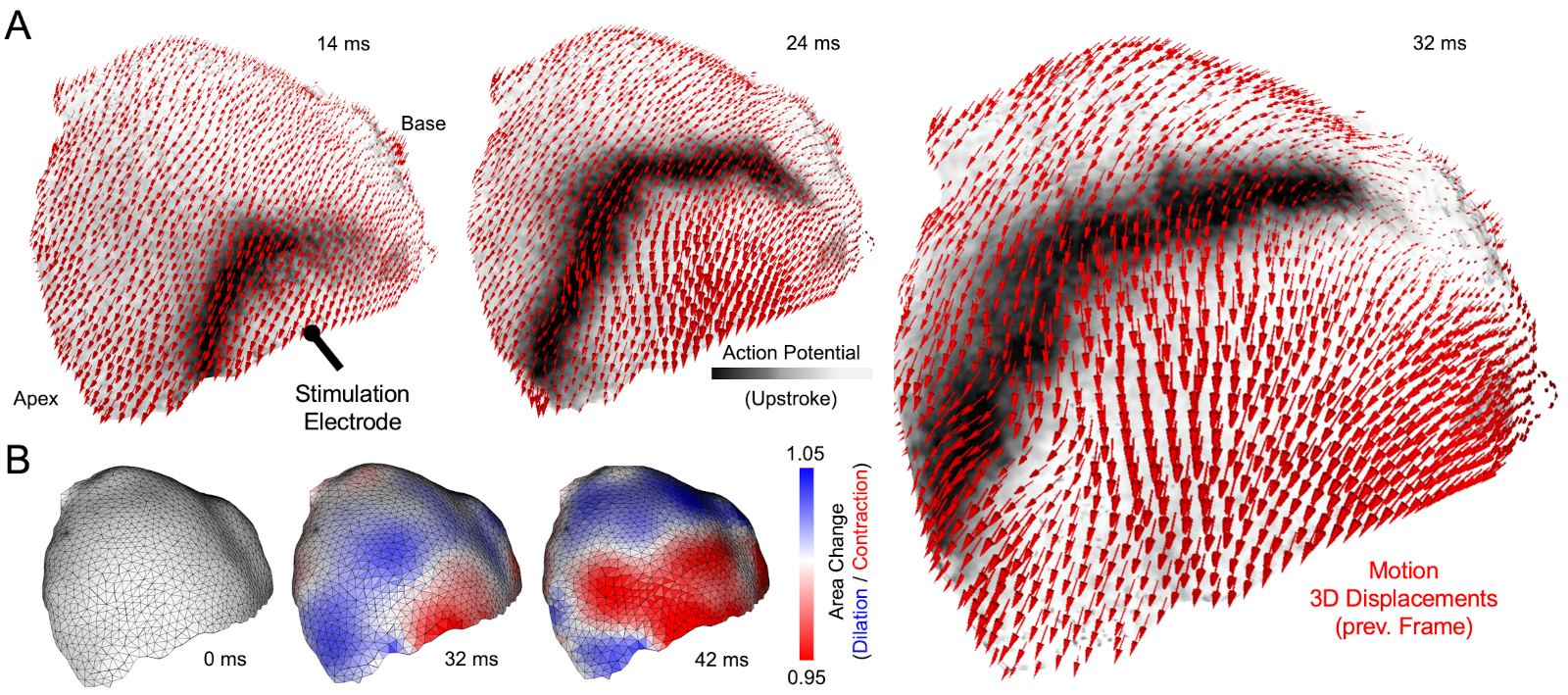}
    \caption{\textbf{Electromechanical wave propagating across part of the ventricular wall.}\\
    \textbf{A)} Action potential wavefront (black: upstroke) propagating across contracting left ventricular surface. The motion and deformation of the tissue is displayed by three-dimensional displacement vectors (red), which indicate instantaneous motion with respect to a previous frame, in this case acquired \SI{4}{\milli \second} earlier. The tissue moves down and towards the approaching wave and then contracts noticeably when the action potential wave front has passed, c.f. Fig.~\ref{fig:SupplementPacing1}.
    \textbf{B)} Area change (per triangle) in response to action potential wave (blue: dilation / area increase, red: contraction / area decrease). The area change was computed with respect to the mechanical configuration and the triangular size at \SI{0}{\milli \second} in percent. The ventricular surface contracts behind the action potential wavefront, see also Supplementary Video \ref{video:pacing2}.}
    \label{fig:SupplementPacing2}
\end{figure*}

\begin{figure*}[!ht]
    \centering
    \includegraphics[clip, trim=0.0cm 0.0cm 0.0cm 0.0cm, width=0.95\textwidth]{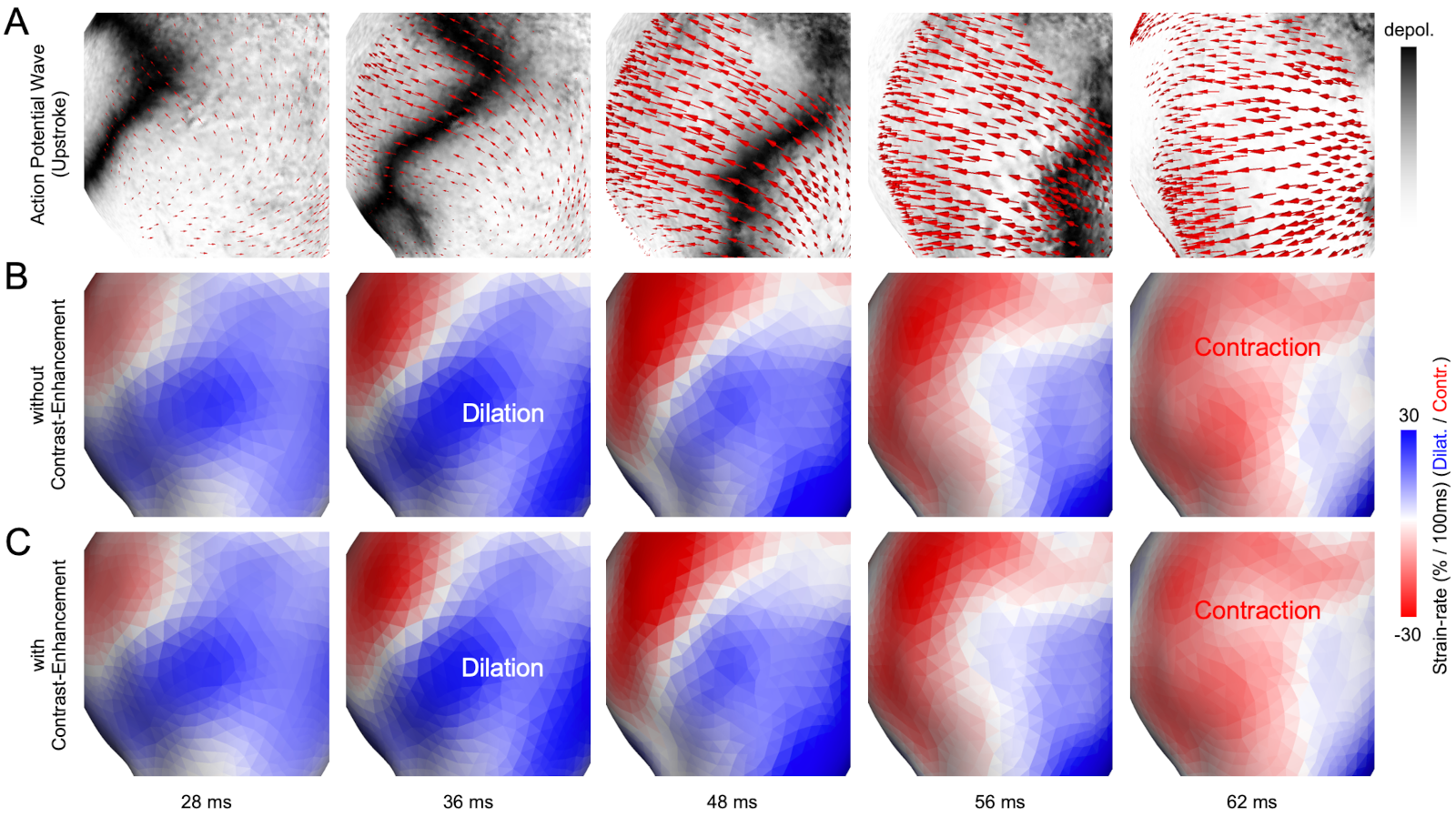}
    \caption{\textbf{Validation of tracking robustness.}\\
    Tracking of ventricular surface with and without contrast-enhancement as described in Christoph \& Luther (2018), see also Fig.~\ref{fig:Pacing2} and Supplementary Video \ref{video:pacing1}. 
    The videos were acquired with continuos green illumination. 
    Tracking with or without contrast-enhancement yields nearly identical strain-rate patterns, which is an indication that the tracking is not sensitive to the fluorescent signal.
    In previous work, we demonstrated that 2D motion tracking algorithm can be suceptible to intensity modulations caused by the fluorescent dyes, see Christoph \& Luther (2018) or Lebert et al. (2022).
    \textbf{A} Action potential upstroke (black) and wave front propagating across ventricles during pacing (red vectors: motion of tissue), see also Fig.~\ref{fig:Pacing2}.
    \textbf{B} Strain-rate pattern computed when tracking was performed without contrast-enhancement.
    \textbf{C} Strain-rate pattern computed when tracking was performed with contrast-enhancement (on the contrast-enhanced video data instead of the original video data). 
    The two strain-rate patterns with and without contrast-enhancement are visually nearly indistinguishable. 
    Contrast-enhancement was introduced in Christoph \& Luther (2018) to inhibit tracking artifacs. 
    Tracking artifacts can potentially arise when the tracking algorithm inadvertently tracks the movement of the action potential wave instead of the motion of the tissue. 
    Our tracking algorithm does not exhibit such tracking artifacts, as demonstrated by the two nearly identical strain-rate patterns.
    Here, strain-rate was approximated by rate of triangular area change (blue: dilation or area increase, red: contraction or area decrease, about $\pm$ 30\% per \SI{100}{\milli \second}).
    }
    \label{fig:SupplementCrosstalk}
\end{figure*}
%  CONFIGURE NEW SINGLE-PAGE FORMAT 

\onecolumn % go back to one column
\fancyhead{} % make sure we get no headers
\renewcommand{\floatpagefraction}{0.1}
\lfoot[\bSupInf]{\dAuthor}
\rfoot[\dAuthor]{\cSupInf}
\newpage

\captionsetup*{format=largeformat} % make figure legend slightly larger than in the paper
\setcounter{figure}{0} % reset figure counter for Supplementary Video Figures
\makeatletter 
\renewcommand{\thefigure}{SV\@arabic\c@figure} % make Figure legend start with Figure S
\makeatother

%  MAIN TEXT 
\newpage
\section*{Supplementary Videos}
The videos as well as interactive 3D renderings are available at \href{https://cardiacvision.ucsf.edu/videos/3d-optical-mapping/}{https://cardiacvision.ucsf.edu/videos/3d-optical-mapping/}.

\begin{SCfigure*}[50][!ht]
    \includegraphics[width=0.15\linewidth]{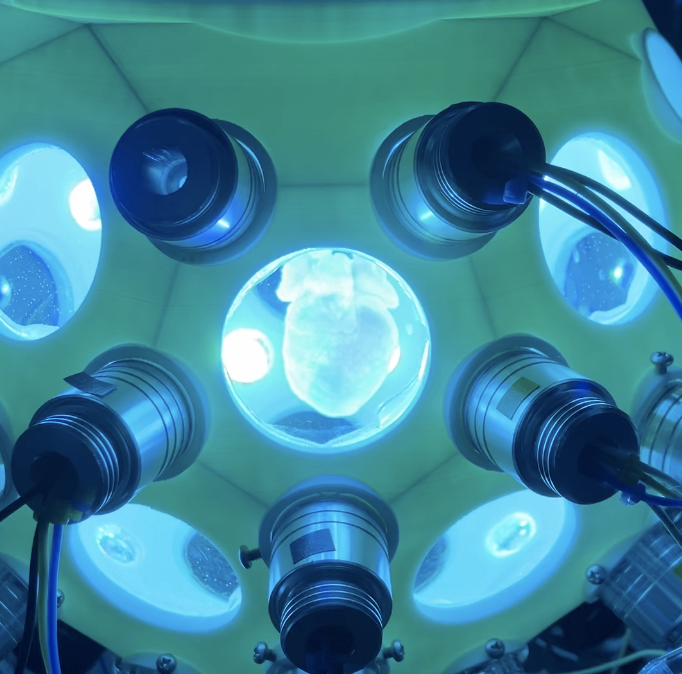}
    \caption{\textbf{Supplementary Video 1.}\\
    Intact isolated rabbit heart in soccer ball-shaped imaging chamber with 24 windows during voltage-sensitive panoramic optical mapping with 12 high-speed cameras, see also Fig.~\ref{fig:Setup}. The voltage-sensitive dye is excited by up to 48 RGBW (red-green-blue-white) light-emitting diodes (LEDs). 
    The positioning of the LEDS allows even illumination and excitation of the tissue and fluorescent dye.
    }
    \label{video:SV1}
\end{SCfigure*}

\begin{SCfigure*}[50][!ht]
    \includegraphics[width=0.15\linewidth]{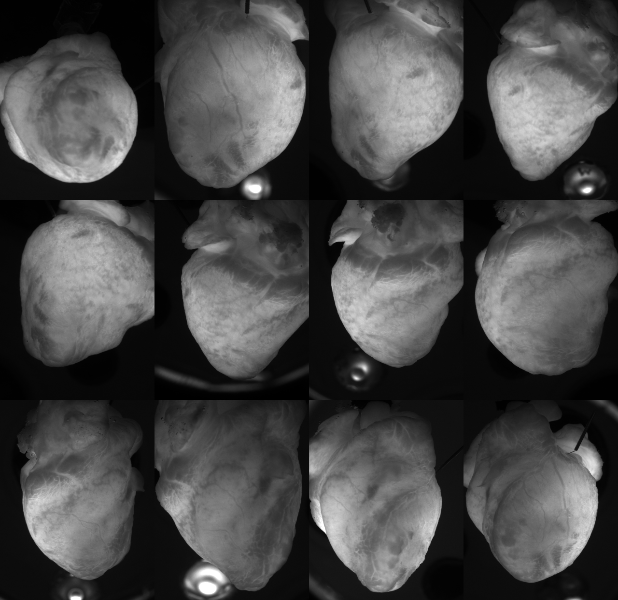}
    \caption{\textbf{Supplementary Video 2.}\\
    Creation of 3D model of contracting heart using panoramic optical mapping data obtained with 12 calibrated high-speed cameras (left), see also Fig.~\ref{fig:Setup}D). The 3D model (right) is texturized using the grayscale video data from the cameras. In this example, only the ventricles are reconstructed because the cameras are pointed mainly at the ventricles using the configuration shown in Fig.~\ref{fig:Setup}E), see also Supplementary Video \ref{video:SV3}. The voltage-sensitive optical mapping videos show intensity fluctuations representing action potential waves. The imaging was performed with continuous green illumination to excite the voltage-sensitive fluorescent dye Di-4-ANEPPS. 
    }
    \label{video:SV2}
\end{SCfigure*}

\begin{SCfigure*}[50][!ht]
    \includegraphics[width=0.15\linewidth]{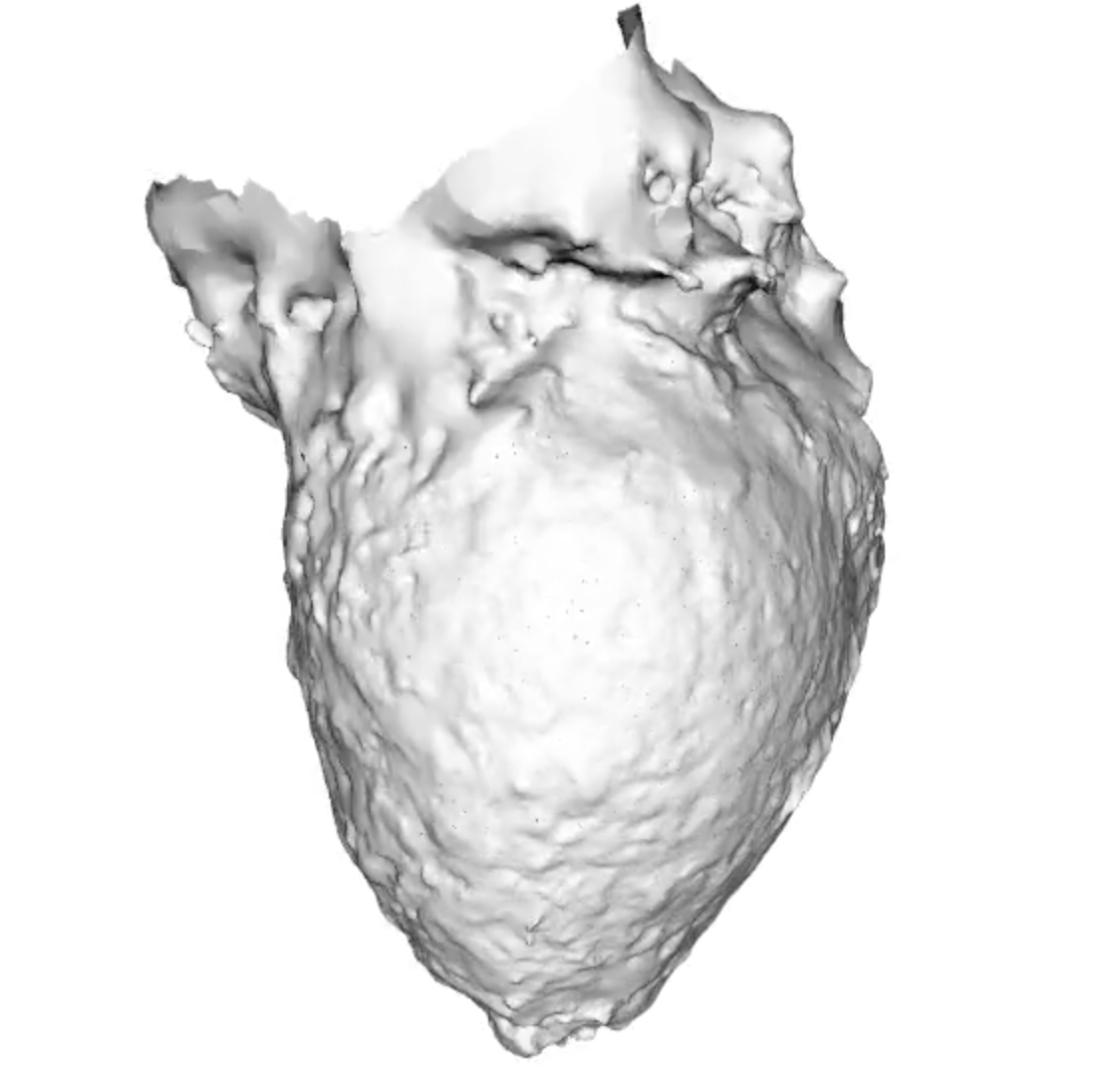}
    \caption{\textbf{Supplementary Video 3.}\\
    Panoramic voltage-sensitive imaging and tracking of 3D heart surface using multi-camera optical mapping. The mesh-based tracking technique utilizes a template mesh (black wireframe mesh) and searches per vertex for movements of the surrounding tissue in the camera images. The search is guided by a sequence of static raw meshes (gray/white). The raw meshes were computed prior to the tracking using a photogrammetry-like technique, see also Fig.~\ref{fig:Fig2}. The moving mesh is then texturized using the grayscale video data, see also Supplementary Video \ref{video:SV2}.
    }
    \label{video:SV3}
\end{SCfigure*}

\begin{SCfigure*}[50][!ht]
    \includegraphics[width=0.15\linewidth]{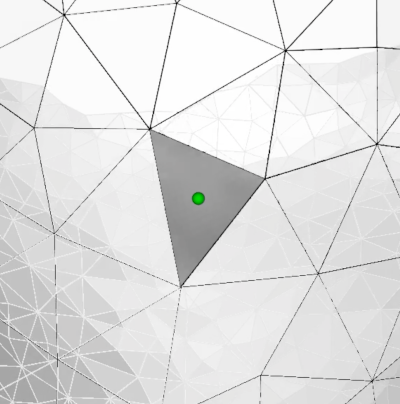}
    \caption{\textbf{Supplementary Video 4.}\\
    Optical traces are sampled from the center of each triangle, here shown separately for the green and blue channels with ratiometric imaging data. 
    Same triangle as shown in Fig.~\ref{fig:Setup}F).
    Each texturized triangle represents a segment of the epicardial surface moving through 3D space (the rest of the epicardial surface is represented by a wireframe mesh).
    The first half of the video shows the triangle from a stationary viewpoint.
    The second half shows the triangle from the perspective of a co-moving camera that hovers over the triangle.
    The video illustrates the effectiveness of motion tracking and co-moving signal analysis. 
    Motion artifacts are effectively inhibited when the grayscale pattern on the triangle becomes stationary (in its local reference frame), see also Fig.~\ref{fig:Fig2}.
    }
    \label{video:SV4}
\end{SCfigure*}

\begin{SCfigure*}[50][!ht]
    \includegraphics[width=0.15\linewidth]{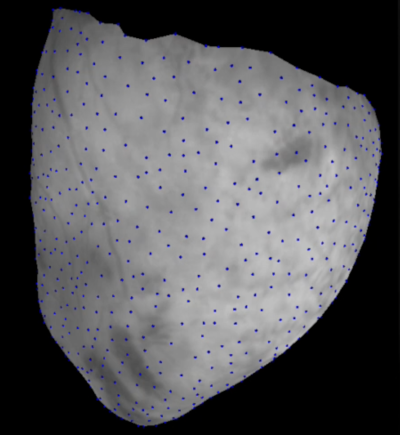}
    \caption{\textbf{Supplementary Video 5.}\\
    Comparison of (left) movements of mesh vertices (blue) drawn as spheres onto surface of reconstructed, tracked, and texturized 3D mesh vs. (right) mesh vertices projected into one of the camera images, see also Figs.~\ref{fig:Fig2}E) and Supplementary Video \ref{video:SV2}.
    Both visualizations confirm that the vertices follow the motion of the heart.
    }
    \label{video:SV5}
\end{SCfigure*}

\begin{SCfigure*}[50][!ht]
    \includegraphics[width=0.15\linewidth]{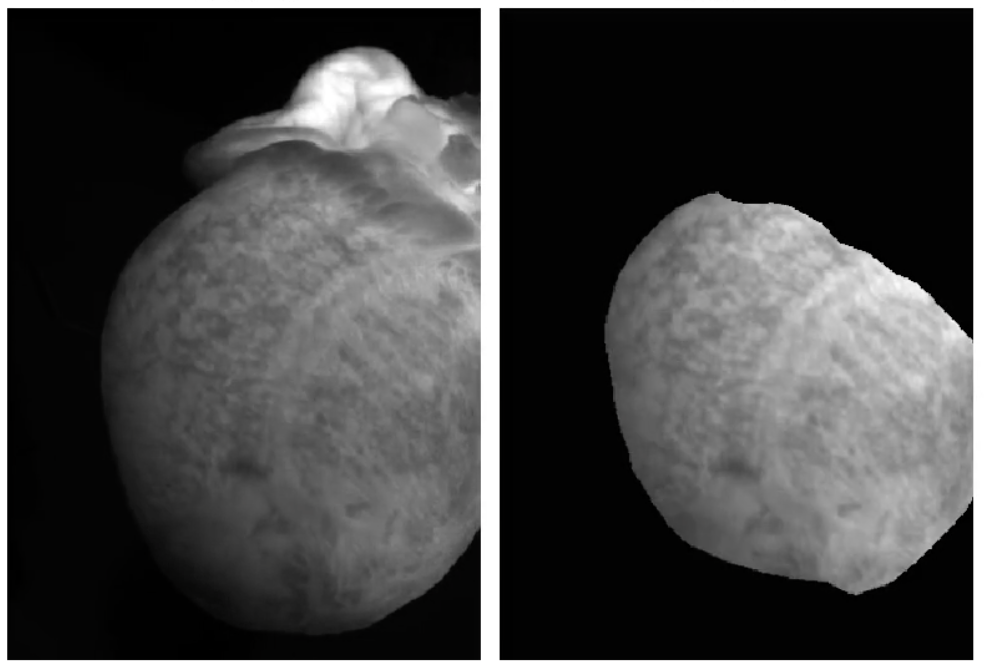}
    \caption{\textbf{Supplementary Video 6.}\\
    Comparison of 2D vs. 3D tracking. Left: Original video (blue channel during ratiometric imaging) showing contracting heart during pacing. Center: Tracked and motion-stabilized video after single-view 2D tracking performed with optimap using the Farneb\"ack tracking algorithm. With the 2D tracking alone it is not possible to fully track and compensate the motion. Correspondingly, the video exhibits strong tracking and warping artifacts. Right: With cooperative multi-view 3D tracking, the motion and video can be tracked and stabilized (except boundary pixels which are discarded at later stages during the processing).
    }
    \label{video:2Dvs3D}
\end{SCfigure*}

\begin{SCfigure*}[50][!ht]
    \includegraphics[width=0.15\linewidth]{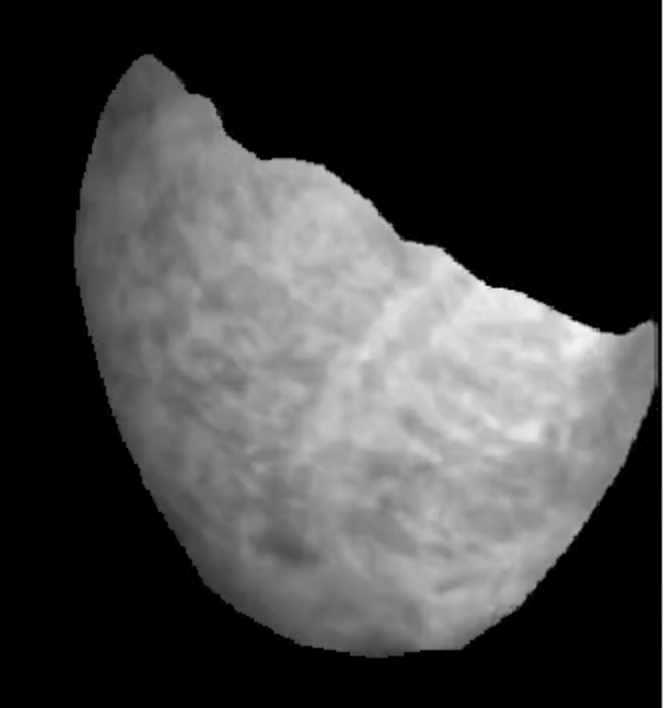}
    \caption{\textbf{Supplementary Video 7.}\\
    Comparison of the effect of 3D tracking alone vs. 3D tracking with additional 2D tracking onto motion-stabilized optical maps. Left: Due to the lower spatial resolution of the 3D mesh, the tracked and motion-stabilized videos exhibit residual motion when using only the 3D tracking data to warp the videos. Right: To increase the efficacy of motion-stabilization and motion artifact inhibition, the already tracked and motion-stabilized videos were subsequently tracked and motion-stabilized a second time using 2D tracking. Same data as in Supplementary Video \ref{video:2Dvs3D}.
    }
    \label{video:2Dvs3Dwarping}
\end{SCfigure*}

\begin{SCfigure*}[50][!ht]
    \includegraphics[width=0.15\linewidth]{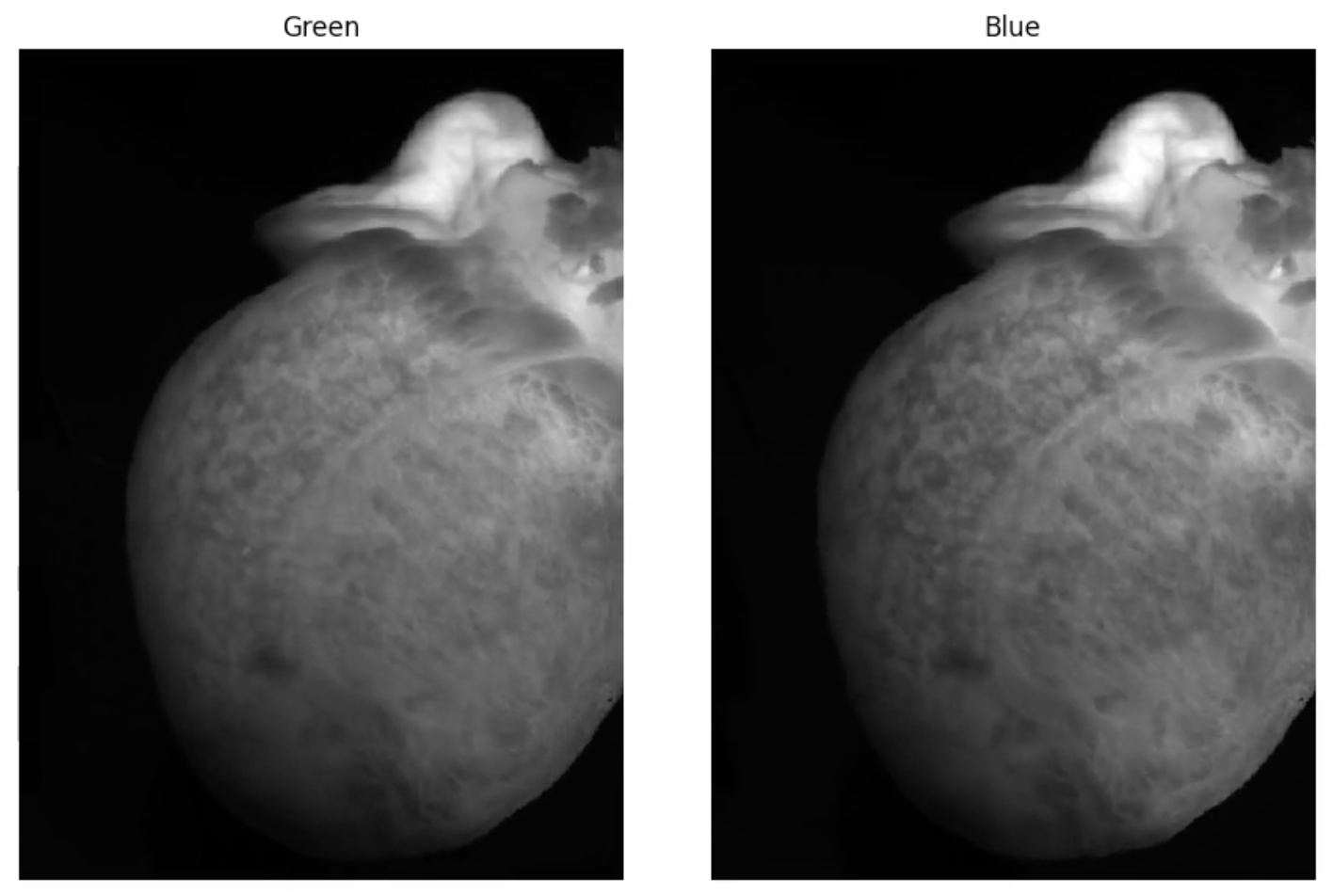}
    \caption{\textbf{Supplementary Video 8.}\\
    Ratiometric optical mapping video with separate green (left) and blue (right) channels showing contracting isolated rabbit heart during ventricular pacing, see also Fig.~\ref{fig:SupplementRatiometry}. The voltage-sensitive fluorescent signal emerges in the green channel, while the blue channel comprises only background signal. With ratiometric data, we used the blue channel to track the tissue.
    }
    \label{video:green-blue-video}
\end{SCfigure*}

\begin{SCfigure*}[50][!ht]
    \includegraphics[width=0.15\linewidth]{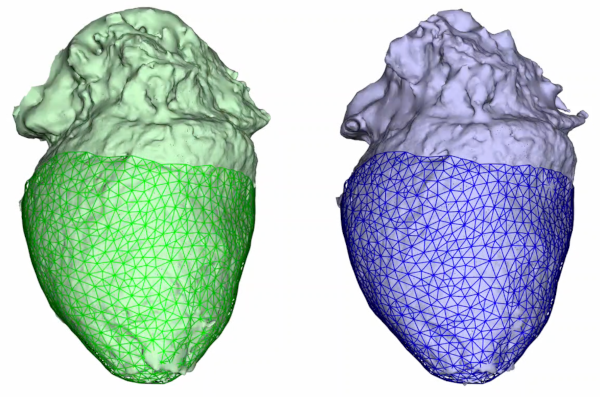}
    \caption{\textbf{Supplementary Video 9.}\\
    Side-by-side comparison of 3D reconstruction of moving ventricular surface from two-channel (green, blue) ratiometric optical mapping data. The voltage-sensitive optical mapping data was acquired with Di-4-ANEPPs and excitation ratiometry with alternating green and blue excitation resulting in 2 sets of videos (green and blue), see also Fig.~\ref{fig:SupplementRatiometry}. From the two sets of videos we created 2 separate 3D reconstructions (green and blue). The visualization shows the dynamic meshes (wireframe) and static meshes (surface).
    }
    \label{video:green-blue-video-3D}
\end{SCfigure*}

\begin{SCfigure*}[50][!ht]
    \includegraphics[width=0.15\linewidth]{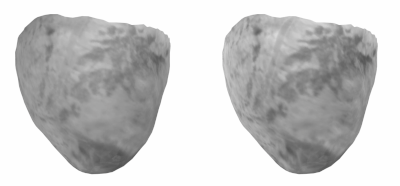}
    \caption{\textbf{Supplementary Video 10.}\\
    Green (left) and blue (right) 3D optical maps showing moving ventricular surface from two-channel (green, blue) ratiometric 3D optical mapping data, see also Figs.\ref{fig:Setup}, \ref{fig:Sinus3} and \ref{fig:SupplementRatiometry}. The green optical map reports the transmembrane potential changes due to the voltage-sensitive staining with Di-4-ANEPPs. The blue optical map is neutral and only exhibits illumination changes as the heart moves through a 3D light-field produced by the LEDs.
    }
    \label{video:green-blue-video-3D-texture}
\end{SCfigure*}

\begin{SCfigure*}[50][!ht]
    \includegraphics[width=0.15\linewidth]{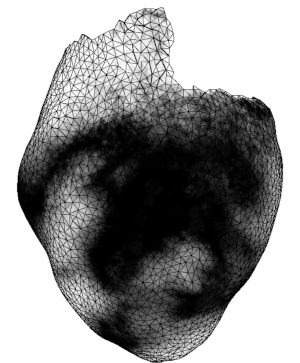}
    \caption{\textbf{Supplementary Video 11.}\\
    Electrical activation of the ventricles in a beating isolated rabbit heart during sinus rhythm imaged using voltage-sensitive multi-camera panoramic optical mapping. The tissue was tracked in 3D space, which allowed us to measure the action potential wavefront in a co-moving frame of reference, as well as an electrical activation map and local tissue strain (as a measure of triangular area change). The imaging was performed with continuous green illumination (non-ratiometric). 
    }
    \label{video:sinus}
\end{SCfigure*}

\begin{SCfigure*}[50][!ht]
    \includegraphics[width=0.15\linewidth]{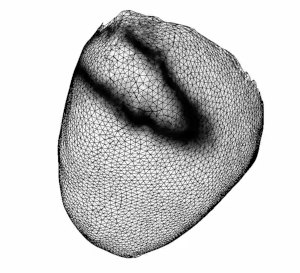}
    \caption{\textbf{Supplementary Video 12.}\\
    Focal action potential wavefront (black: upstroke) propagating across ventricles during ventricular pacing, see also Figs.~\ref{fig:Pacing1}-\ref{fig:Pacing2}. The wave was induced using a flexible, thin electrode, see also Figs.~\ref{fig:SupplementTrackingElectrode} and \ref{fig:SupplementCoverage}A,B). The ventricles deform in response to the electrical activation (black wireframe mesh). Our data shows that the electrical activation and mechanical deformation are strongly correlated. The imaging was performed with continuous green illumination (non-ratiometric). 
    }
    \label{video:pacing1}
\end{SCfigure*}

\begin{SCfigure*}[50][!ht]
    \includegraphics[width=0.15\linewidth]{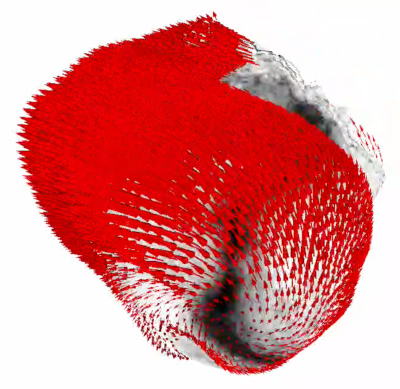}
    \caption{\textbf{Supplementary Video 13.}\\
    Focal / planar action potential wavefront (black: upstroke) propagating across ventricles during ventricular pacing, see also Figs.~\ref{fig:Pacing1}-\ref{fig:Pacing2}. The wave was induced using a flexible, thin electrode, see also Figs.~\ref{fig:SupplementDepthMaps} and \ref{fig:SupplementTrackingElectrode}. The ventricles move (red displacement vectors) and deform in response to the electrical activation wave. Our data shows that the electrical activation and mechanical motion and deformation are strongly correlated. The imaging was performed with continuous green illumination (non-ratiometric). 
    }
    \label{video:pacing1vectors}
\end{SCfigure*}

\begin{SCfigure*}[50][!ht]
    \includegraphics[width=0.15\linewidth]{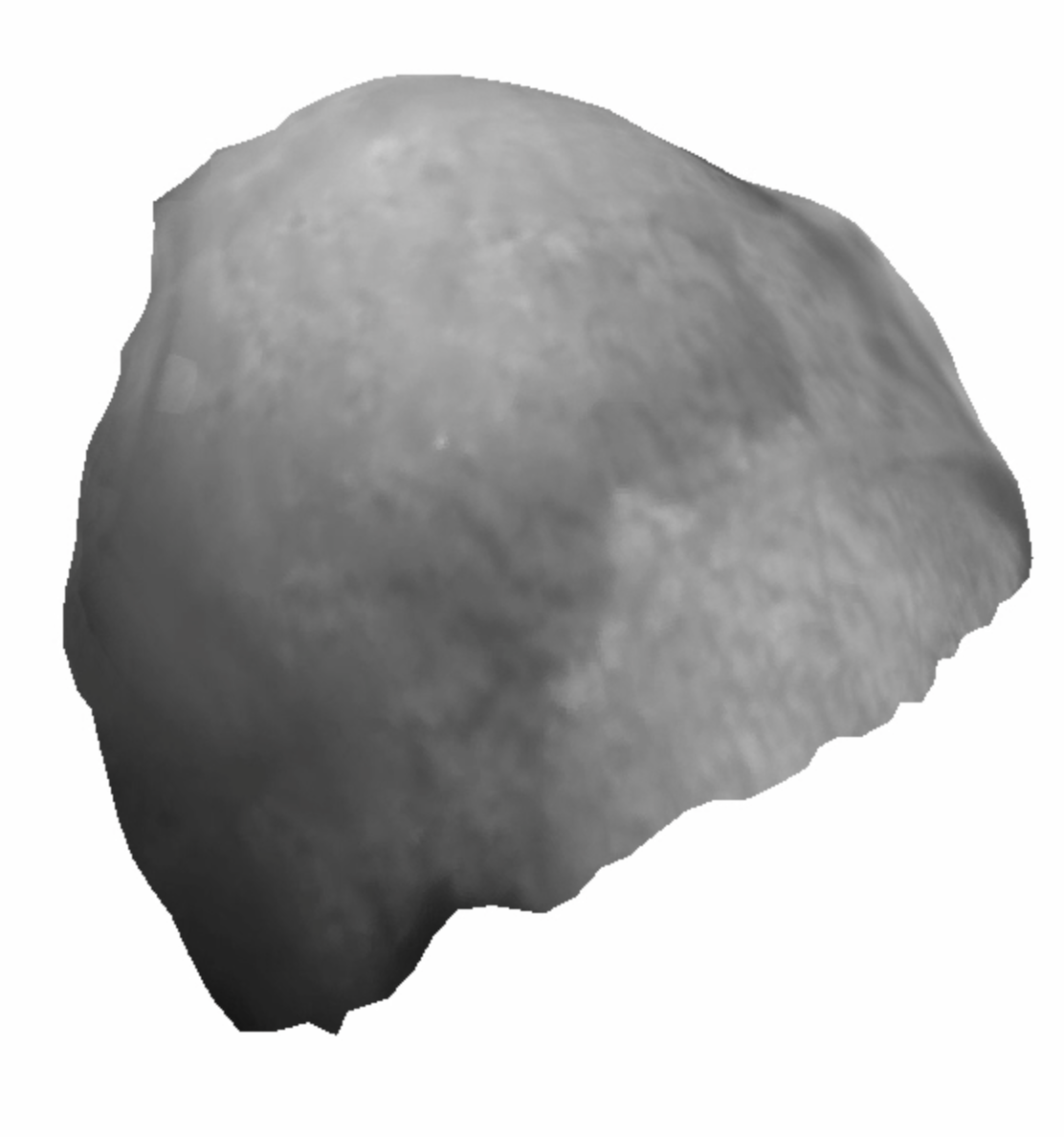}
    \caption{\textbf{Supplementary Video 14.}\\
    Focal / planar action potential wavefront (black: upstroke) propagating across ventricles during ventricular pacing, see also Figs.~\ref{fig:Pacing1}-\ref{fig:Pacing2} and \ref{fig:SupplementPacing1}-\ref{fig:SupplementPacing2}. The wave was induced using a flexible, thin electrode, see also Figs.~\ref{fig:SupplementDepthMaps} and \ref{fig:SupplementTrackingElectrode}. The ventricles move and deform in response to the electrical activation wave and exhibit a very characteristic strain-rate pattern, c.f. Fig.~\ref{fig:Pacing2}. Our data shows that the electrical activation and mechanical strain are strongly correlated. The imaging was performed with continuous green illumination (non-ratiometric). 
    }
    \label{video:pacing2}
\end{SCfigure*}

\begin{SCfigure*}[50][!ht]
    \includegraphics[width=0.15\linewidth]{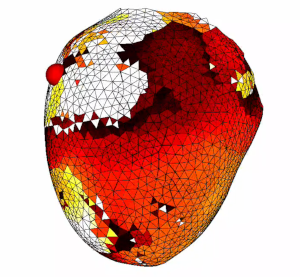}
    \caption{\textbf{Supplementary Video 15.}\\
    Electrical and mechanical activation maps during ventricular pacing, see also Fig.~\ref{fig:Pacing2}E,F). The activation maps are correlated across large parts of the ventricles. Mechanical activation times cannot be computed everywhere (white triangles), especially surrounding the pacing electrode (red sphere). Mechanical activation times were computed using the method shown in Fig.~\ref{fig:Pacing2}B,D).
    }
    \label{video:SV13}
\end{SCfigure*}

\begin{SCfigure*}[50][!ht]
    \includegraphics[width=0.15\linewidth]{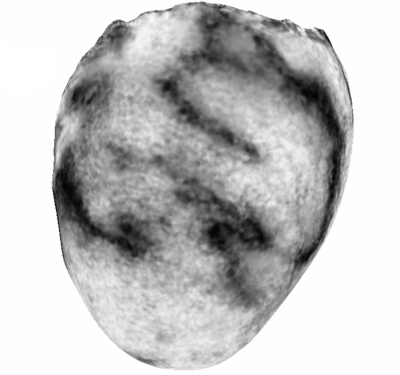}
    \caption{\textbf{Supplementary Video 16.}\\
    Ventricular fibrillation in a (residually) contracting isolated rabbit heart imaged using panoramic voltage-sensitive optical mapping combined with numerical 3D motion tracking, see also Figs.~\ref{fig:VF1} and \ref{fig:VF2}A). The imaging was performed with continuous green illumination. The 3D video shows spiral-like action potential vortex waves moving across the ventricular surface. This example exhibits less motion than the example shown in Supplementary Video \ref{video:VF2}. The optical signals were amplified using a pixel-wise sliding-window normalization.
    }
    \label{video:VF1}
\end{SCfigure*}

\begin{SCfigure*}[50][!ht]
    \includegraphics[width=0.15\linewidth]{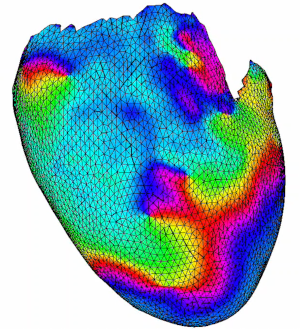}
    \caption{\textbf{Supplementary Video 17.}\\
    Ventricular fibrillation in a (residually) contracting isolated rabbit heart imaged using panoramic voltage-sensitive optical mapping combined with numerical 3D motion tracking, see also Fig.~\ref{fig:VF2}B,C). The imaging was performed with continuous green illumination. The 3D video shows spiral-like action potential vortex waves moving across the ventricular surface. This example exhibits more motion than the example shown in Supplementary Video \ref{video:VF1}. The optical signals were amplified using a pixel-wise sliding-window normalization.
    }
    \label{video:VF2}
\end{SCfigure*}

\end{document}